\documentclass[fleqn,usenatbib]{mnras}
\usepackage{newtxtext,newtxmath}
\usepackage[T1]{fontenc}
\usepackage{mathtools}
\usepackage{cuted}
\usepackage[normalem]{ulem}

\title[Albedo variegation on Comet 67P]{Albedo variegation on Comet 67P/Churyumov--Gerasimenko}

\author[Bj\"{o}rn J. R. Davidsson et al.]{
Bj\"{o}rn J. R. Davidsson,$^{1}$\thanks{E-mail: bjorn.davidsson@jpl.nasa.gov}
Bonnie J. Buratti$^{2}$
and Michael D. Hicks$^{3}$\\
\\
$^{1}$Jet Propulsion Laboratory, California Institute of Technology,  M/S 183--401, 4800 Oak Grove Drive, Pasadena, CA 91109, USA\\
$^{2}$Jet Propulsion Laboratory, California Institute of Technology,  M/S 183--601, 4800 Oak Grove Drive, Pasadena, CA 91109, USA\\
$^{3}$Jet Propulsion Laboratory, California Institute of Technology,  M/S 183--301, 4800 Oak Grove Drive, Pasadena, CA 91109, USA\\
}

\date{Accepted 2022 August 25. Received 2022 August 02; in original form 2022 May 17}

\pubyear{2022}

\begin{document}
\label{firstpage}
\pagerange{\pageref{firstpage}--\pageref{lastpage}}
\maketitle

\begin{abstract}
We here study the level of albedo variegation on the nucleus of Comet 67P/Churyumov--Gerasimenko. This is done by fitting the parameters 
of a standard photometric phase function model to disk--average radiance factor data in images acquired by the \emph{Rosetta}/OSIRIS Narrow Angle Camera 
in the orange filter. Local discrepancies between the observed radiance factor and the disk--average solution are interpreted as a proxy $\mathcal{W}$ 
of the local single--scattering albedo. We find a wide range $0.02 \stackrel{<}{_{\sim}}\mathcal{W}\stackrel{<}{_{\sim}} 0.09$ around an average 
of $\mathcal{W}=0.055$. The observed albedo variegation is strongly correlated with nucleus morphology -- smooth terrain is brighter, and consolidated 
terrain is darker, than average. Furthermore, we find that smooth terrain darken prior to morphological changes, and that stratigraphically low terrain 
(with respect to the centre of each nucleus lobe) is brighter than stratigraphically high terrain. We propose that the observed albedo variegation is 
due to differences in porosity and the coherent effect: compaction causes small brighter particles to act collectively as larger optically effective particles, 
that are darker. Accordingly, we consider the dark consolidated terrain materials more compacted than smooth terrain materials, and darkening of the latter is 
due to subsidence. 
\end{abstract}


\section{Introduction} \label{sec_intro}

The ESA \emph{Rosetta} spacecraft orbited its primary target Comet 67P/Churyumov--Gerasimenko (hereafter, 67P) from 
August 2014 to September 2016 \citep{glassmeieretal07,tayloretal17}. The scientific camera system OSIRIS \citep{kelleretal07} documented 
the nucleus and coma with unprecedented spatial resolution in numerous spectral bands. The basic physical and morphological properties 
of the nucleus, derived from OSIRIS imaging, are summarised by \citet{sierksetal15} and \citet{thomasetal15a}. Considering the small 
dimensions of the nucleus ($\sim 4\,\mathrm{km}$ diameter), its surface is surprisingly diverse. Detailed analysis of various landforms and 
their properties \citep{thomasetal15a,elmaarryetal15,elmaarryetal16} resulted in the definition and characterisation of numerous named 
morphological units, as summarised by \citet{thomasetal18}. 

The nucleus consists of a large and a small lobe (sometimes referred to in the literature as the `body' and `head', respectively), joined at a thin neck. 
The northern hemisphere of both lobes contains numerous systems of terraces. The terrace cliffs (prominent examples include Hathor, Aswan, and Serqet) 
consist of heavily fractured material known as `consolidated terrain', that appears rock--like in images because of its faceted and angular texture. 
Terrace plateaus, and other horizontal surfaces with respect to the local gravity field (represented by Hapi, Imhotep, and large portions of Seth, Ash, and Ma'at) 
constitute rather featureless `smooth terrain', that appears sand--like because low--resolution images rarely reveal morphological details other than dunes and other 
gently undulating structures. High--resolution images show that smooth terrains are formed by piles of millimetre--decimetre--sized chunks \citep{mottolaetal15,pajolaetal17b}. 
Ubiquitous outcropping pieces of consolidated terrain throughout the northern hemisphere suggest such terrain is not confined to cliffs but is present everywhere, 
though often covered by a thin (metres) layer of smooth--terrain material. However, the southern hemisphere almost exclusively consists of consolidated terrain, and 
smooth--terrain patches are few and small. 

The ultimate goal of studying the nucleus morphology is to understand the processes that govern the evolution of comet nuclei subjected to 
solar heating. Understanding solar--driven evolution (i.~e., comet activity) is a prerequisite when attempting to define and characterise more 
primordial properties of the nucleus, that inform on its formation and early evolution (e.~g., whether large--scale collisional processing has taken place, or if 
comets preserve more primitive structures). Such primordial properties feed into higher--level questions about the environments where comets formed and 
evolved, with consequences for our understanding of protoplanetary discs and planetary formation.

The photometric properties of the nucleus offer information that is complementary to morphology studies by providing insights on composition and micro--physical 
characteristics of the surface material. Previous studies of OSIRIS spectrophotometry has primarily generated spectral slope maps and RGB 
colour maps \citep[e.~g.,][]{felleretal16,felleretal19,fornasieretal15,fornasieretal16,fornasieretal17,fornasieretal19,fornasieretal21,hasselmannetal17,oklayetal16}. 
Such work has often focused on finding and characterising exposed surface ice, which is rare on the surface that is dominated by refractories. 

However, there are aspects that are not necessarily captured in such maps. Visual inspection of images often reveal differences in intensity of the 
reflected light within or among regions. Though many of those intensity variations likely are caused by changes in illumination and 
viewing conditions across the image, they still appear in portions of the images that have similar incidence, emergence, and phase angles according to the nucleus 
shape model. That suggests that there may exist an intrinsic nucleus albedo variegation. Furthermore, such suspected albedo variegation often appears in regions 
where there are no obvious changes in spectral slope. This suggests that there exists a relatively unexplored source of information in OSIRIS images.

In this paper we take the first steps in exploring albedo variegation in OSIRIS images of Comet 67P, by identifying disk--averaged solutions to a standard 
photometric model (by fitting a nucleus phase function), and then studying the spatial variation of the residuals between the actual image and the average 
photometric model solution. Ideally, such an approach removes intensity variation that merely arises because of illumination and viewing conditions (which may locally 
suppress or enhance the influence of roughness, the opposition effect, and/or the single--particle scattering phase function), and isolates the intrinsic changes in albedo. 

We present our image selection in section~\ref{sec_images} and describe our methodology and light scattering model in section~\ref{sec_method}. 
Results are presented in section~\ref{sec_results} and discussed in section~\ref{sec_discussion}. Finally, our conclusions are summarised in section~\ref{sec_conclusions}.

\section{Image selection} \label{sec_images}

In order to obtain the highest possible resolution, we here rely exclusively on imaging from the OSIRIS Narrow Angle Camera (NAC). Because our 
goal is to construct a nucleus phase function, we need the largest possible phase angle coverage, ranging from zero--phase to 
large ($\alpha\approx 90^{\circ}$) phase angles. Ideally, such a phase function could be used to study local albedo variegation 
in a large number of images, acquired at any phase angle.

In reality, building such a phase function is difficult. Observations at very low phase angles were primarily obtained during three 
near--nucleus flybys, that took place on 2015 February 14, 2015 March 28 and 2016 April 9--10. Most flyby images were acquired with the orange ($649.2\,\mathrm{nm}$), 
blue ($480.7\,\mathrm{nm}$), and red ($743.7\,\mathrm{nm}$) filters, in combination with a neutral density filter ($640.0\,\mathrm{nm}$) with just 5 per cent 
transmission, used to avoid saturation when the spacecraft is very close to the bright nucleus (these filter combinations are referred to as F82, F84, and F88). Unfortunately, these filter combinations were 
rarely used during other phases of the mission. Nevertheless, we managed to find F82 images covering the $0\leq\alpha\leq 70^{\circ}$ region, as given in Table~\ref{tab1}. We here 
exclusively use level 4f images, that provide radiance factors and have been corrected for solar stray light, in--field straylight, and geometric distortion during absolute calibration. 
The radiance factors typically have uncertainties of 1--1.7 per cent due to statistical errors in the radiometric calibration \citep{tubianaetal15b}. We also use the associated derived georeferencing data (level 5 files), in the form of 
incidence, emergence, and phase angles for each on--nucleus pixel. Both types of files have been downloaded from the NASA Planetary Data System, PDS\footnote{\url{https://pds-smallbodies.astro.umd.edu/data\_sb/missions/rosetta/index\_OSIRIS.shtml}}.

\begin{table}
\begin{center}
\begin{tabular}{||l|l|l|l|l||}
\hline
\hline
ID & File name & Phase angle $\alpha$ & $\mathcal{R}$ & $R_{\rm co}$\\
\hline
F82a & n20160409t235945283 & 0--$2.4^{\circ}$ & 0.53 & $4\cdot 10^{-2}$\\ 
F82b & n20160409t233245258 & 2.8--$5.2^{\circ}$ & 0.53 & $4\cdot 10^{-2}$\\ 
F82c & n20160410t012854198 & 8.8--$11.1^{\circ}$ & 0.55 & $2\cdot 10^{-2}$\\ 
F82d & n20160409t214844745 & 13.9--$16.1^{\circ}$ & 0.52 & $2\cdot 10^{-2}$\\ 
F82e & n20140803t111934576 & 40.3--$41.0^{\circ}$ & 5.40 & $8\cdot 10^{-3}$\\ 
F82f & n20150328t131531164 & 43.0--$44.2^{\circ}$ & 0.26 & $5\cdot 10^{-3}$\\ 
F82g & n20150328t161850365 & 57.6--$59.8^{\circ}$ & 0.35 & $5\cdot 10^{-3}$\\ 
F82h & n20150328t063149463 & 67.4--$69.6^{\circ}$ & 0.55 & $5\cdot 10^{-3}$\\ 
\hline 
\hline
\end{tabular}
\caption{This table summarises the OSIRIS NAC F82 filter images used in this work. ID--numbers, only used in this paper, are 
provided for easy reference. The formal PDS archive file name root is provided, to which should be added the tags `id4ff82.img' for 
radiance factor files, and `id50f82.img' for georeferencing meta--data files. The phase angles represented in each image are provided 
(often covering several degrees due to the proximity of \emph{Rosetta} to the nucleus). The average resolution $\mathcal{R}$ ($\mathrm{m\,px^{-1}}$) is provided.  Radiance cut--offs $R_{\rm obs}>R_{\rm co}$ are 
applied to avoid radiation from diffusely lit shadows.}
\label{tab1}
\end{center}
\end{table}

The orange filter was used extensively throughout the Rosetta mission, albeit in combination with a visual far focus plate (then referred to as F22), 
that has a substantially higher transmission ($>95$ per cent) than the neutral density filter. Because radiance factors are calibrated to be 
valid at the central wavelength,  which is the same for F82 and F22, they are photometrically equivalent. We verify that the phase function derived for F82 
images works equally well for F22 images, which allows us to extend the albedo variegation study to F22 images as well. A selection of F22 images for a 
variety of phases ($24^{\circ}\leq\alpha\leq 93^{\circ}$) are listed in Table~\ref{tab2}, of which some are of particular geophysical interest.

\begin{table}
\begin{center}
\begin{tabular}{||l|l|l|l|l||}
\hline
\hline
ID & File name & Phase angle $\alpha$ & $\mathcal{R}$ & $R_{\rm co}$\\
\hline
F22a & n20160409t201351167 & 24.4--$26.7^{\circ}$ & 0.56 & $1\cdot 10^{-2}$\\ 
F22b & n20160410t052428077 & 30.1--$32.4^{\circ}$ & 0.69 & $5\cdot 10^{-3}$\\ 
F22l & n20140830t034253546 & 41.2--$43.4^{\circ}$ & 1.03 & $4\cdot 10^{-4}$\\ 
F22c & n20160608t143426745 & 46.5--$48.8^{\circ}$ & 0.54 & $2\cdot 10^{-3}$\\ 
F22d & n20150320t011747593 & 50.7--$52.7^{\circ}$ & 1.49 & $2\cdot 10^{-3}$\\ 
F22e & n20150228t044349351 & 61.7--$63.7^{\circ}$ & 2.01 & $5\cdot 10^{-4}$\\ 
F22f & n20160127t072519732 & 63.6--$65.1^{\circ}$ & 1.34 & $5\cdot 10^{-4}$\\ 
F22g & n20160210t140154858 & 64.6--$67.8^{\circ}$ & 0.90 & $5\cdot 10^{-4}$\\ 
F22h & n20160210t135354886 & 64.6--$66.8^{\circ}$ & 0.90 & $1\cdot 10^{-4}$\\ 
F22i & n20150328t191247644 & 67.2--$69.5^{\circ}$ & 0.53 & $2\cdot 10^{-4}$\\ 
F22j &  n20160101t051315898 & 88.9--$90.8^{\circ}$ & 1.50 & $2\cdot 10^{-4}$\\ 
F22k & n20141210t062855791 & 90.7--$92.9^{\circ}$& 0.37 & $5\cdot 10^{-4}$ \\ 
\hline 
\hline
\end{tabular}
\caption{OSIRIS NAC F22 filter images used in this work. PDS archive file name roots are here added to `id4ff22.img' (radiance factor) or 
`id50f22.img' (georeference data). See Table~\ref{tab1} caption for further explanations.}
\label{tab2}
\end{center}
\end{table}

Portions of the imaged nucleus are in shadow due to its irregular shape. By definition, direct solar illumination does not enter shadows, 
but the shadows are still lit by diffuse radiation scattered off nearby sun--lit portions of the nucleus. The OSIRIS NAC dynamic range is so 
large, that it easily picks up the multiple--scattered radiation from shadows. When properly stretched, images of shadows readily 
reveal substantial morphological detail.

We need to remove the regions containing (as we call them) 'lit shadows' because the 
applied light--scattering model (see section~\ref{sec_method}) is only defined for directly illuminated portions of the nucleus. 
For this reason, each image was scrutinised individually, and the radiance of lit shadows was estimated in each case. 
We apply the radiance cut--offs $R_{\rm co}$ given in Tables~\ref{tab1} and \ref{tab2}, so that only observed radiance 
factors $R_{\rm obs}>R_{\rm co}$ are applied in this work.

\section{Methodology} \label{sec_method}

We here apply a simple version of the \citet{hapke93} reflectance model, that considers the (shadow--hiding) opposition effect, a double--lobed Henyey--Greenstein 
single--particle phase function, and surface roughness, in addition to the single--scattering albedo of the surface material. This basic 
model, to be described in the following, is then simplified in several stages. Those simplified expressions are used in order to 
fit groups of model parameters consecutively to suitable sub--sets of data, thereby gradually recovering the full complexity of the original model. 

The radiance factor (defined for each on--nucleus pixel of each image) of the full model is given by

\begin{equation} \label{eq:01}
\begin{array}{c}
\displaystyle R_{\rm rough}(w,\,h,\,\xi,\,\bar{\theta},\,i,\,e,\,\alpha)=\frac{w}{4}\frac{\mu_0'(\bar{\theta},\,i,\,e,\,\alpha)S(\bar{\theta},\,i,\,e,\,\alpha)}{\mu_0'+\mu'(\bar{\theta},\,i,\,e,\,\alpha)}\times\\
\\
\displaystyle \left\{[1+B(h,\,\alpha)]p(\xi,\,c,\,\alpha)+H(w,\,\mu_0')H(w,\,\mu')-1\right\}=\\
\\
\displaystyle wD(h,\,\xi,\,\bar{\theta},\,i,\,e,\,\alpha)
\end{array}
\end{equation}
where parameter dependencies only appear once per function for brevity. Here, $i$ is the incidence angle (between the outward surface normal and the Sun), $e$ is 
the emergence angle (between the outward surface normal and the observer), and $\alpha$ is the phase angle (between the Sun and the observer, as seen from the nucleus surface at the pixel). 
The single--scattering albedo $w=P_{\rm S}/P_{\rm E}$ is the ratio between the power $P_{\rm S}$ scattered by the particle  and the total power $P_{\rm E}=P_{\rm S}+P_{\rm A}$ affected by the particle, 
where $P_{\rm A}$ is the power absorbed by the particle. The opposition effect term is here given by
\begin{equation} \label{eq:02}
B(h,\,\alpha)=B_0\left[1+\frac{\tan(\alpha/2)}{h}\right]^{-1},
\end{equation}
that is dependent on the angular width of the opposition effect $h$, the opposition effect amplitude $B_0$, and phase angle. Comet 67P has no detectable 
coherent--backscattering opposition effect \citep{fornasieretal15,hasselmannetal17}, and for the shadow--hiding opposition effect the amplitude is constrained to $0\leq B_0\leq 1$. 
Other mechanisms that may elevate $B_0$ above unity \citep[intrinsic opposition effect of monomers bigger than the wavelength that forms aggregates; glory for spherical particles; 
internal reflections for optically perfect and transparent grains;][]{hapke93}, are not expected in cometary material because the monomers are sub--micron, highly irregular, 
dark and opaque, based on MIDAS and COSIMA documentation of resolved particles \citep{bentleyetal16,hilchenbachetal17}. Finally, amplitude reduction caused by internal 
scattering \citep{hapke93} is not expected either, hence we assume $B_0=1$.

The single--particle phase function, parameterised by the cosine asymmetry factor $\xi=-\langle\cos\alpha\rangle$ is here taken as \citep{hapke12}
\begin{equation} \label{eq:03}
\begin{array}{c}
\displaystyle p(\xi,\,c,\,\alpha)=\frac{1+c}{2}\frac{1-(\xi/c)^2}{[1+2(\xi/c)\cos\alpha+(\xi/c)^2]^{3/2}}+\\
\\
\displaystyle \frac{1-c}{2}\frac{1-(\xi/c)^2}{[1-2(\xi/c)\cos\alpha+(\xi/c)^2]^{3/2}}.\\
\end{array}
\end{equation}
Here, on the right--hand side, the first  term is the back--scattering lobe (dominating when $\alpha<90^{\circ}$ and $\xi<0$), and the second term is the forward--scattering lobe, with their 
relative weights regulated by the parameter $c$. We apply the two--stream approximation to the Chandrasekhar function,
\begin{equation} \label{eq:04}
H(w,\,x)=\frac{1+2x}{1+2x\sqrt{1-w}}.
\end{equation}
Roughness is parameterised by $\bar{\theta}$ and acts on the radiance factor through the functions $\mu_0'$, $\mu'$, and $S$. These very 
complex mathematical expressions are provided in appendix~\ref{appendix01}. Finally, equation~(\ref{eq:01}) defines the function $D=R_{\rm rough}/w$, 
to be discussed further later on. 

The first, and most important, simplification that can be made for equation~(\ref{eq:01}) concerns pixels for which roughness effects are 
negligible. In such cases,
\begin{equation} \label{eq:05}
\left\{\begin{array}{l}
\displaystyle \lim_{\bar{\theta}\to 0} \mu_0'(\bar{\theta},\,i,\,e,\,\alpha) \rightarrow \mu_0(i)=\cos i\\
\\
\displaystyle \lim_{\bar{\theta}\to 0} \mu'(\bar{\theta},\,i,\,e,\,\alpha) \rightarrow \mu(e)=\cos e\\
\\
\displaystyle \lim_{\bar{\theta}\to 0}S(\bar{\theta},\,i,\,e,\,\alpha) \rightarrow 1\\
\\
\displaystyle \lim_{\bar{\theta}\to 0}R_{\rm rough} \rightarrow R_{\rm flat}.\\
\end{array}\right.
\end{equation}
In such cases, equation~(\ref{eq:01}) simplifies to 
\begin{equation} \label{eq:06}
R_{\rm flat}=\frac{w}{4}\frac{\mu_0}{\mu_0+\mu}\left\{[1+B(h,\,\alpha)]p(\xi,\,c,\,\alpha)+H(w,\,\mu_0)H(w,\,\mu)-1\right\}
\end{equation}
Equations~(\ref{eq:01}) and (\ref{eq:06}) allow us to directly calculate the level of dimming due to roughness (reduction of $R_{\rm rough}$ with 
respect to $R_{\rm flat}$), for a given pixel at a given $\bar{\theta}$. Importantly, this makes it possible to identify pixels in an image that 
suffers negligible dimming, even when the $\bar{\theta}$ value is significant. For such a sub--set of pixels, $R_{\rm flat}$ would closely 
approximate $R_{\rm rough}$, even when the surface is very rough. Such conditions prevail virtually globally when $\alpha\stackrel{<}{_{\sim}} 20^{\circ}$. 
However, even at $\alpha\approx 90^{\circ}$ one may have $R_{\rm flat}\approx R_{\rm rough}$, for certain combinations of $i$ and $e$.

Two further simplifications can be made: 1) when $w\ll 1$, then $H(w,\,\mu_0)H(w,\,\mu)\approx 1$; 2) for $\alpha\stackrel{<}{_{\sim}} 90^{\circ}$ the 
forward--scattering contribution to the single--particle phase function is very small, thus $c=1$ can be applied. The corresponding radiance factor 
is now
\begin{equation} \label{eq:07}
 R_{\rm approx}(w,\,h,\,\xi,i,\,e,\,\alpha)=\frac{w}{4}\frac{\mu_0}{\mu_0+\mu}[1+B(h,\,\alpha)]p(\xi,\,\alpha,\,c=1).
\end{equation}
This is a case where the photometric function is separable into a Lommel--Seeliger disk function, and a phase function that carries 
the full phase angle dependence \citep[see, e.~g.,][]{schroederetal13}. Equation~(\ref{eq:07}) can trivially be re--arranged as
\begin{equation} \label{eq:08}
\frac{4(\mu_0+\mu)R_{\rm approx}}{\mu_0}=w[1+B(h,\,\alpha)]p(\xi,\,\alpha,\,c=1).
\end{equation}
The significance of equation~(\ref{eq:08}) is evident once we realise that the right--hand side now only depends on a single geometric 
parameter, the phase angle $\alpha$, and the three physical parameters $\{w,\,h,\,\xi\}$. Inspired by the left--hand side we may use 
the observed radiance factor $R_{\rm obs}$, along with known $\{i,\,e\}$--values for each pixel to generate a $\tilde{Q}_{\rm obs}$--value 
for a given pixel,
\begin{equation} \label{eq:09}
\tilde{Q}_{\rm obs}=\frac{4(\cos i+\cos e)R_{\rm obs}}{\cos i}.
\end{equation}
In essence, we here make a Lommel--Seeliger photometric correction to the observed radiance factors, in order to 
refer all pixels to the same reference viewing and illumination geometry. Furthermore, we may group pixels having very similar $\alpha$--values into bins, and average the corresponding $\tilde{Q}_{\rm obs}$--values 
within each bin. Thereby, an empirical function $Q_{\rm obs}(\alpha)$ is obtained that solely depends on the phase angle $\alpha$,
\begin{equation} \label{eq:10}
Q_{\rm obs}(\alpha)=\langle\tilde{Q}_{\rm obs}\rangle_{\alpha}=\Big\langle\frac{4R_{\rm obs}(\cos i+\cos e)}{\cos i}\Big\rangle_{\alpha}.
\end{equation}
At this point, the right--hand side of equation~(\ref{eq:08}) is used to formulate a function $Q_{\rm fit}=Q_{\rm fit}(\alpha)$ of the 
phase angle, 
\begin{equation} \label{eq:11}
Q_{\rm fit}=w_a[1+B(h_a,\,\alpha)]p(\xi_a,\,\alpha,\,c=1)
\end{equation}
that ideally should equate the empirical function, i.~e. $Q_{\rm fit}(\alpha)\approx Q_{\rm obs}(\alpha)$, as well as possible, for some parameter 
combination $\{w_a,\,h_a,\,\xi_a\}$. That parameter combination would be taken as the best available \emph{disk--averaged} single--scattering albedo, 
opposition--width parameter, and cosine asymmetry factor of the nucleus. 

\begin{figure*}
\centering
\begin{tabular}{cc}
\scalebox{0.15}{\includegraphics{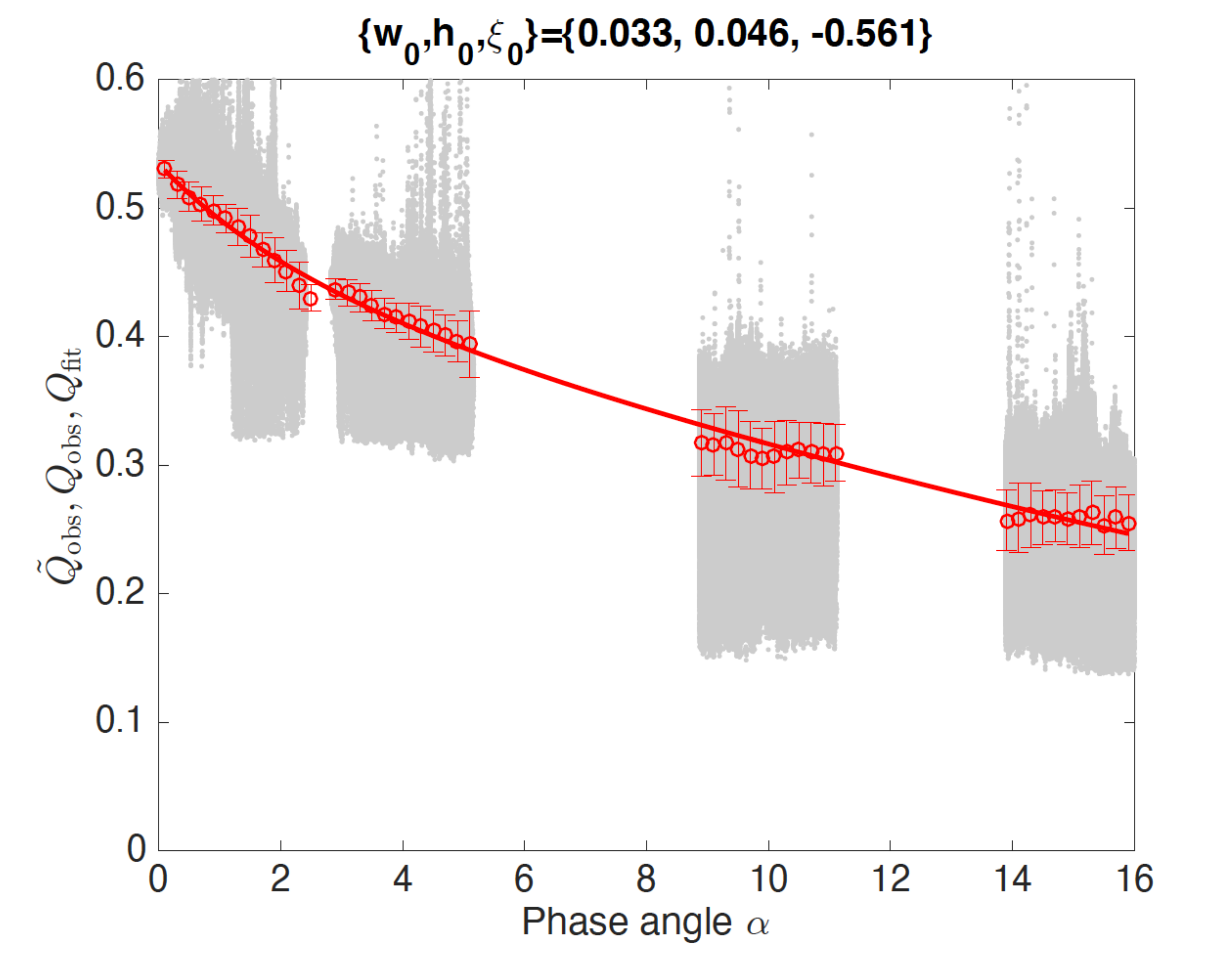}} & \scalebox{0.15}{\includegraphics{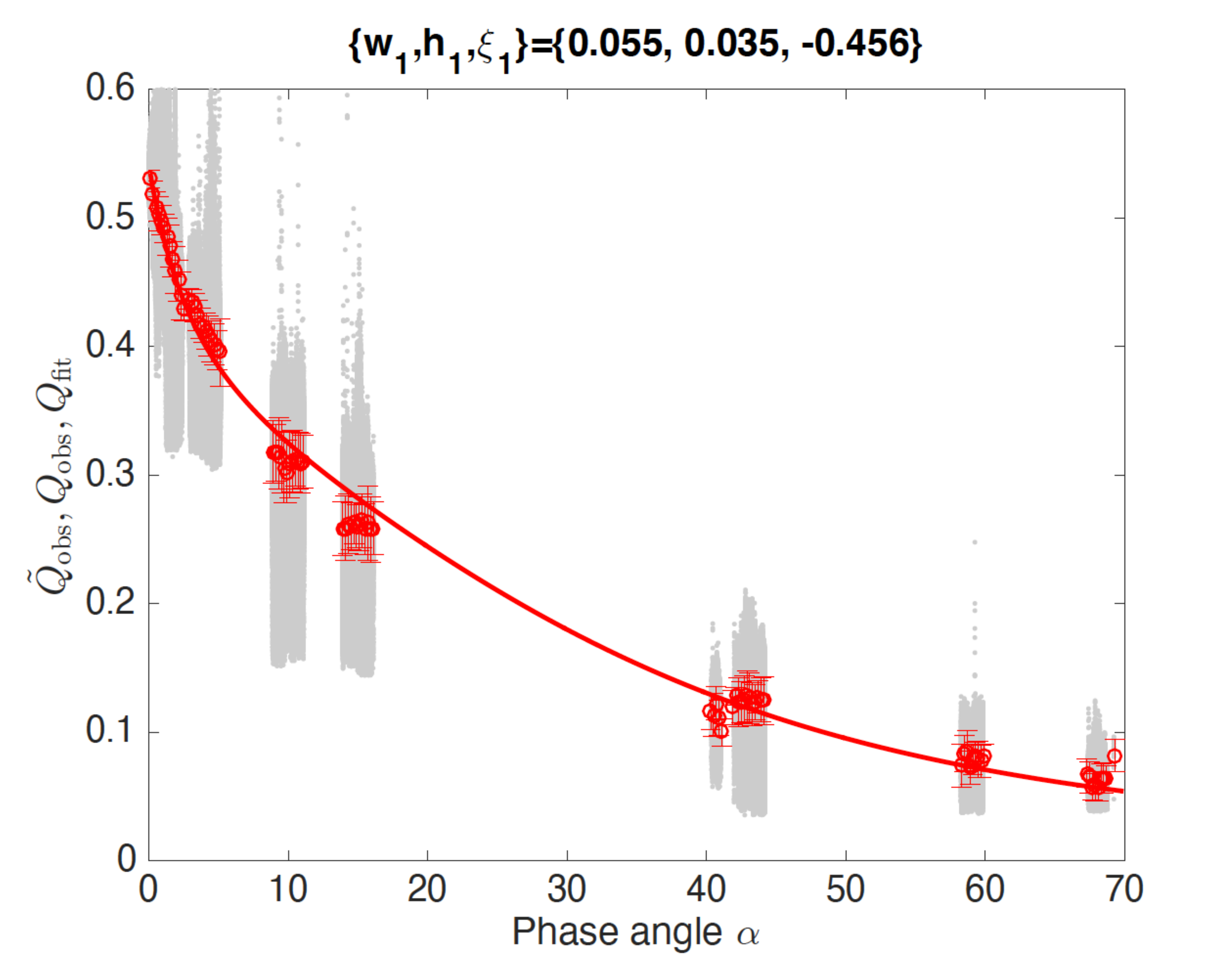}}\\
\scalebox{0.15}{\includegraphics{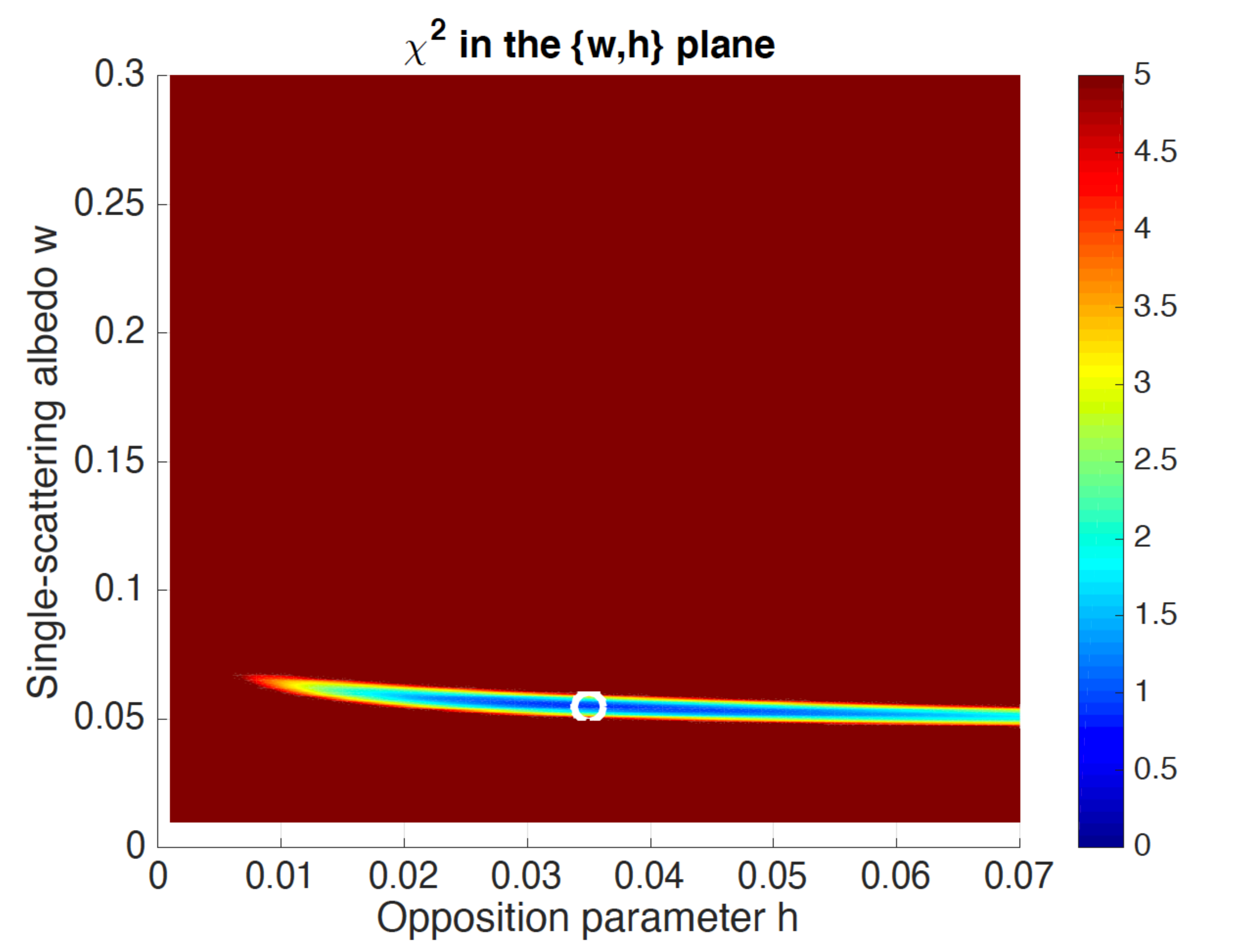}} & \scalebox{0.15}{\includegraphics{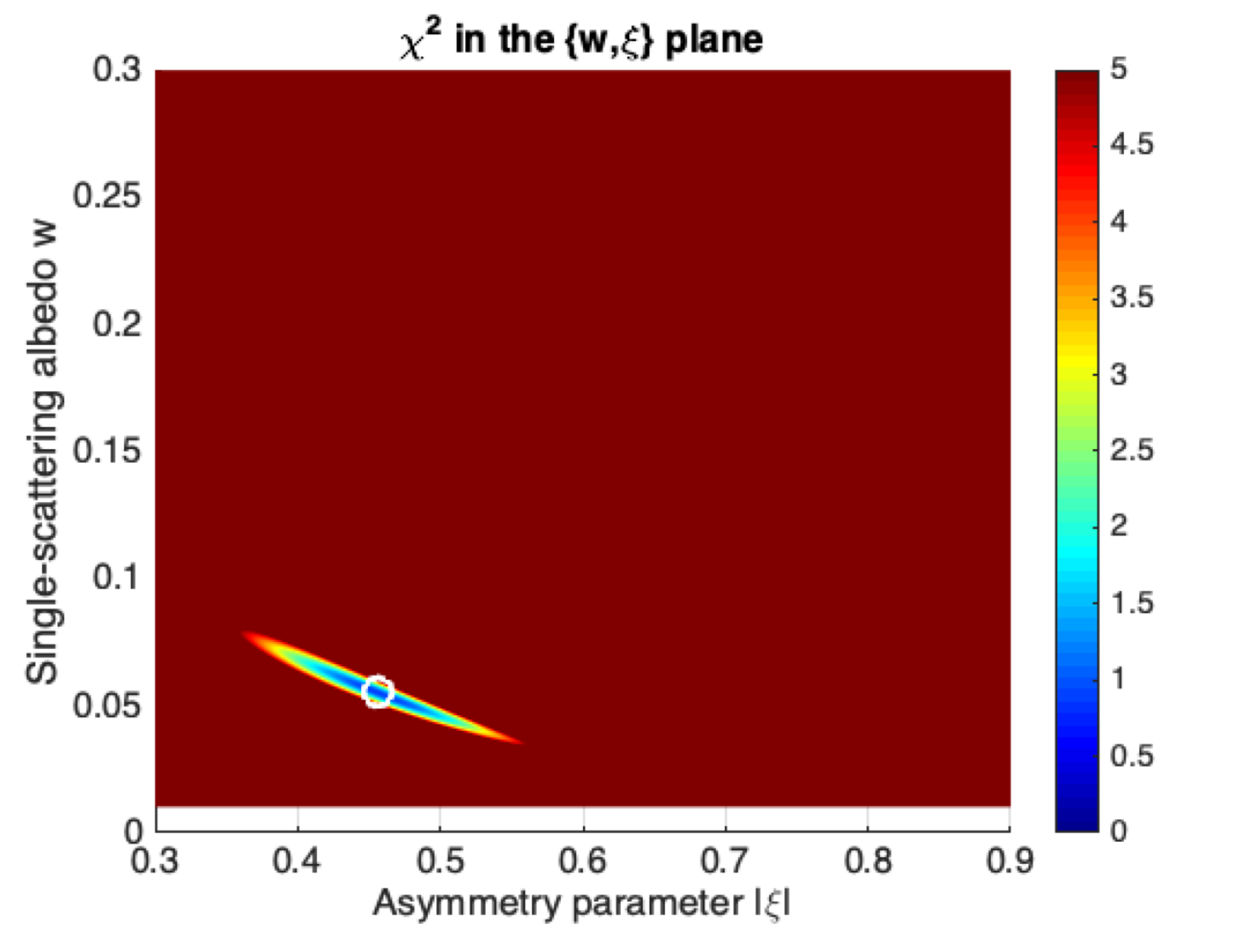}}\\
\end{tabular}
     \caption{\emph{Upper left:} The $\tilde{Q}_{\rm obs}$ values in images F82a--d according to equation~(\ref{eq:09}), as function of phase angle $\alpha$, are shown as grey dots. When binning these data with 
$\Delta\alpha=0.2^{\circ}$ resolution, the red circle averages and red bar standard deviations are obtained, called $Q_{\rm obs}(\alpha)$. The theoretical function given by equation~(\ref{eq:11}) achieves a $\chi^2$ minimum with 
respect to $Q_{\rm obs}(\alpha)$ for $\{w_0,\,h_0,\,\xi_0\}=\{0.033,\,0.046,\,-0.561\}$, here shown as a red curve called $Q_{\rm fit}(\alpha)$. \emph{Upper right:} The $\tilde{Q}_{\rm obs}$ values (grey dots) and 
$Q_{\rm obs}(\alpha)$ averages with standard deviations (red circles and bars) are here shown for all F82a--h images, limited to pixels for which surface roughness would dim the radiance factor by at most 2 per cent if 
$\bar{\theta}=25^{\circ}$. The best--fit model $Q_{\rm fit}(\alpha)$, for which $\{w_1,\,h_1,\,\xi_1\}=\{0.055,\,0.035,\,-0.456\}$ is seen as a red curve. \emph{Lower left:} the $\chi^2$ function in the $\{w,\,h\}$--plane 
at $\xi=-0.456$ for the final $\{w_1,\,h_1,\,\xi_1\}$ fit. \emph{Lower right:} the $\chi^2$ function in the $\{w,\,\xi\}$--plane at $h=0.035$ for the final $\{w_1,\,h_1,\,\xi_1\}$ fit.}
     \label{fig_4pic_8pic_fits}
\end{figure*}

With this mathematical background, we now summarise our procedure for retrieving the disk--averaged nucleus parameters, and 
creating the single--scattering albedo proxy maps. We apply these steps to the F82 filter images in Table~\ref{tab1}.

\begin{enumerate}
\item Calculate $Q_{\rm obs}(\alpha)$ for a limited data set for which roughness effects likely are insignificant ($\alpha\leq 16^{\circ}$, $i\leq 60^{\circ}$, $e\leq 60^{\circ}$), according to equation~(\ref{eq:10}). 
Fit a function $Q_{\rm fit}(\alpha)$ to $Q_{\rm obs}(\alpha)$, as defined by equation~(\ref{eq:11}), thereby obtaining a \emph{first estimate} of the disk--averaged parameters ($a=0$), or $\{w_0,\,h_0,\,\xi_0\}$.
\item For the  $\{w_0,\,h_0,\,\xi_0\}$ parameter combination and an assumed high level of roughness ($\bar{\theta}=25^{\circ}$), identify pixels in images on the $0^{\circ}\leq\alpha\leq 70^{\circ}$ phase 
angle interval, for which a radiance factor reduction of $\leq 2$ per cent due to roughness is expected (by comparing $R_{\rm flat}$ from equation~\ref{eq:06} to $R_{\rm rough}$ from equation~\ref{eq:01}). Call this subset $S_1$.
\item For subset $S_1$, calculate a new $Q_{\rm obs}(\alpha)$ on the $0^{\circ}\leq\alpha\leq 70^{\circ}$ phase angle interval, and fit a new function $Q_{\rm fit}(\alpha)$ to $Q_{\rm obs}(\alpha)$, 
resulting in a final disk--averaged parameters set ($a=1$), or $\{w_1,\,h_1,\,\xi_1\}$.
\item For the $\{w_1,\,h_1,\,\xi_1\}$ fit and an assumed  $\bar{\theta}=25^{\circ}$, identify pixels in high--phase images where roughness effects 
are expected to be significant (resulting in a radiance factor reduction of, e.~g., 10--30 per cent for $R_{\rm rough}$ compared to $R_{\rm flat}$). Call this subset $S_2$.
\item For subset $S_2$, and while applying the $\{w_1,\,h_1,\,\xi_1\}$ parameter combinations, determine the best--fit disk--averaged $\bar{\theta}_1$ value,
by finding the best possible match between $R_{\rm rough}$ from equation~(\ref{eq:01}) to the observed radiance factors $R_{\rm obs}$, on a pixel--by--pixel basis. 
\item With the best available average solution $\{h_1,\,\xi_1,\,\bar{\theta}_1\}$ applied to all pixels above the lit--shadow threshold, calculate 
the function $D$ in equation~(\ref{eq:01}). By analogy to $w=R_{\rm rough}/D$ from the theoretical expression, calculate a local (pixel--by--pixel) 
single--scattering albedo proxy $\mathcal{W}=R_{\rm obs}/D$ by requiring that the model $\{\mathcal{W},\,h_1,\,\xi_1,\,\bar{\theta}_1\}$ radiance factor 
matches the observed one exactly. We caution that $\mathcal{W}$ is identical to the local single--scattering albedo only if 
all other photometric properties are uniform across the cometary surface. Because $\mathcal{W}$ may absorb potential regional variability in $\{h,\,\xi,\,\bar{\theta}\}$, we 
refer to it as a proxy.
\end{enumerate}

These steps will be further explained, motivated, and illustrated in section~\ref{sec_results}, as the procedure is being implemented and the 
impact of various assumptions are being investigated. By comparing the filter F82 fit  $Q_{\rm fit}(\alpha)$ for the $\{w_1,\,h_1,\,\xi_1\}$ parameters, 
to a $Q_{\rm obs}(\alpha)$ function prepared from filter F22 data, in a similar fashion as in step (iv) above, we verify that the F82 phase function can be 
extended to the F22 data set. If this is the case, step (vi) can be applied to F22 images as well.

\section{Results} \label{sec_results}

\subsection{F82 filter images} \label{sec_results_F82}

In order to determine the disk--averaged $\{w,\,h,\,\xi\}$--values as well as possible, we first need to identify pixels in the images 
that are not significantly affected by surface roughness. One can identify such pixels by comparing $R_{\rm rough}$ (evaluated for some assumed 
high $\bar{\theta}$ value) with $R_{\rm flat}$, requiring that $R_{\rm rough}\approx R_{\rm flat}$. However, evaluating $R_{\rm flat}$ and 
$R_{\rm rough}$ requires that we already have a reasonable estimate of $\{w,\,h,\,\xi\}$. We obtain such an estimate by only considering 
images F82a--d (having $\alpha\leq 16.1^{\circ}$), and excluding pixels for which $i\geq 60^{\circ}$, $e\geq 60^{\circ}$, and $R_{\rm obs}\leq R_{\rm co}$. 
For each such pixel, we evaluate $\tilde{Q}_{\rm obs}$ according to equation~(\ref{eq:09}).

The $\tilde{Q}_{\rm obs}$ values, plotted as functions of the phase angle $\alpha$, are 
seen as grey dots in the upper left panel of Fig.~\ref{fig_4pic_8pic_fits}. The applied cut--offs $R_{\rm co}$ are evident. Because of the way phase functions are normalised, 
strong back--scattering asymmetry yields $p(\alpha=0)\gg 1$. That is why $\tilde{Q}_{\rm obs}$ in the upper left panel of 
Fig.~\ref{fig_4pic_8pic_fits} has numerical values much higher than the single--scattering albedo of a few per cent expected for dark cometary material.

We divide the phase angle interval into $\Delta\alpha=0.2^{\circ}$ wide bins, and average the $\tilde{Q}_{\rm obs}$ values within each bin, 
thereby obtaining $Q_{\rm obs}$. The average values for each bin are seen as red circles in the upper panel of Fig.~\ref{fig_4pic_8pic_fits} 
and the red bars show the standard deviation of the $\tilde{Q}_{\rm obs}$--values within each bin. These $Q_{\rm obs}$ values (see equation~\ref{eq:10}) line up 
rather well, and the bars are small compared to the range of $Q_{\rm obs}$. Note that the grey areas are saturated in the figure and do not convey 
the degree of concentration as well as the standard deviation bars do. The low degree of scatter among red circles and the small standard deviations suggest that the following concepts indeed are 
meaningful: 1) $Q_{\rm obs}=Q_{\rm obs}(\alpha)$, i.~e., this quantity is primarily a function of phase angle; 2)  most pixels closely adhere to a 
common photometric behaviour, described by the disk--averaged properties. Nevertheless, numerous outliers exist, suggesting that there exist 
intrinsic deviations from the baseline photometric properties, e.~g., in the form of albedo variegation.

Next, equation~(\ref{eq:11}) was evaluated for all combinations of $w$--values on the range $0.01\leq w\leq 0.3$ with resolution $\Delta w=0.001$, 
$h$--values on the range $0.001\leq h\leq 0.07$ with resolution $\Delta h=0.001$, and $\xi$--values on the range $-0.9\leq \xi\leq -0.3$ with resolution $\Delta\xi=0.001$. 
For each of these 12,242,370 different $Q_{\rm fit}=Q_{\rm fit}(w,\,h,\,\xi,\,\alpha)$ curves, the $\chi^2$ residuals with respect to the $Q_{\rm obs}$ bin values were 
calculated, i.~e.,  a brute--force grid--search for the minimum was made. The smallest $\chi^2$--value was encountered for $\{w_0,\,h_0,\,\xi_0\}=\{0.033,\,0.046,\,-0.561\}$, that we adopt as our first \emph{preliminary} 
$a=0$ disk--averaged photometric solution. The corresponding $Q_{\rm fit}=Q_{\rm fit}(w_0,\,h_0,\,\xi_0,\,\alpha)$ is seen as a red curve in the upper left panel of 
Fig.~\ref{fig_4pic_8pic_fits}. We consider this curve a good representation of the empirical data. This completes step (i) in section~\ref{sec_method}.

The purpose of this preliminary evaluation of the disk--averaged photometric properties is twofold: 1) we want to be able to evaluate $R_{\rm rough}$ and $R_{\rm flat}$ (equations~\ref{eq:01} and \ref{eq:06}) 
for a reasonable region of the parameter space, with the goal of identifying pixels that are only weakly affected by surface roughness; 2) we want to verify that our assumption $H(w,\,\mu_0)H(w,\,\mu)\approx 1$ is 
valid. Starting with the second point, we compared the magnitude of the expression $H(w_0,\,\mu_0)H(w_0,\,\mu)-1$ with that of $\{1+B(h_0,\,\alpha)\}p(\xi_0,\,\alpha,\,c=1)$, pixel--by--pixel for all F82 images in 
Table~\ref{tab1}. We find that the highest local value of this ratio is 0.17 per cent for F28a, increasing slightly with $\alpha$ to peak value of 2.4 per cent for F28h. Therefore, the first term within curled brackets in 
equation~(\ref{eq:06}) completely dominates the sum of the following two, hence they can be neglected. 

Next, we assume a level of roughness ($\bar{\theta}=25^{\circ}$) that is similar to previously fitted values for 67P: $\bar{\theta}=15.1$--$16.5^{\circ}$ \citep{felleretal16}; 
$\bar{\theta}=21$--$33^{\circ}$ \citep{hasselmannetal17}; $\bar{\theta}=28^{\circ}$ \citep{fornasieretal15}. We then require that $R_{\rm rough}$ is lowered by at most 2 per cent relative $R_{\rm flat}$, due to dimming caused by 
surface roughness. Furthermore, we require $i<85^{\circ}$, $e<70^{\circ}$, and $R_{\rm obs}>R_{\rm co}$. For images F82a--d, we find that (2.3--4.1)$\cdot 10^6$ pixels (out of a total of 
$4.19\cdot 10^6$ per image) pass this test. This confirms our previous assumption that roughness effects are minimal for those images. For the highest--phase images, F82f--h, the 
number of pixels with quasi--flat behaviour drops from $7.3\cdot 10^5$ to $2.4\cdot 10^4$ with increasing $\alpha$. Although these numbers are low, it is encouraging that portions of those 
images (sub--set $S_1$) do have negligible surface roughness effects, and therefore can be applied for the final $\{w,\,h,\,\xi\}$ determination. This completes step (ii) in section~\ref{sec_method}.

The roughness--insensitive $\tilde{Q}_{\rm obs}(\alpha)$ F82 $S_1$ data for the full $0\leq \alpha\leq 70^{\circ}$ interval are seen as grey dots in the upper right panel of Fig.~\ref{fig_4pic_8pic_fits}. 
We applied the same binning resolution to obtain $Q_{\rm obs}(\alpha)$, and the same test combinations of $\{w,\,h,\,\xi\}$ to obtain $Q_{\rm fit}(\alpha)$, as done previously 
(seen as red circles and red curve, respectively). Our final ($a=1$) best estimate of the disk--average parameters is $\{w_1,\,h_1,\,\xi_1\}=\{0.055,\,0.035,\,-0.456\}$. The lower panels in 
Fig.~\ref{fig_4pic_8pic_fits} show $\chi^2$ as function of $\{w,\,h\}$ (left) and $\{w,\,\xi\}$ (right). Both $w_1$ and $\xi_1$ are tightly constrained (as evidenced by the narrow low--$\chi^2$ regions in those 
dimensions). The $h$--value is less certain, but is securely confined to the $0.02\leq h\leq 0.05$ interval. 

That model $Q_{\rm fit}(\alpha)$ passes within the standard deviation margins of all bins $Q_{\rm obs}(\alpha)$. In some cases, the curve does not 
intercept the nominal $Q_{\rm obs}(\alpha)$ values, but differences are small. One possibility is that this version of the \citet{hapke93} photometric model is not fully adequate 
to describe the data. However, there could be physically real deviations, that would be difficult for any model to match perfectly. The most evident example is image F82e at $40.3\leq\alpha\leq 41.0^{\circ}$ 
that is somewhat displaced towards lower radiance factors, with respect to image F82f at $42.0\leq\alpha\leq 44.9^{\circ}$. That could be due to a real nucleus brightening between August 2014 and 
March 2015 \citep[the spectral slope is known to have decreased somewhat in this part of the orbit;][]{fornasieretal16}. Alternatively, or in addition, the fields of view are sometimes rather small, which 
could introduce a bias with respect to the disk--averaged properties due to local conditions. In any case, such effects are very small compared to the full range of measured radiance factors (shown by the grey dots). 
Nevertheless, we caution that some systematic differences among the resulting $\mathcal{W}$ maps do exist: it is safe to compare \emph{relative albedo differences within images}, but the \emph{absolute albedo values 
may be inaccurate} (preventing straight--forward comparison between images).

We did attempt to consider $c\not=1$ solutions, because forward scattering may start to play a minor role at the highest considered phase angles. However, using a double--lobed Henyey--Greenstein 
single--particle phase function did not improve the overall fit further, therefore we consider $\{w_1,\,h_1,\,\xi_1\}=\{0.055,\,0.035,\,-0.456\}$ our final solution. This completes step (iii) in section~\ref{sec_method}.

\begin{figure}
\scalebox{0.3}{\includegraphics{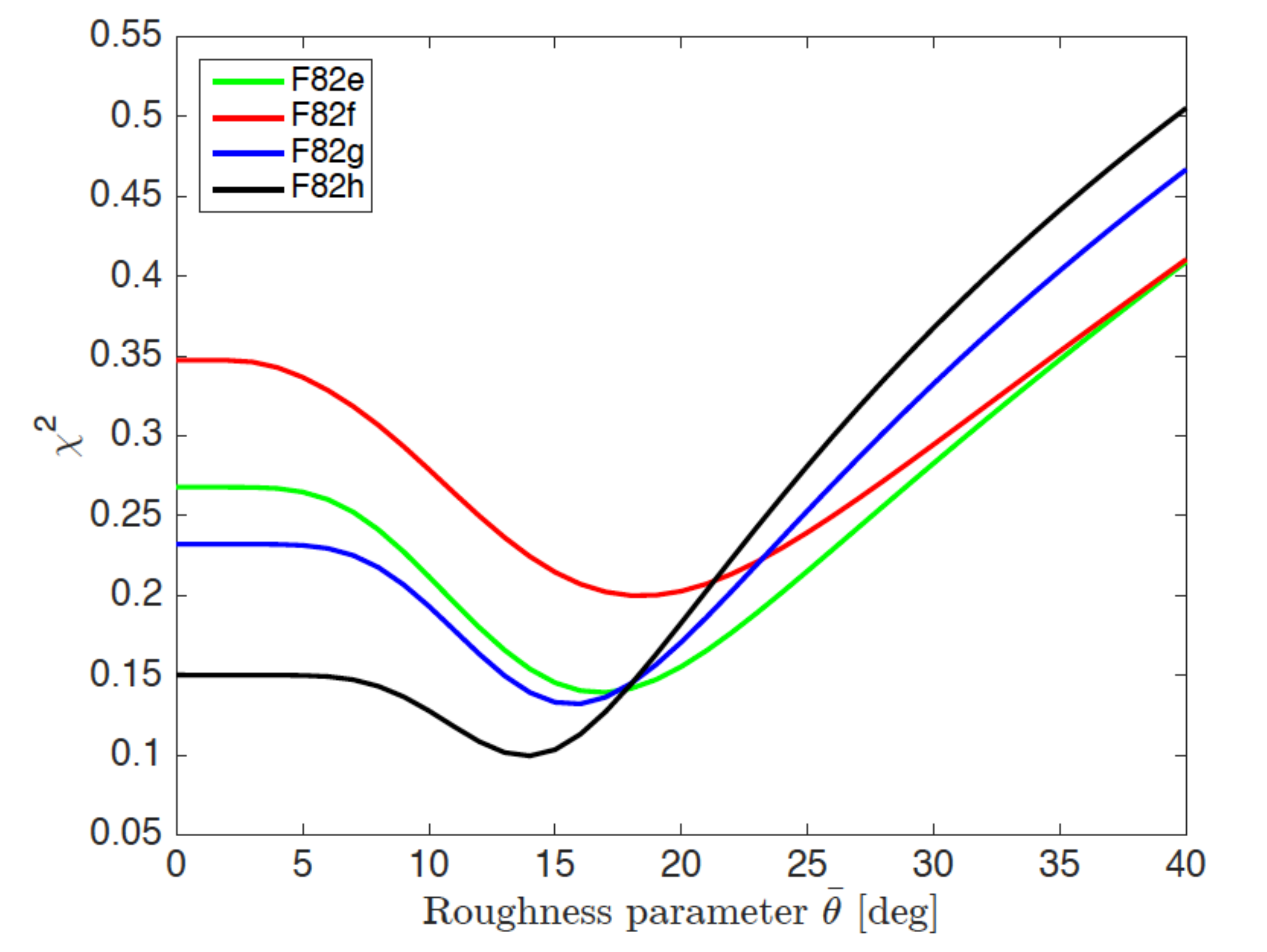}}
     \caption{The global $\chi^2$ residuals between $R_{\rm obs}$ and the photometric model given by equation~(\ref{eq:01}), as function of the 
level of surface roughness $\bar{\theta}$, while assuming $\{w_1,\,h_1,\,\xi_1\}=\{0.055,\,0.035,\,-0.456\}$, and limiting the considered pixels to 
the sub--set $S_2$ (pixels that would have $\geq 30$ per cent dimming of $R_{\rm rough}(\bar{\theta}=25^{\circ})$ with respect to $R_{\rm flat}$).}
     \label{fig_thetabar}
\end{figure}

The next step is to try to determine a disk--averaged value for the surface roughness parameter $\bar{\theta}$. At this point, we can no longer rely on $\alpha$--binning, but need to consider 
data and models pixel--by--pixel for each image. We first considered each value on the $0\leq\bar{\theta}\leq 40^{\circ}$ interval with $\Delta\bar{\theta}=1^{\circ}$ resolution, calculated $R_{\rm rough}$ 
in each case according to equation~(\ref{eq:01}), while always using  $\{w_1,\,h_1,\,\xi_1\}$, then calculated the $\chi^2$ with respect to the measured values $R_{\rm obs}$ for all pixels with 
$i<85^{\circ}$, $e<70^{\circ}$, and $R_{\rm obs}>R_{\rm co}$. Note, that we also account for the $H$--function terms at this stage. However, none of the images had a convincing $\chi^2$ minimum. 
The reason for this is that the large dispersion of radiance factor 
values about the average (illustrated by the grey dots) is not primarily caused by roughness. First, roughness can never brighten the nucleus with respect to the average solution. The grey dots 
above the red curve, that sometimes are 60 per cent brighter than the average solution, can therefore only be explained by intrinsic brightness enhancements, e.~g., an unusually high single--scattering albedo. 
Second, if there are large local albedo increases relative to the average, we likely also have large local albedo reductions. In fact, we have pixels that are up to 60 per cent dimmer than expected from the 
average photometric solutions, and surface roughness rarely manages to reduce radiance factors by more than 20--40 per cent even at $\bar{\theta}=25^{\circ}$. In fact, the vast majority of pixels in these 
images are expected to dim by at most 5--10 per cent, even at the largest phase angles. The flat $\chi^2=\chi^2(\bar{\theta})$ curves (not shown here) therefore indicate that surface roughness effects simply drown in a more 
substantial single--scattering albedo variegation. 

\begin{figure*}
\centering
\begin{tabular}{cc}
\scalebox{0.4}{\includegraphics{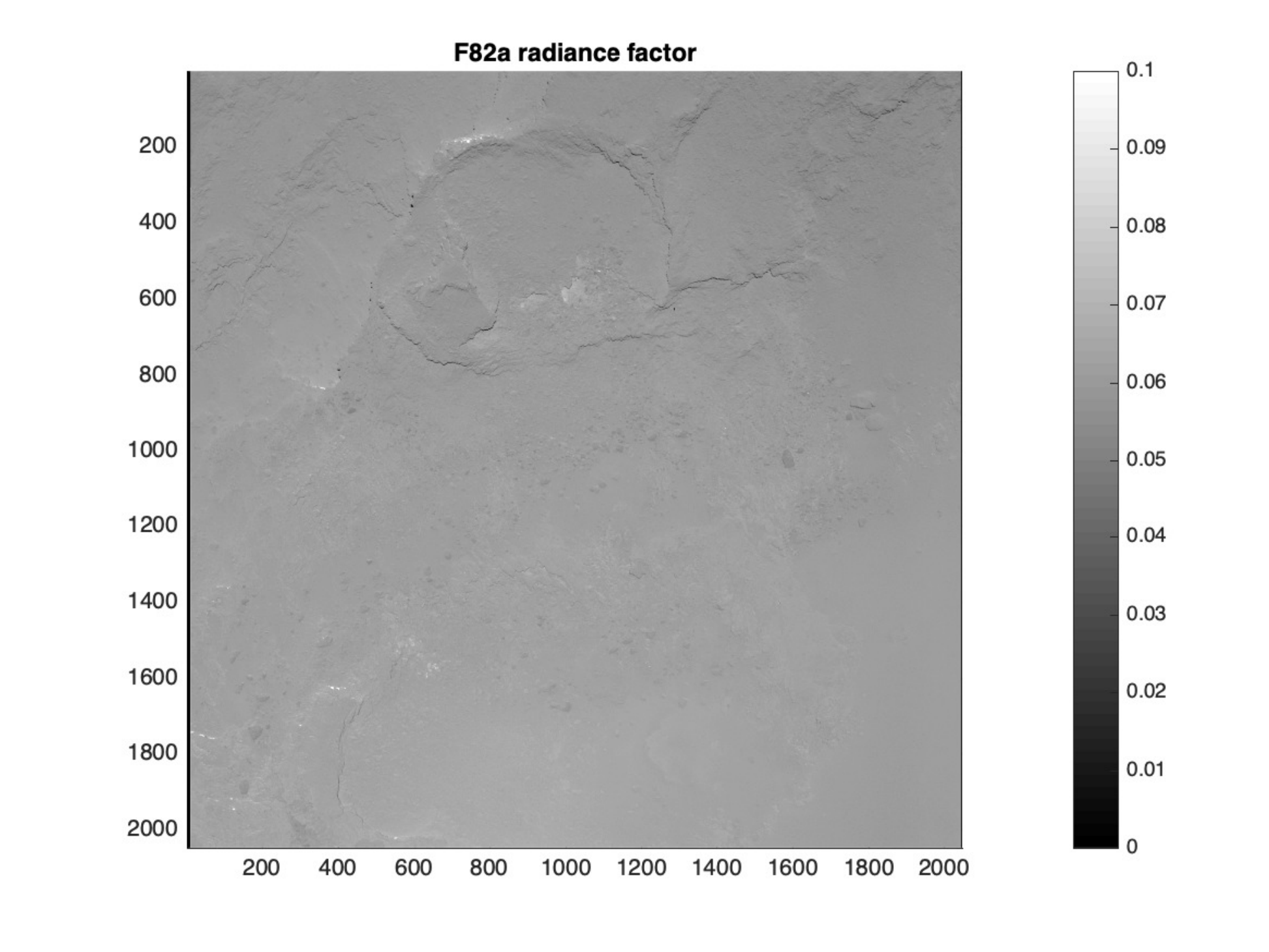}} & \scalebox{0.4}{\includegraphics{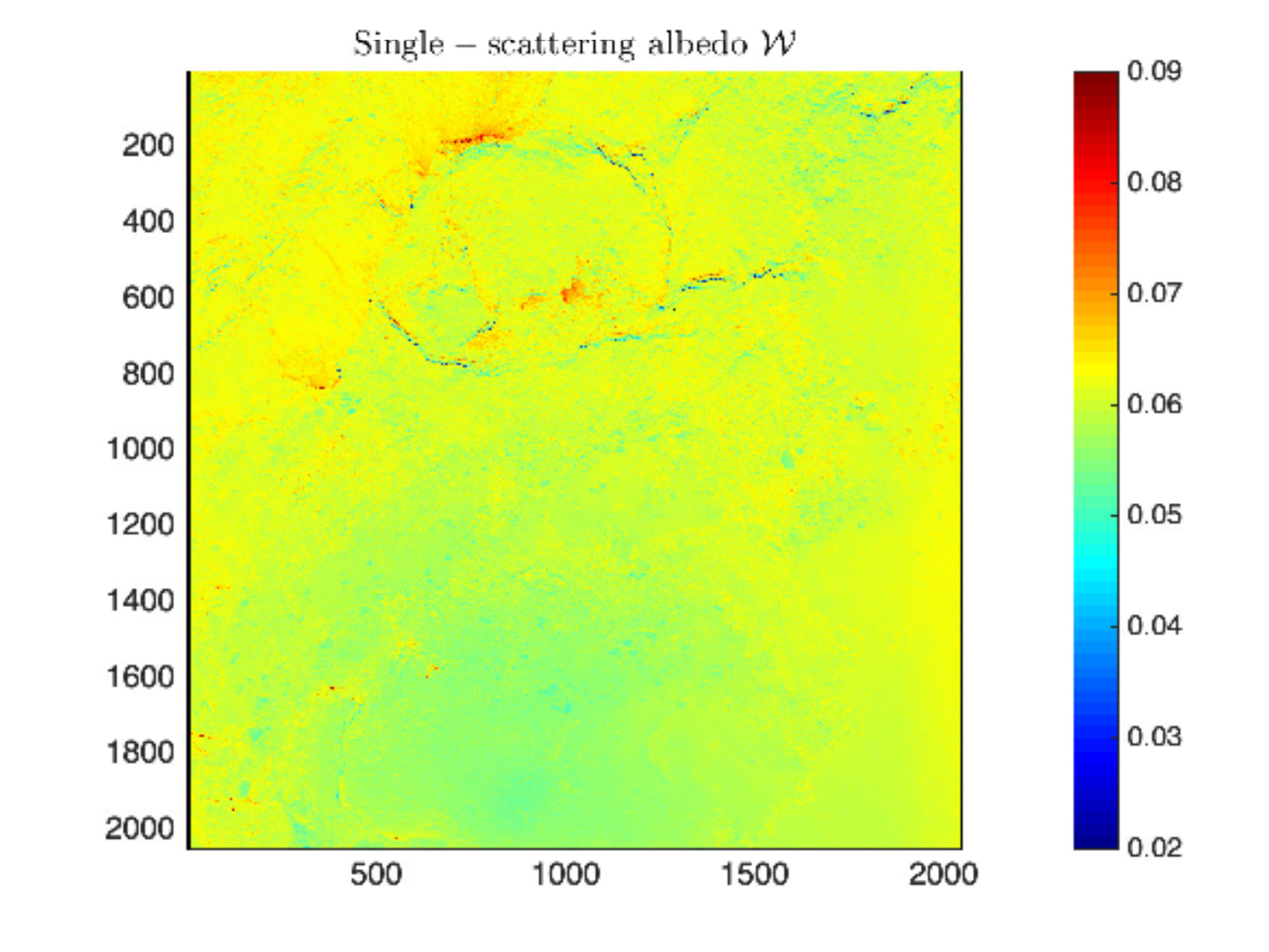}}\\
\scalebox{0.4}{\includegraphics{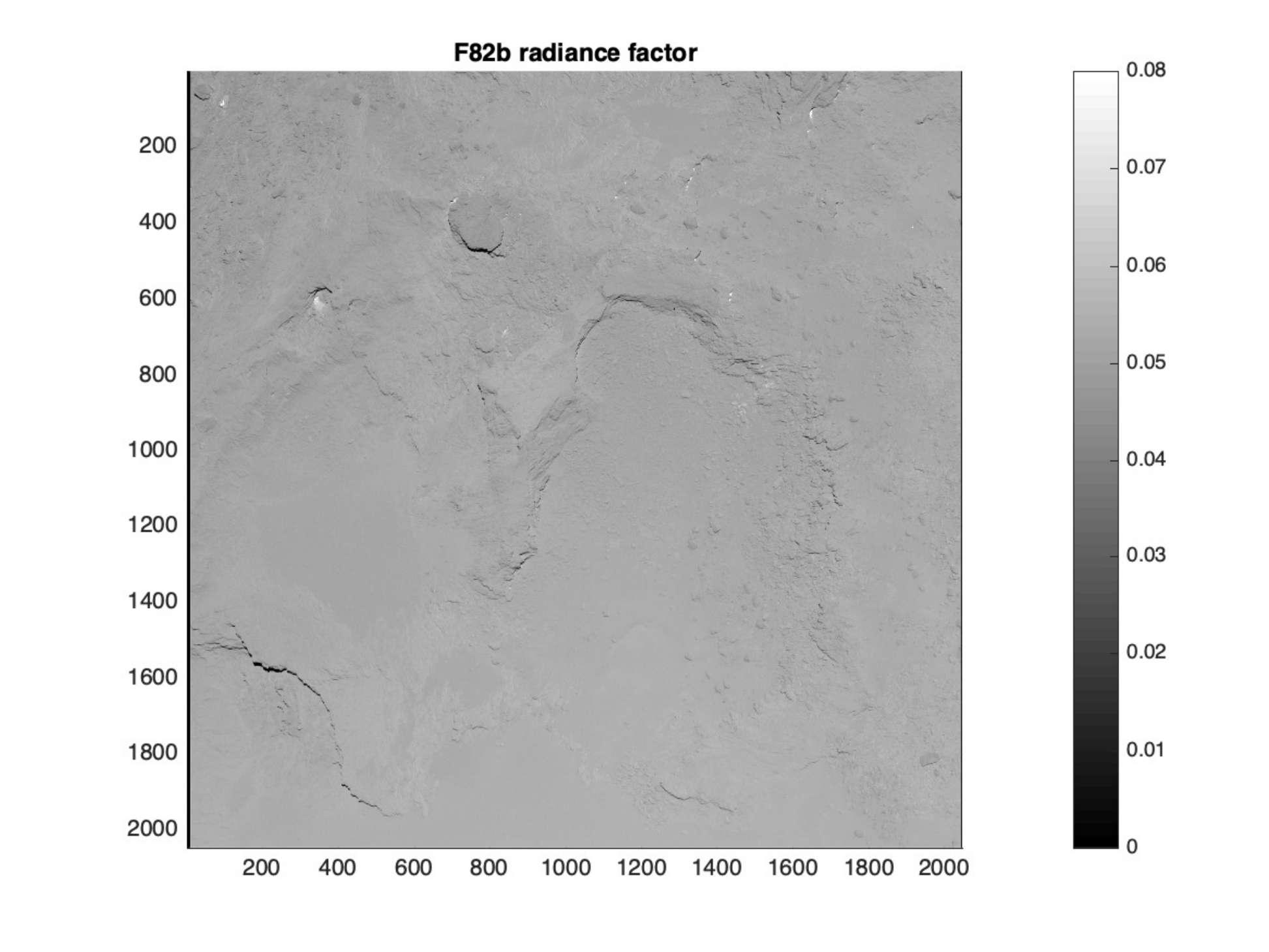}} & \scalebox{0.4}{\includegraphics{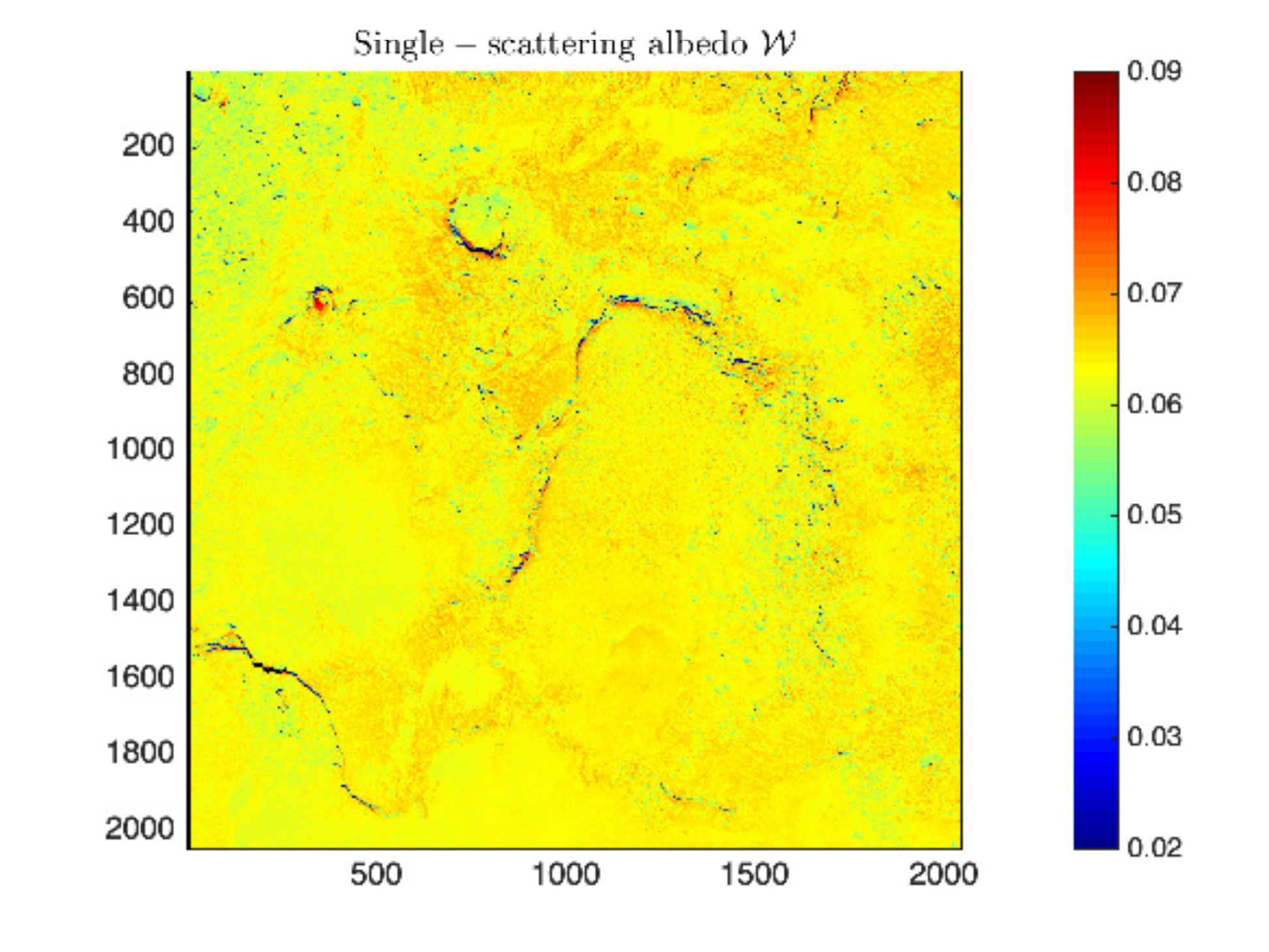}}\\
\end{tabular}
     \caption{\emph{Upper panels:} F82a at $0^{\circ}\leq\alpha\leq 2.4^{\circ}$, showing parts of Ash, Khepry, and Imhotep \citep[terraces 4, 3, and 1 in the right panel of Fig.~3 in][for context]{felleretal19}. 
\emph{Lower panels:} F82b at $2.8^{\circ}\leq\alpha\leq 5.2^{\circ}$, showing parts of Imhotep \citep[terraces 1 and 2 in the right panel of Fig.~3 in][]{felleretal19}.}
     \label{fig_image_albedo_F82_01}
\end{figure*}

To handle this situation, we attempted a $\bar{\theta}$--fit for a sub--set of pixels, for which we expect to have the strongest surface roughness effects. Assuming $\bar{\theta}=25^{\circ}$, we 
selected a sub--set $S_2$ of pixels, for which $R_{\rm rough}$ would be at least 30 per cent dimmer than $R_{\rm flat}$ (when considering the $\{w_1,\,h_1,\,\xi_1\}$ parameter combination). We then 
repeated the $\bar{\theta}$ search as described above, but limited to this sub--set $S_2$. The results for images F82e--h are seen in Fig.~\ref{fig_thetabar}.

By focusing on pixels for which surface roughness effects are expected to be large, we substantially reduce the effect of albedo `noise', and see clear minima in the $\chi^2=\chi^2(\bar{\theta})$ curves. 
These indicate that $\bar{\theta}_1=16.2^{\circ} \pm 1.7^{\circ}$. This completes steps (iv)--(v) in section~\ref{sec_method}.

\begin{figure*}
\centering
\begin{tabular}{cc}
\scalebox{0.4}{\includegraphics{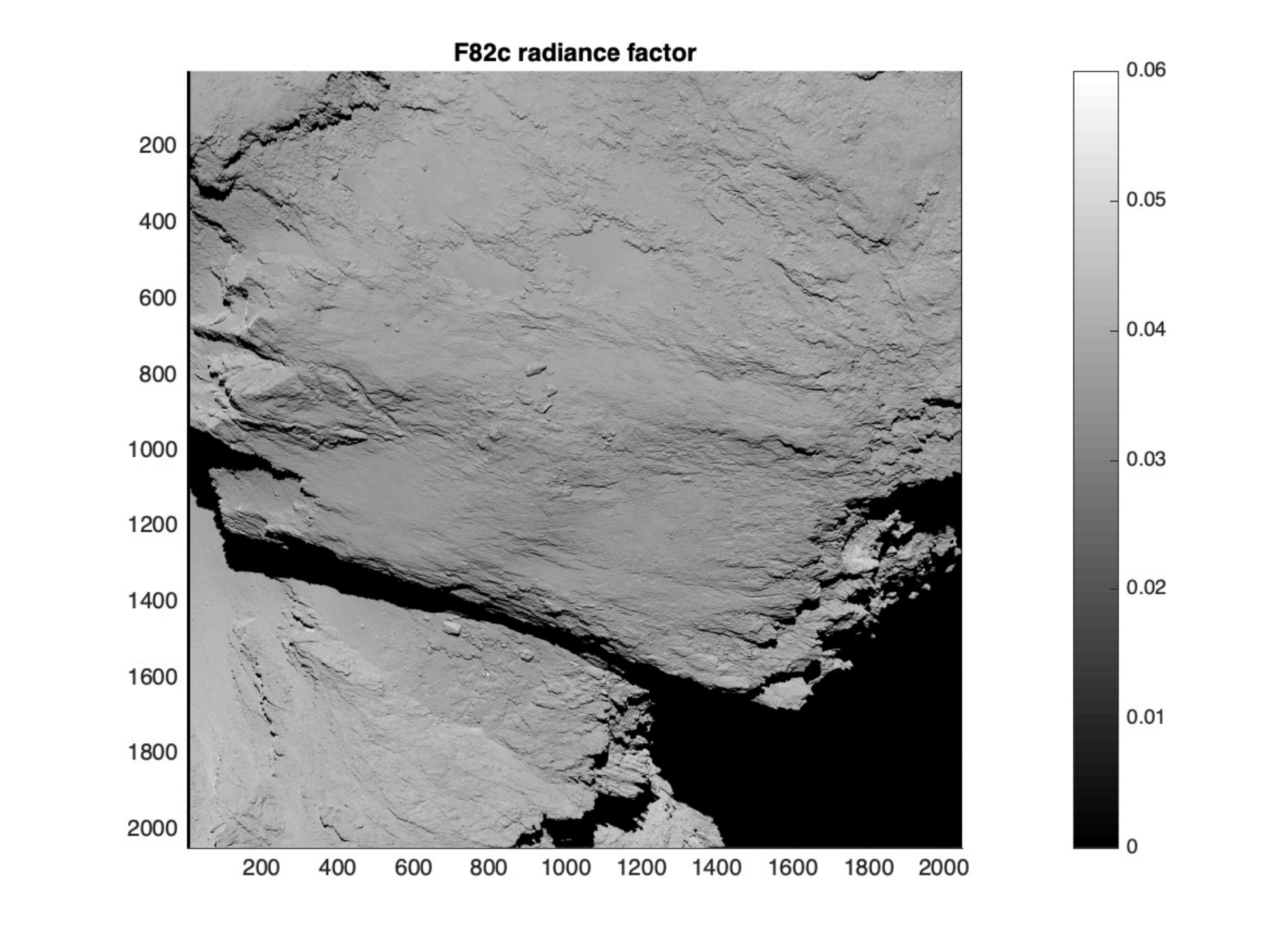}} & \scalebox{0.4}{\includegraphics{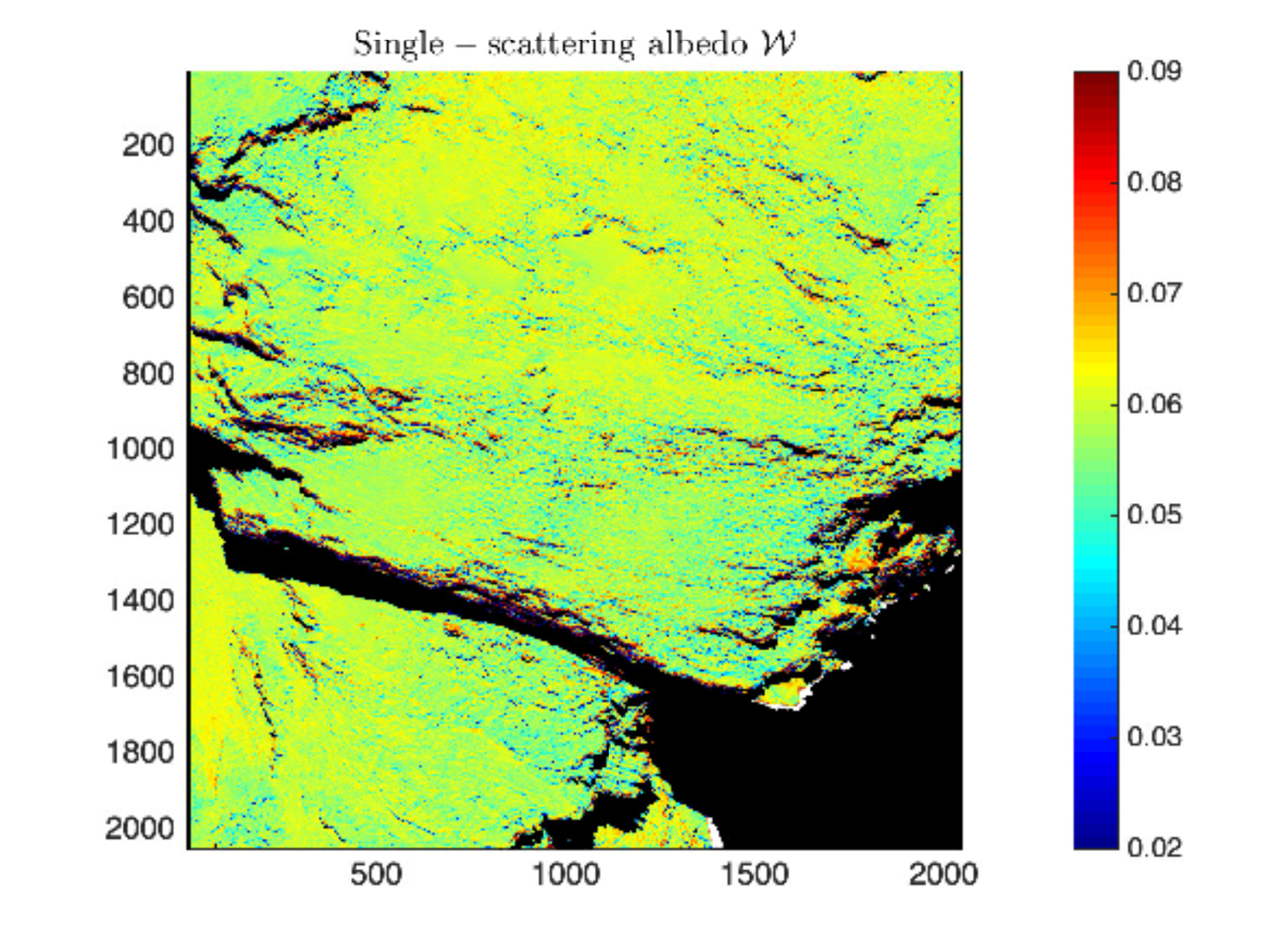}}\\
\scalebox{0.4}{\includegraphics{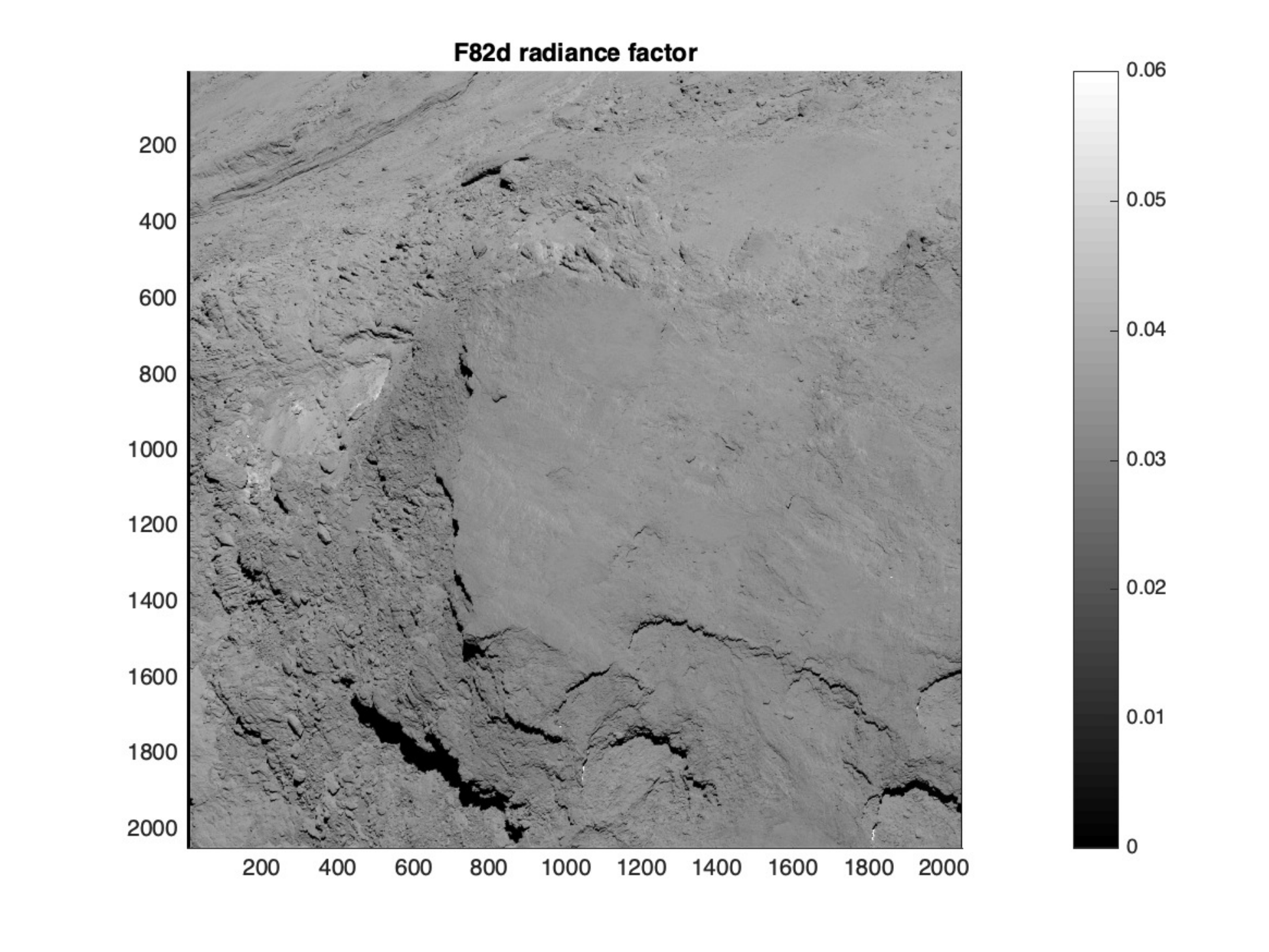}} & \scalebox{0.4}{\includegraphics{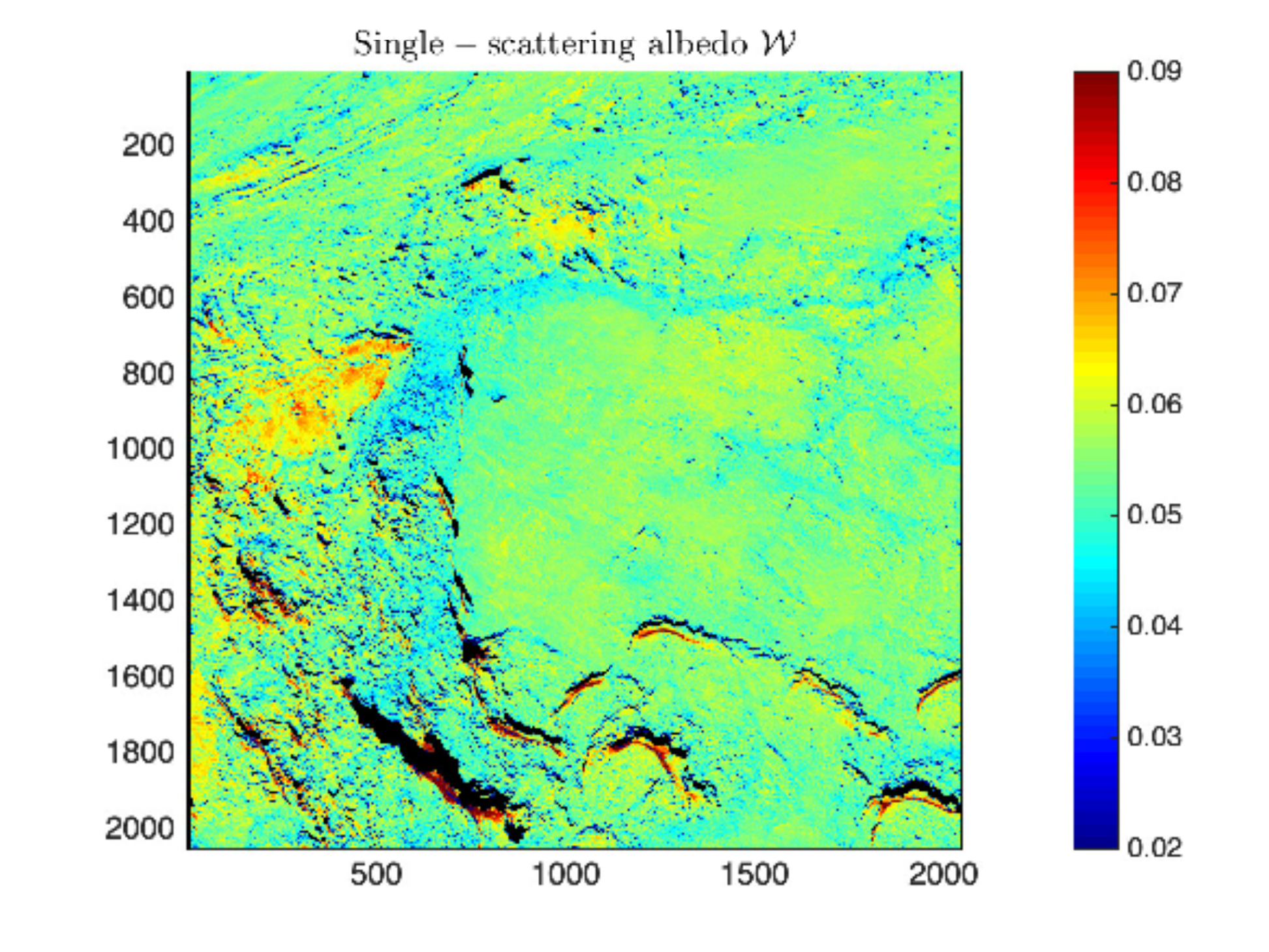}}\\
\end{tabular}
     \caption{\emph{Upper panels:} F82c at $8.9^{\circ}\leq\alpha\leq 11.1^{\circ}$, showing Khepry and Aker \citep[see the right part of Fig.~1 in][for context]{thomasetal15a}. 
\emph{Lower panels:} F82d at $13.9^{\circ}\leq\alpha\leq 16.1^{\circ}$ primarily shows Apis, surrounded by Imhotep (top), Khonsu (left), and 
Atum (bottom), see the right part of Fig.~8 in \citet{thomasetal18} for context.}
     \label{fig_image_albedo_F82_02}
\end{figure*}

We now have all parameters needed to evaluate the function $D=D(h_1,\,\xi_1,\,\bar{\theta}_1,\,i,\,e,\,\alpha)$ for each pixel in each image. We apply it to define a single--scattering albedo 
proxy $\mathcal{W}=R_{\rm obs}/D$. If the version of the \citet{hapke93} scattering model applied here had been exact, and if $\{h_1,\,\xi_1,\,\bar{\theta}_1\}$ truly had been well--determined 
\emph{constants} (i.~e., the comet surface had been uniform in terms of those properties), then $\mathcal{W}$ would have been the local real single--scattering albedo. However, despite 
the fact that our scattering function has a qualitative and quantitative behaviour that 
is similar to that of the data, we cannot claim that it is a perfect representation of reality. Furthermore, we do not know the extent of local variability in the parameters $\{h,\,\xi,\,\bar{\theta}\}$. 
All potential shortcomings and photometric variability about the disk--averaged solution are absorbed by $\mathcal{W}$, hence we call it a proxy to the single--scattering albedo, 
rather than the albedo itself. The quantity $\mathcal{W}$ is physically meaningful if its variegation is not random but displays large--scale patterns, if those patterns correlate in a systematic 
manner with nucleus morphology, and if such correlations are repeating in images acquired for a variety of illumination and viewing conditions. If these conditions are met, then $\mathcal{W}$ 
can be used to identify characteristic photometrical behaviour for various terrain types, though detailed physical explanations as to why they differ may require further study.

\begin{figure*}
\centering
\begin{tabular}{cc}
\scalebox{0.4}{\includegraphics{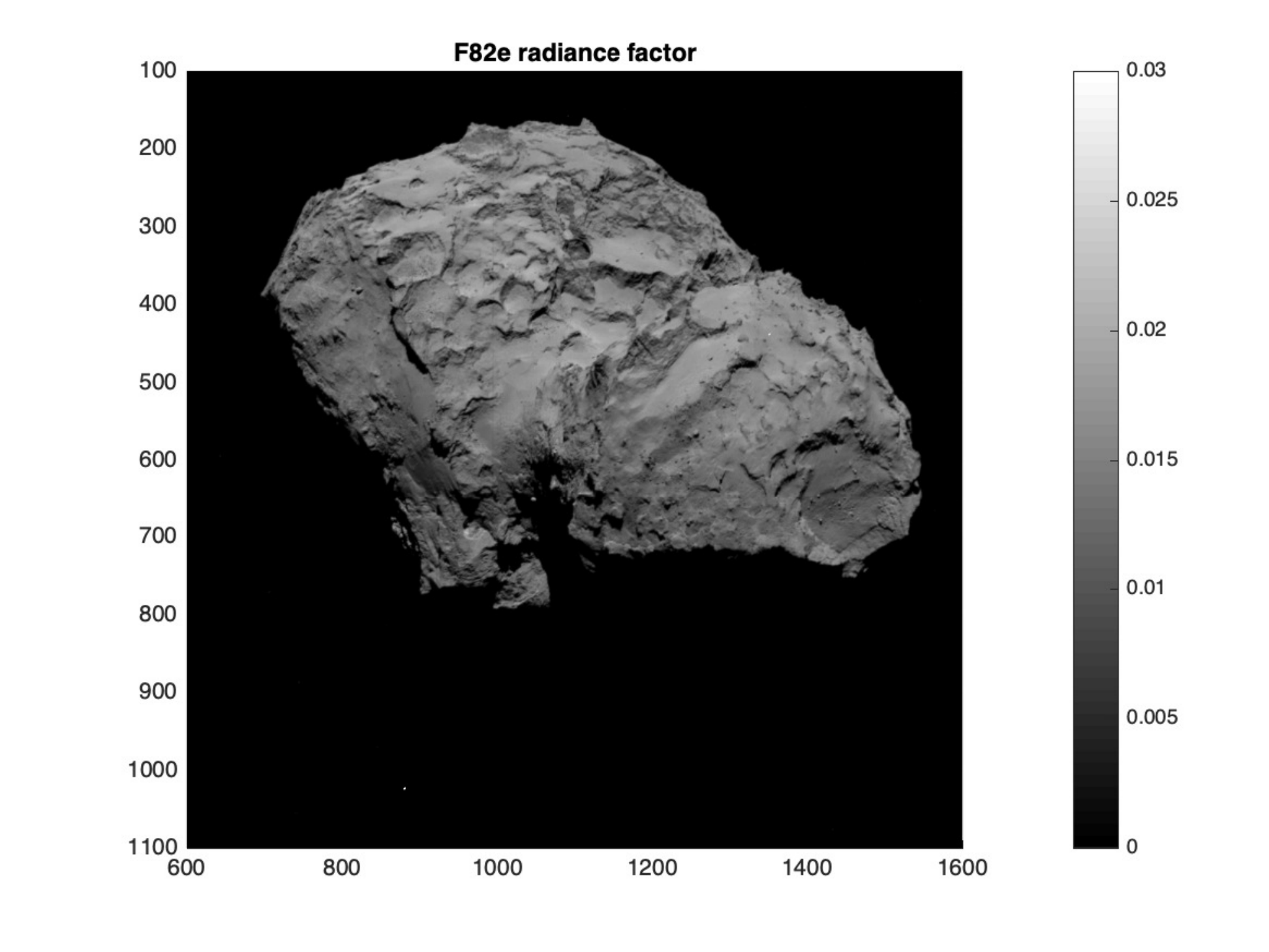}} & \scalebox{0.4}{\includegraphics{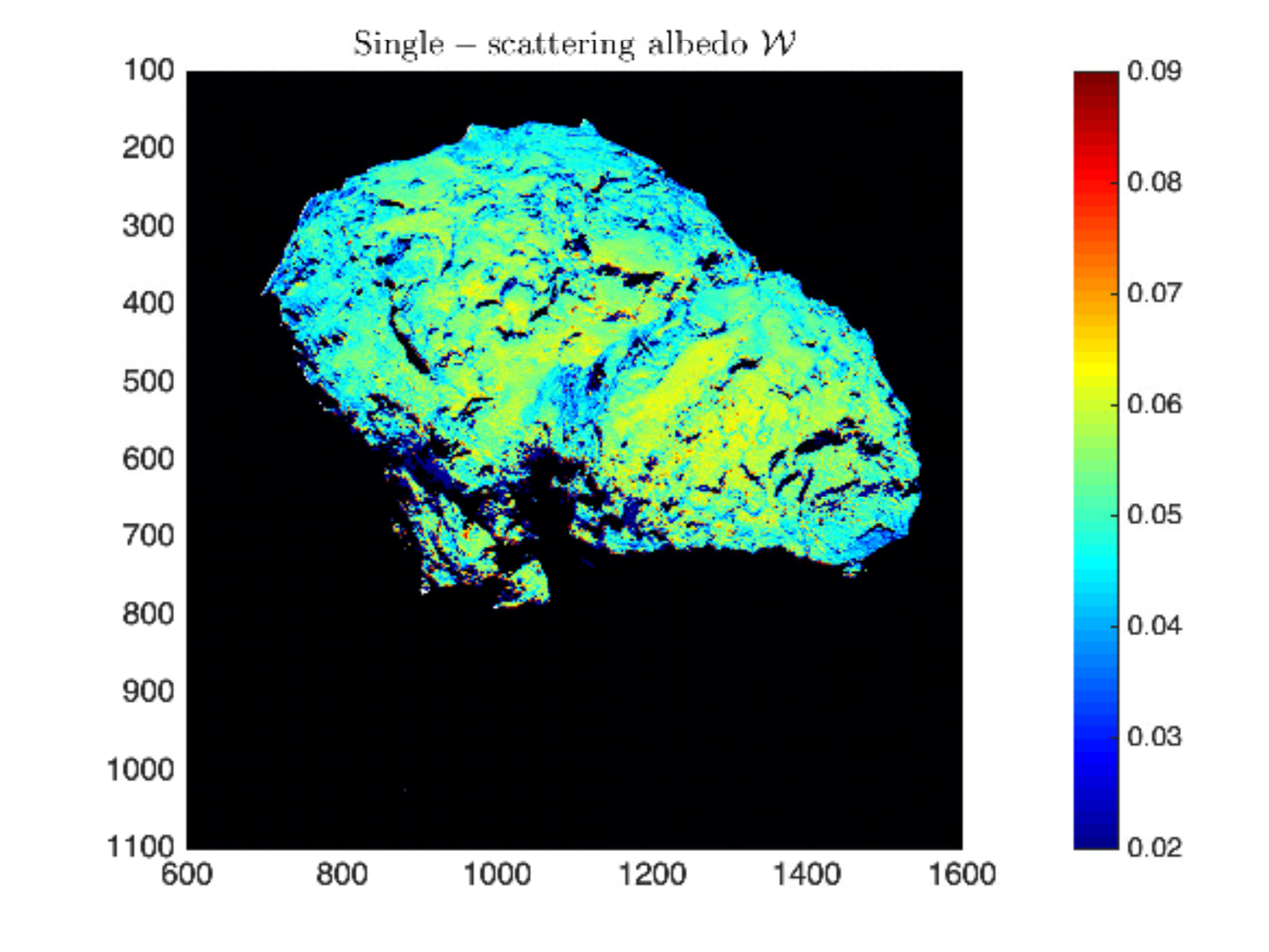}}\\
\scalebox{0.4}{\includegraphics{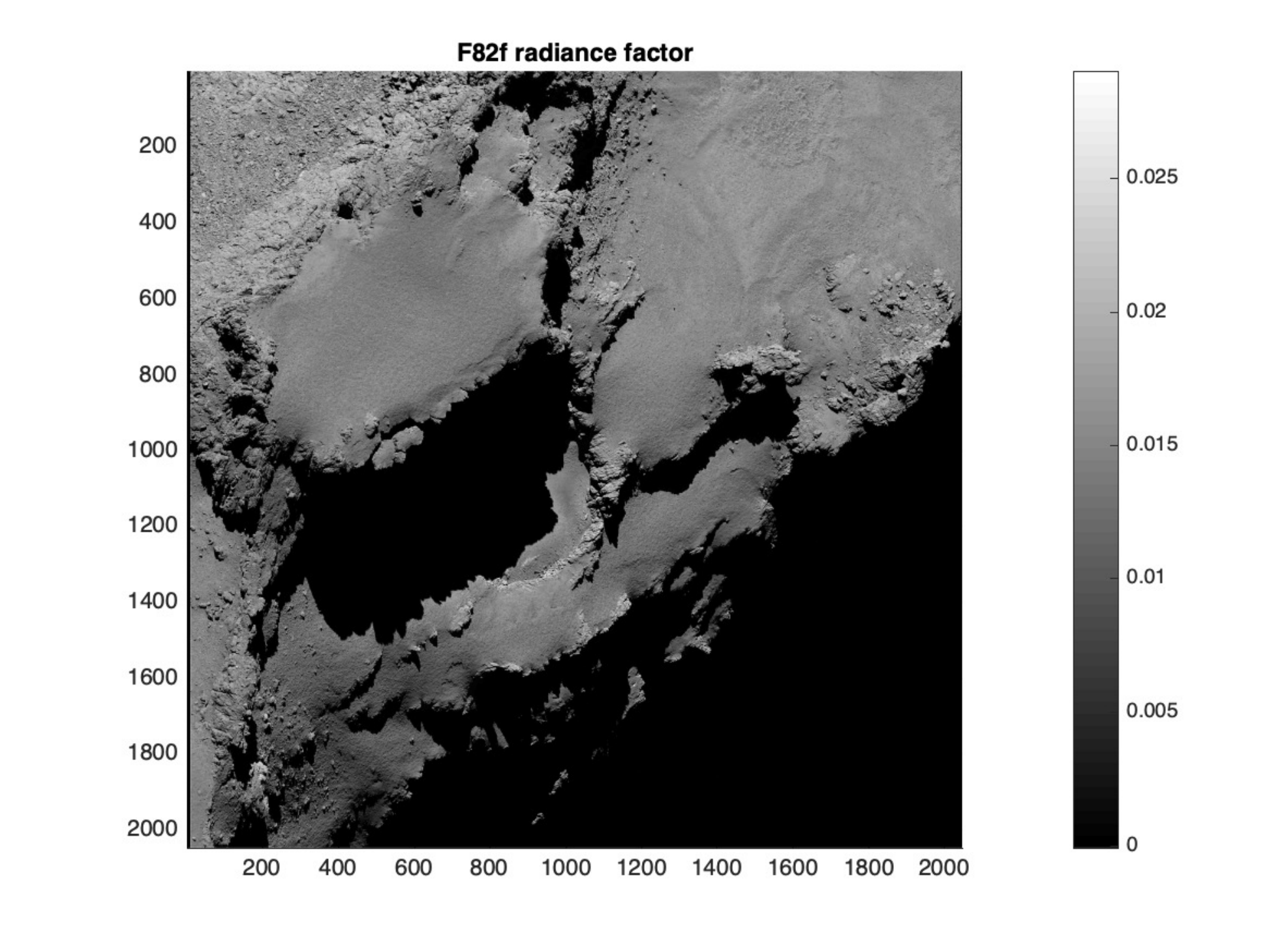}} & \scalebox{0.4}{\includegraphics{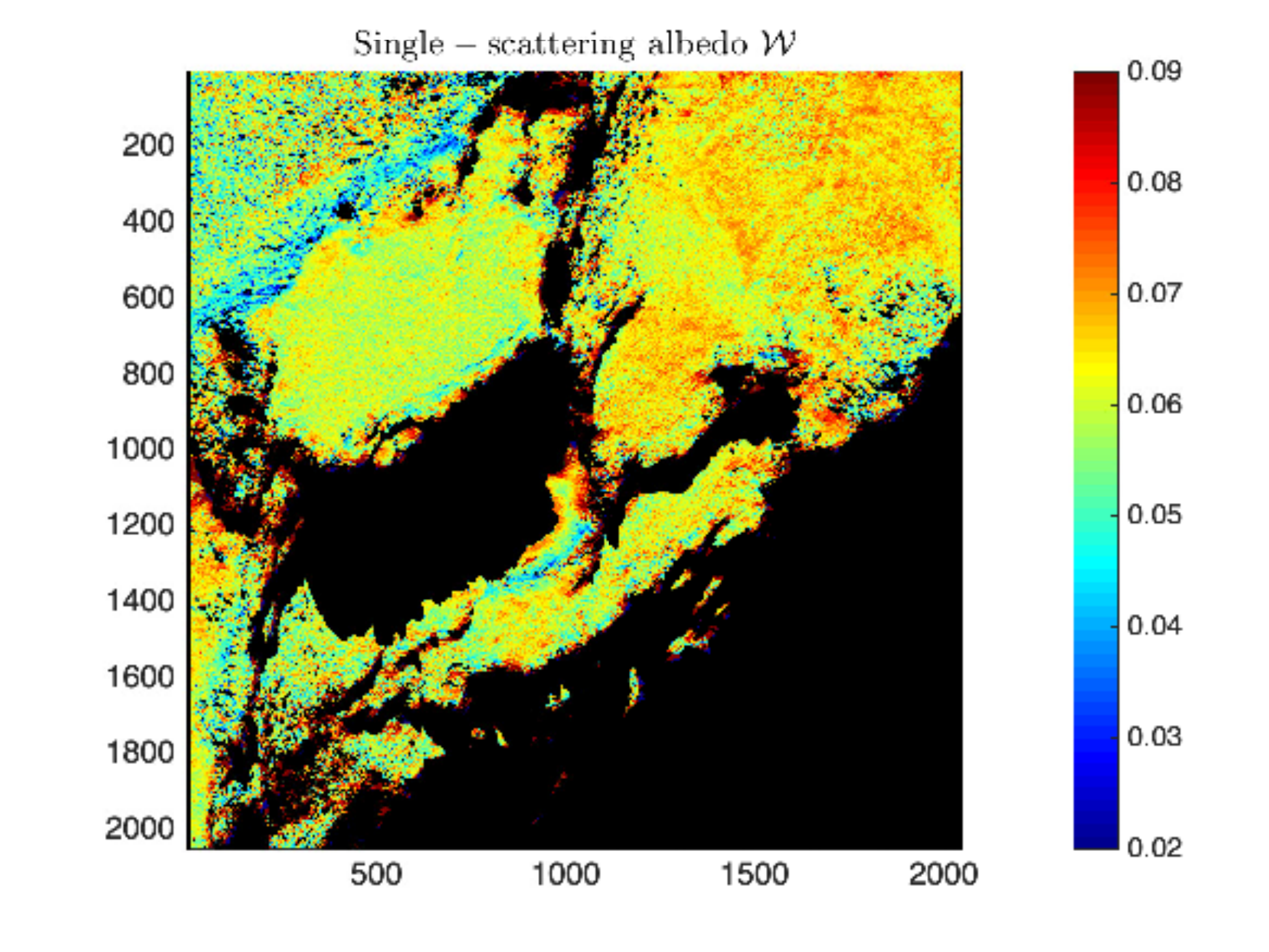}}\\
\end{tabular}
     \caption{\emph{Upper panels:} F82e at $40.3^{\circ}\leq\alpha\leq 41.0^{\circ}$, showing a large fraction of the northern hemisphere \citep[see the middle panel of Fig.~1 in][for context]{elmaarryetal15}. 
\emph{Lower panels:} F82f at $42.0^{\circ}\leq\alpha\leq 44.2^{\circ}$ shows Babi, see Fig.~8c in \citet{elmaarryetal15}.}
     \label{fig_image_albedo_F82_03}
\end{figure*}

\begin{figure*}
\centering
\begin{tabular}{cc}
\scalebox{0.4}{\includegraphics{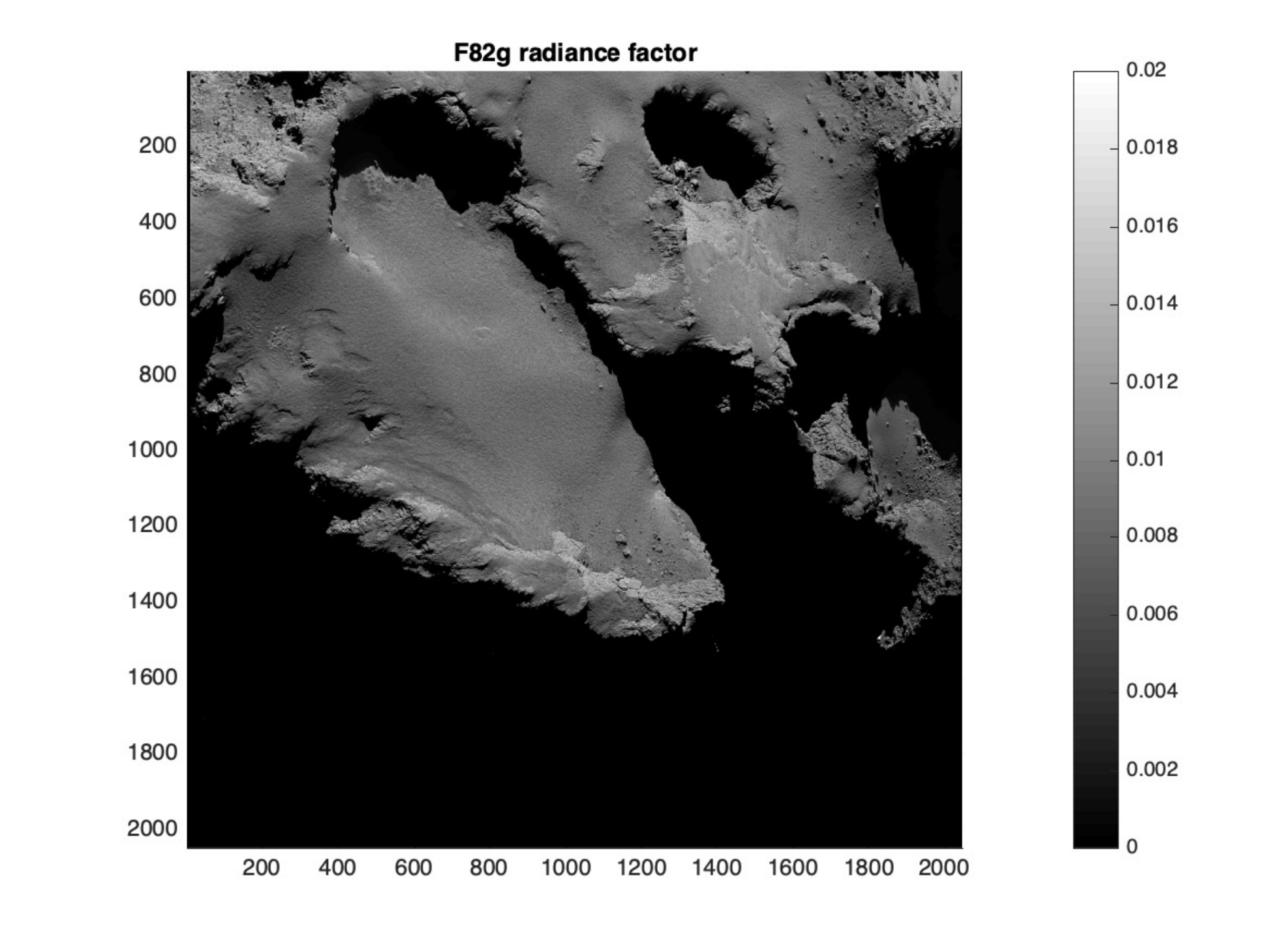}} & \scalebox{0.4}{\includegraphics{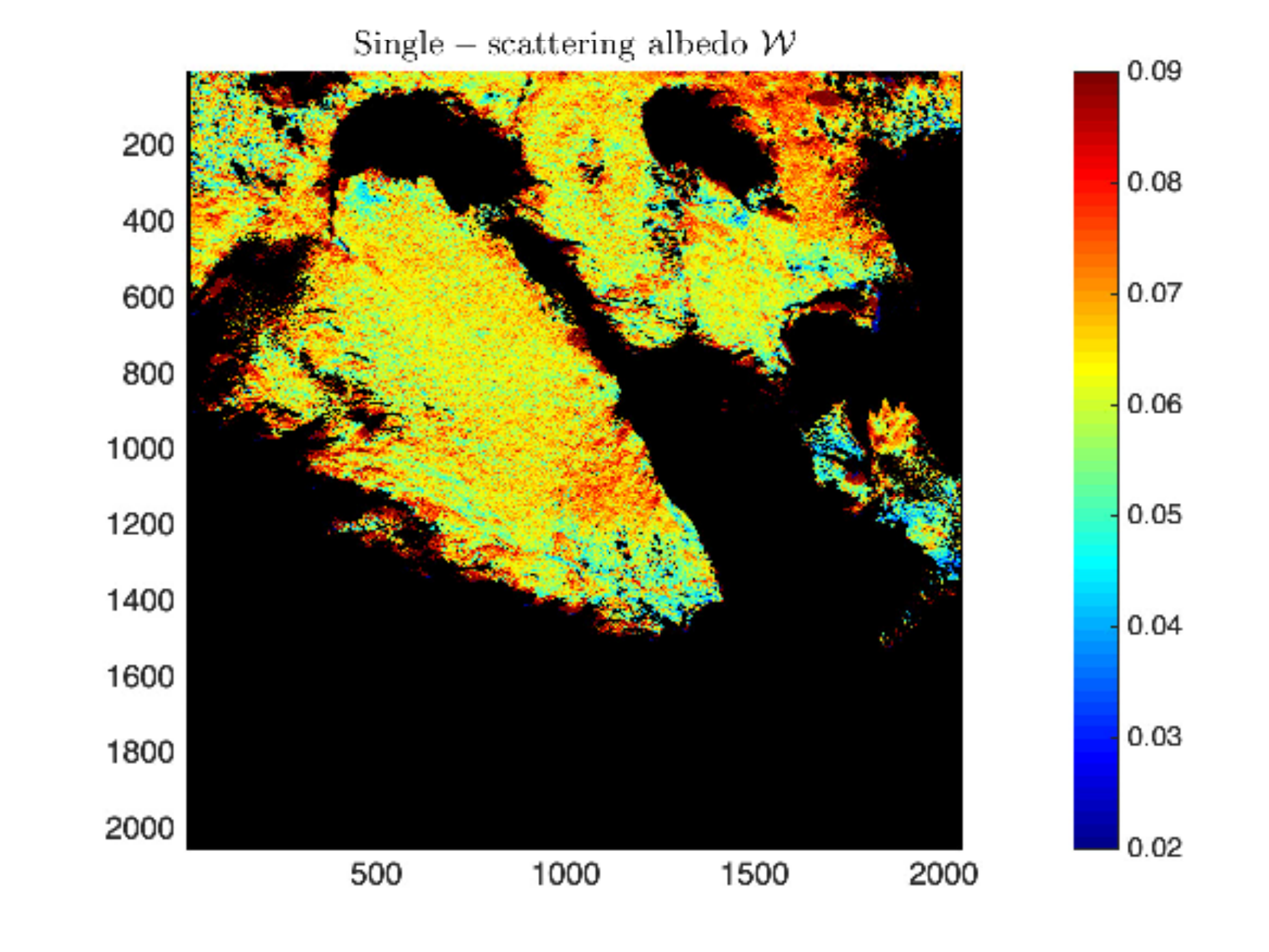}}\\
\scalebox{0.4}{\includegraphics{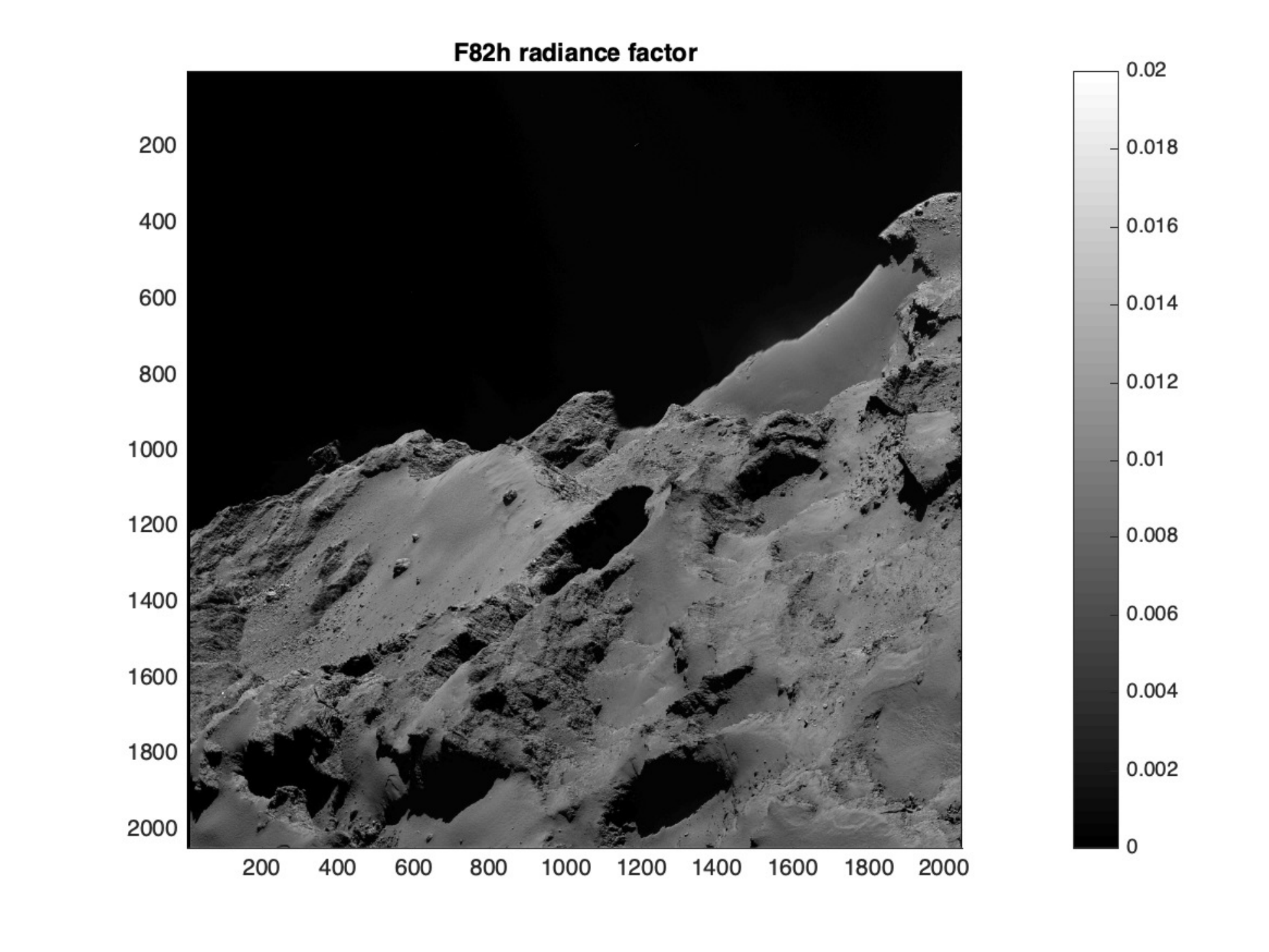}} & \scalebox{0.4}{\includegraphics{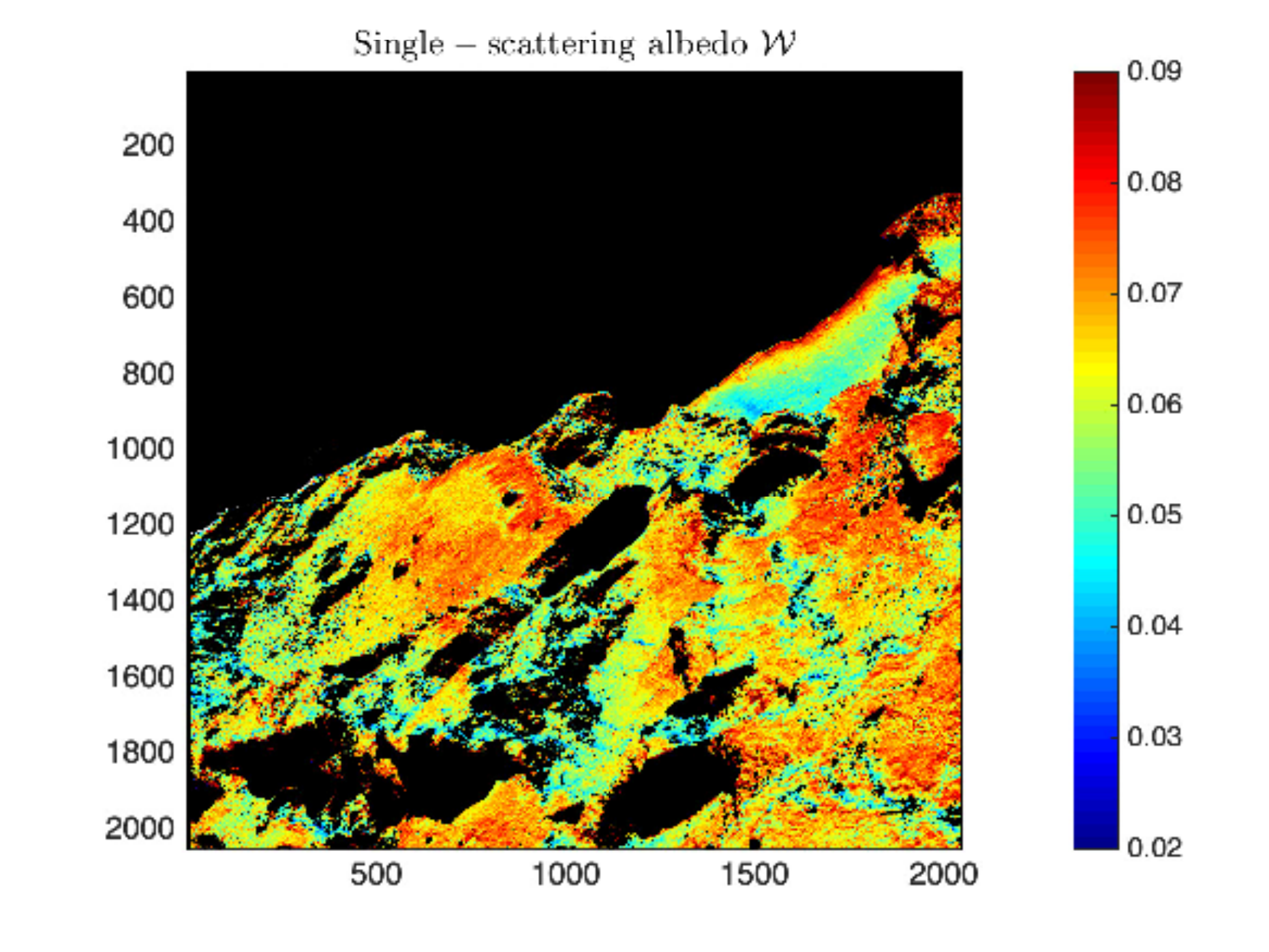}}\\
\end{tabular}
     \caption{\emph{Upper panels:} F82g at $57.6^{\circ}\leq\alpha\leq 59.8^{\circ}$, showing pits in Ma'at on the small lobe \citep[see Fig.~1e in][for context]{huetal17}. 
\emph{Lower panels:} F82h at $67.4^{\circ}\leq\alpha\leq 69.6^{\circ}$ shows Seth with a fraction of Hapi in the background \citep[see Fig.~1 (right) in][for context]{pajolaetal19}.}
     \label{fig_image_albedo_F82_04}
\end{figure*}

Figures~\ref{fig_image_albedo_F82_01}--\ref{fig_image_albedo_F82_04} show all F82 images in Table~\ref{tab1}, with the radiance factor images in the left panels, 
and the corresponding $\mathcal{W}$ maps in the right panels, arranged in order of increasing phase angle. With the applied colour coding, the disk--average single--scattering 
albedo ($w_1=0.055$) is green, so that yellow--orange--red traces brighter terrain and turquoise--blue traces darker terrain. F82a (upper Fig.~\ref{fig_image_albedo_F82_01} panels) 
shows a circular feature in Ash, a boulder--strewn field in Khepry, and a basin in Imhotep \citep[terraces 4, 3, and 1 in the right panel of Fig.~3 in][]{felleretal19}. Also 
see an RGB colour version and a spectral slope version of F82a in Fig.~4 (right) in \citet{felleretal19}, as well as their Fig.~2 for context.  F82b (lower Fig.~\ref{fig_image_albedo_F82_01} panels) 
shows two Imhotep basins \citep[terraces 1 and 2 in the right panel of Fig.~3 in][]{felleretal19}, with RGB colour and spectral slope versions in Fig.~4 (left) in \citet{felleretal19}. 
Regions that are whitish in RGB pictures also have relatively small (more neutral) spectral slopes. These regions coincide with the high--$\mathcal{W}$ parts of F82a--b, which is 
encouraging because it suggests that our $\mathcal{W}$ is not random but does trace previously established photometric variability.

F82c (upper Fig.~\ref{fig_image_albedo_F82_02} panels) shows the boundary between Khepry (top) and Aker (centre), see the right part of Fig.~1 in \citet{thomasetal15a} for context. 
F82d (lower Fig.~\ref{fig_image_albedo_F82_02} panels) is a face--on view of the flat--faced Apis region, surrounded by partial and very oblique views of Imhotep (top), Khonsu (left), and 
Atum (bottom), see the right part of Fig.~8 in \citet{thomasetal18} for context. Khepry, Aker, and Apis consist of consolidated terrain, partially covered by fine--grained deposits 
similar to the smooth airfall terrains. Clean consolidated terrain tends to be darker (blue--green on the $\mathcal{W}$ colour scale), which is best seen on the steep cliff 
just left of Apis, that is incapable of accumulating dust. Patches where fine--grained deposits are thicker (revealed by the smooth fields in Khepry and Aker) tend to be 
brighter (yellowish). Note a quasi--circular feature in Khonsu that is unusually bright (reddish).

The tendency that consolidated terrain is darker (bluish) than smooth terrain (yellowish) is clearly seen in F82e (upper Fig.~\ref{fig_image_albedo_F82_03}), that constitutes 
a panoramic view of the northern hemisphere \citep[see the middle panel of Fig.~1 in][for context]{elmaarryetal15}. Dark/blue regions include Atum, the large 
Seth cliff (bordering Ash, above the Aswan plateau), the vertical part of Serqet, and vertical rims of the Hatmehit depression, that are all dominated by consolidated terrain. 
Brighter/yellow regions are common on horizontal surfaces in Seth (including the Aswan plateau) on the large lobe, as well as in Ma'at and the horizontal portions of Serqet on the 
small lobe, that all are covered by thick airfall deposits. F82f (lower Fig.~\ref{fig_image_albedo_F82_03}) shows a close--up of smooth terrain in Babi \citep[see Fig.~8 in][particularly panel c]{elmaarryetal15}, 
that also is relatively bright (yellow to red). Note, however, the darker/bluish strip in the upper left corner of F82f that constitutes the deposit-free walls of the Aten depression.

Image F82g (upper Fig.~\ref{fig_image_albedo_F82_04}) shows a portion of the Ma'at region with two pits \citep[see Fig.~1e in][for context]{huetal17}. The left pit contains a `honeycomb' 
feature labelled MAT02 by \citet{huetal17}. Honeycombs are densely pitted features revealed when overlying fine--grained material is ejected to space when the comet approaches its perihelion. 
The pit surroundings are covered by thick airfall deposits, recognisable in the $\mathcal{W}$ map by their brighter--than--average yellow/red colour. Interestingly, the honeycomb feature 
at the pit centre is one of the darker/bluer features in F82g. Its $\mathcal{W}$ is similar to outcropping consolidated material, visible below the right pit, as well as at the tip of the V--shaped 
feature below the left pit. This suggests that this honeycomb constitutes consolidated material as well. If water ice sublimation at the honeycomb is responsible for the ejection of 
airfall material that revealed its presence, then this water ice does not seem to be exposed on the surface, because then we would have expected it to be intensively red on the 
$\mathcal{W}$ map, like the isolated bright features in Fig.~\ref{fig_image_albedo_F82_01}. 

Finally, image F82h (lower Fig.~\ref{fig_image_albedo_F82_04}) primarily shows Seth, with the Aswan plateau seen prominently to the centre--left, and a small part of the 
smooth terrain of Hapi is seen to the centre--right \citep[see Fig.~1 (right) in][for context]{pajolaetal19}. The high--$\mathcal{W}$ (red) terrain primarily coincides with smooth terrain (e.~g., the Aswan plateau itself). 
Interestingly, the smooth terrain of Hapi is uncharacteristically dark. The emergence and incidence angles are very similar at Hapi and Aswan ($e\approx 60^{\circ}$ and $i\approx 50^{\circ}$), 
so if the roughness correction is wrong, it should impact both terrains in a similar way. This suggests that there are real photometric differences among different types of smooth terrain. Only one 
part of this piece of Hapi is distinctively bright (red): a narrow strip running along the shadow. We note that \emph{Rosetta}/VIRTIS observations presented by \citet{desanctisetal15} show $3\,\mathrm{\mu m}$ 
absorption lines in thin slices of Hapi terrain that just emerges from long shadows into illumination, corresponding to estimated volumetric water ice abundances of 5--14 per cent. This is interpreted as frost that 
condensed on the shadowed surface, and is removed after $\sim 20\,\mathrm{min}$ of exposure to sunlight \citep[the water abundance elsewhere in Hapi is below the detection limit of 
$\sim 1$ per cent;][]{desanctisetal15}. We propose that the thin high--$\mathcal{W}$ albedo anomaly along the shadow through Hapi in Fig.~\ref{fig_image_albedo_F82_04} (bottom) is due to exposed temporary water frost.

\subsection{F22 filter images} \label{sec_results_F22}

\begin{figure}
\scalebox{0.2}{\includegraphics{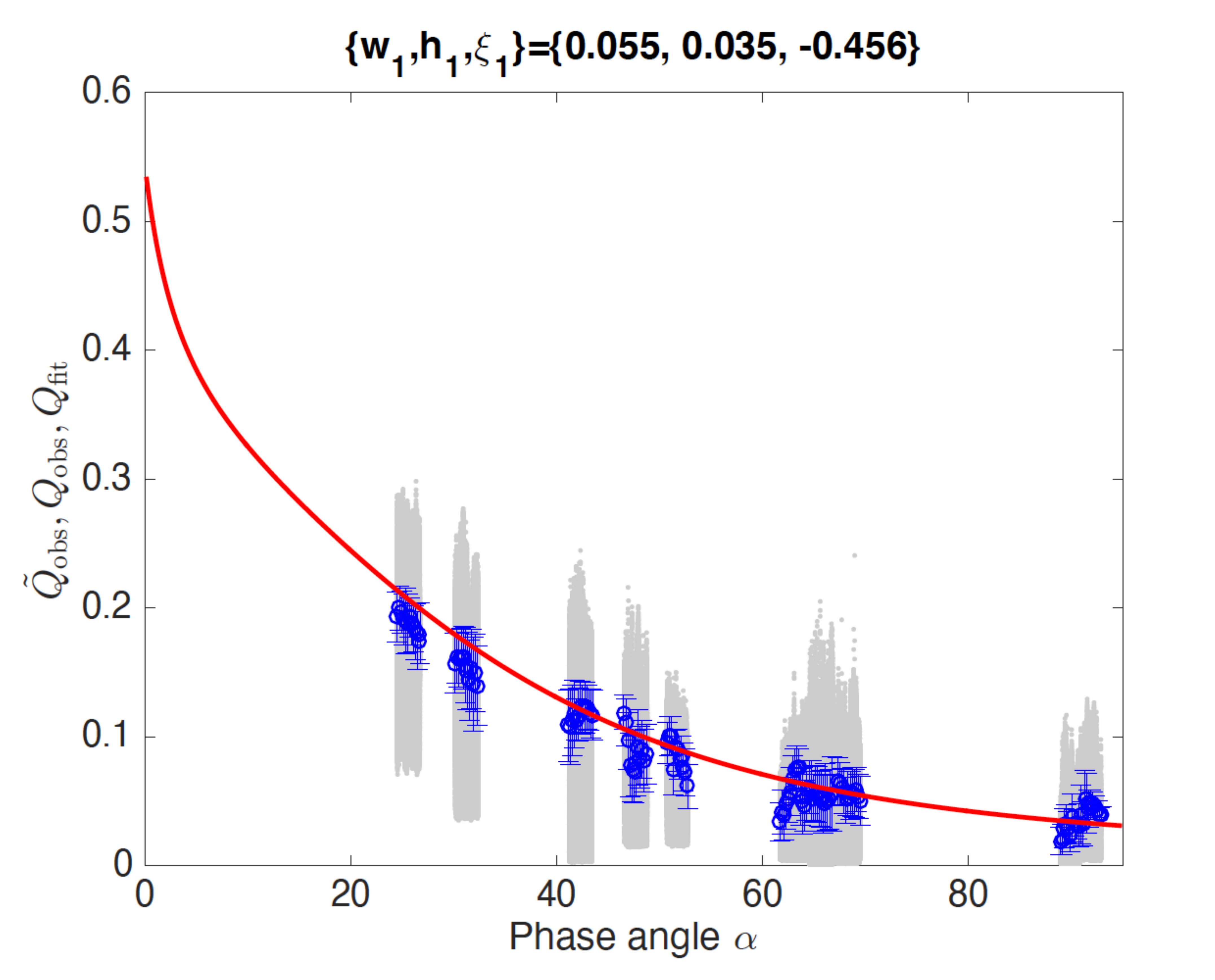}}
     \caption{The $\tilde{Q}_{\rm obs}$ values (grey dots) and $Q_{\rm obs}(\alpha)$ averages with standard deviations (red circles and bars) are here shown for all F22a--k images, 
limited to pixels for which surface roughness would dim the radiance factor by at most 2 per cent if $\bar{\theta}=25^{\circ}$. The F82a--h best--fit model $Q_{\rm fit}(\alpha)$, for 
which $\{w_1,\,h_1,\,\xi_1\}=\{0.055,\,0.035,\,-0.456\}$ is over--plotted the data as a red curve. }
     \label{fig_F22_F82fit}
\end{figure}

The combination of the OSIRIS NAC orange filter with the visual far focus plate (F22) is expected to be photometrically equivalent to the 
orange filter with the neutral density filter (F82) after calibration. Therefore, one should be able to use the same disk--average solution for 
both (the lack of low--$\alpha$ F22 images prevents an independently obtained $Q_{\rm fit}(\alpha)$ for that filter). We selected twelve F22 
images (Table~\ref{tab2}) with a combined phase angle coverage of $24.4^{\circ}\leq\alpha\leq 92.9^{\circ}$ and prepared $\tilde{Q}_{\rm obs(\alpha)}$ values and
$Q_{\rm obs(\alpha)}$ bins using the same procedure as for F82 images. These quantities are seen in Fig.~\ref{fig_F22_F82fit}, 
where we also over--plot the same $Q_{\rm fit}(\alpha)$ function as seen in Fig.~\ref{fig_4pic_8pic_fits} (upper right). We consider the F82 $Q_{\rm fit}(\alpha)$ 
a reasonable representation of F22 $Q_{\rm obs}(\alpha)$ values. We therefore proceed by defining $D$--functions for $\{w_1,\,h_1,\,\xi_1,\,\bar{\theta}_1\}$, and calculate 
the single--scattering albedo proxy $\mathcal{W}$ also for F22 images.

In the following, these $\mathcal{W}$ maps are used to study four different phenomena: 1) the photometric anomaly in Hapi; 2) the Aswan cliff collapse site; 3) albedo differences between 
consolidated and brittle materials on the large lobe; 4) albedo variegation of consolidated material on the southern hemisphere. 

\begin{figure*}
\centering
\begin{tabular}{cc}
\scalebox{0.4}{\includegraphics{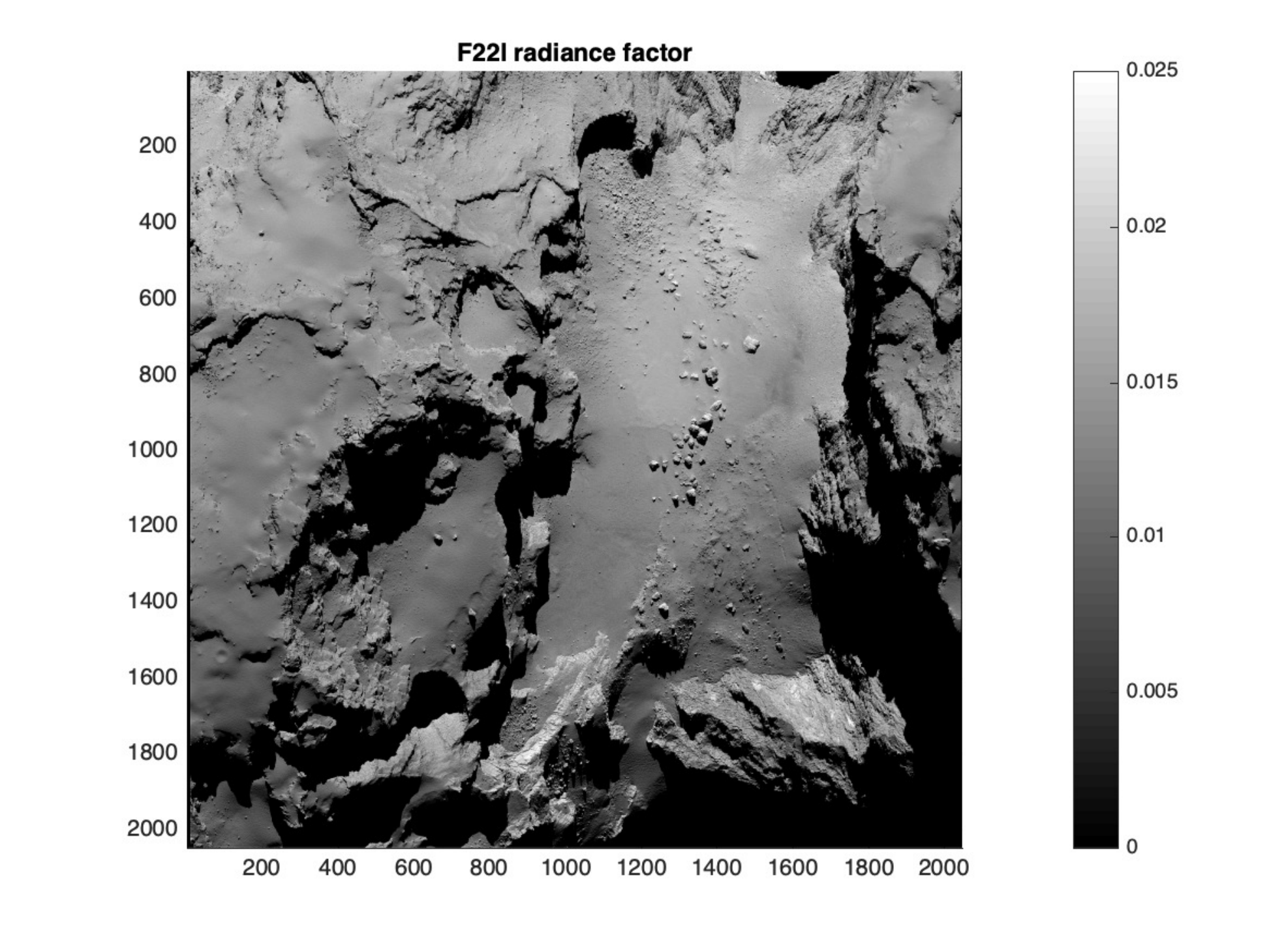}} & \scalebox{0.4}{\includegraphics{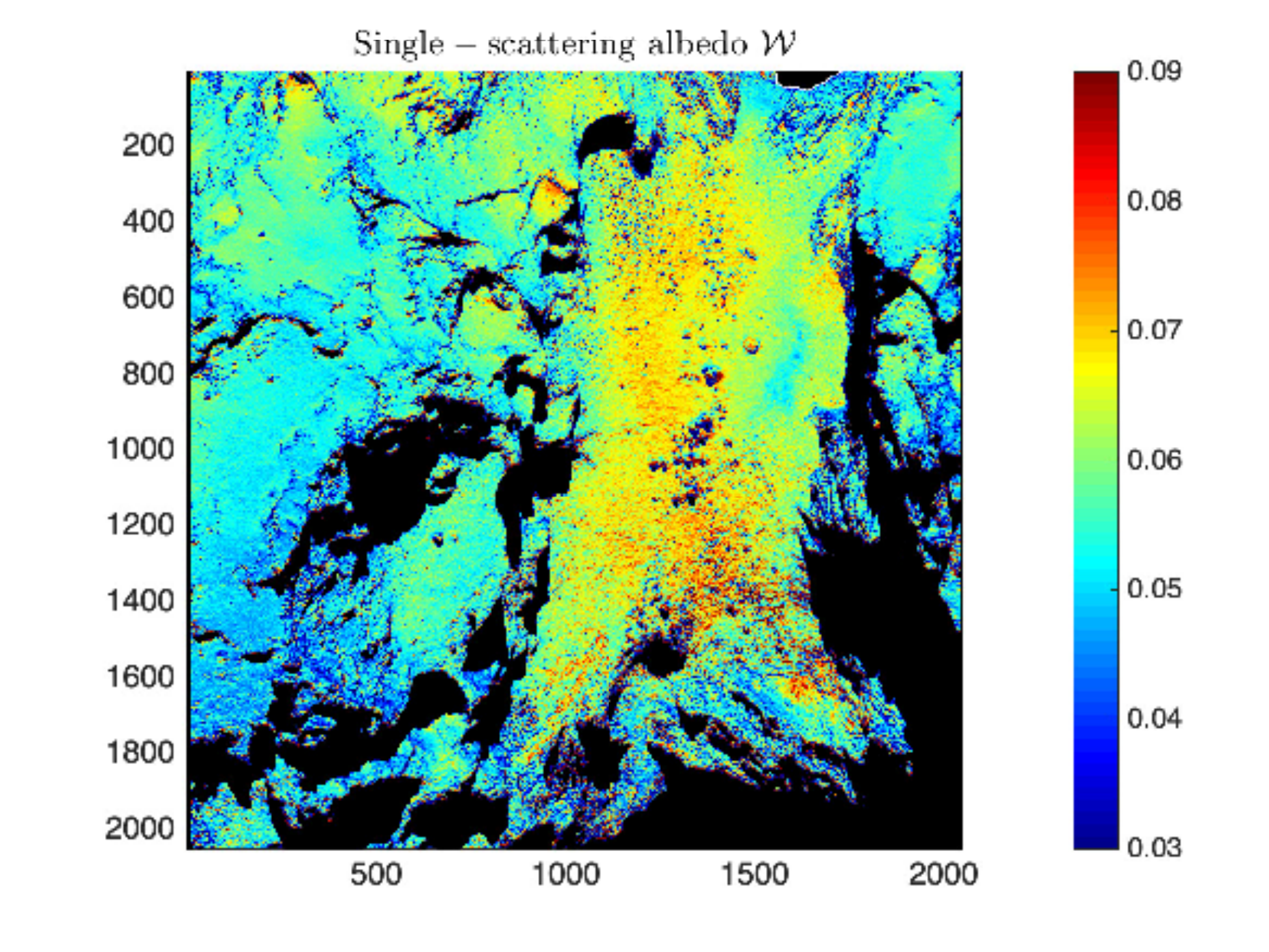}}\\
\scalebox{0.4}{\includegraphics{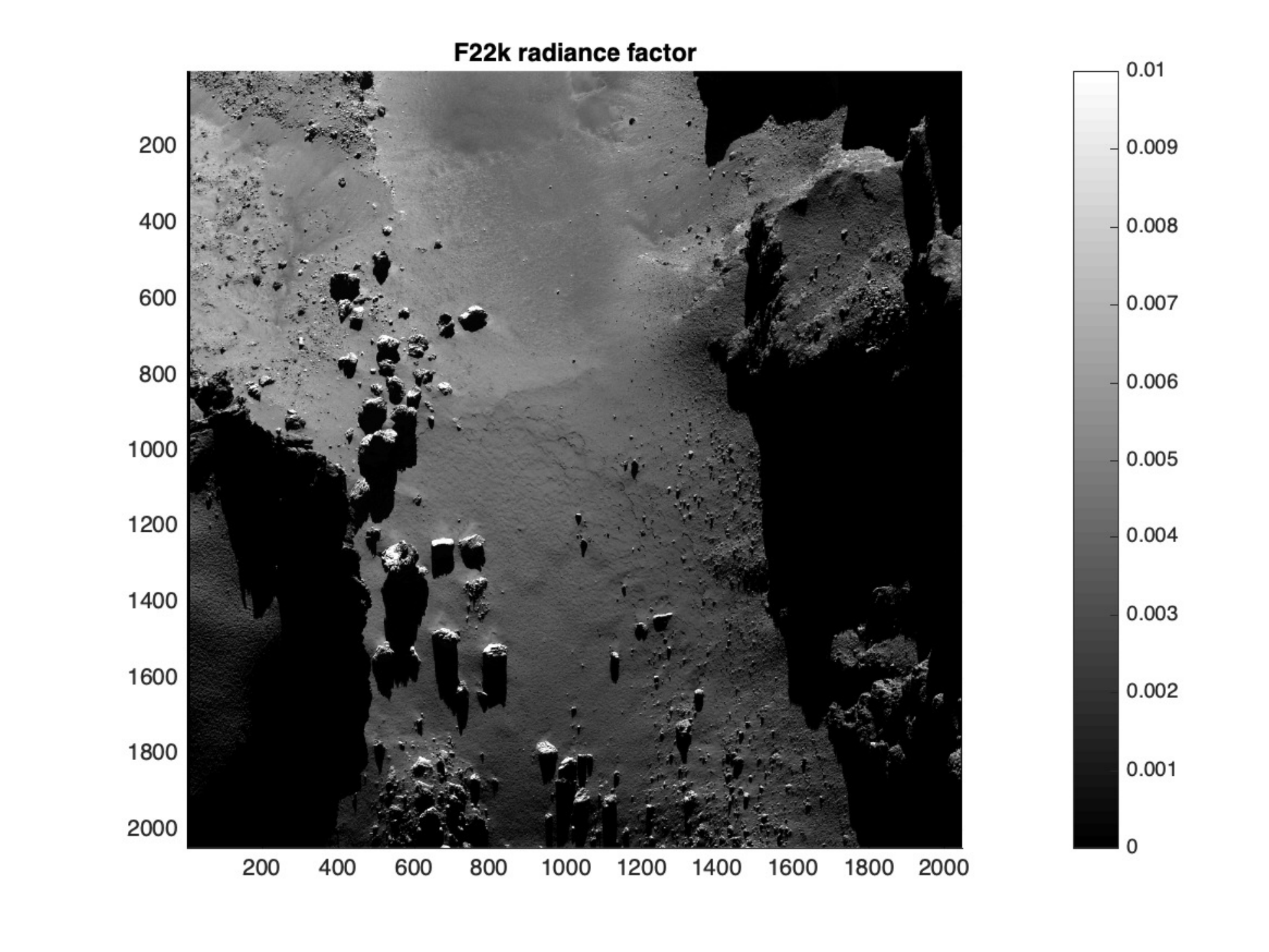}} & \scalebox{0.4}{\includegraphics{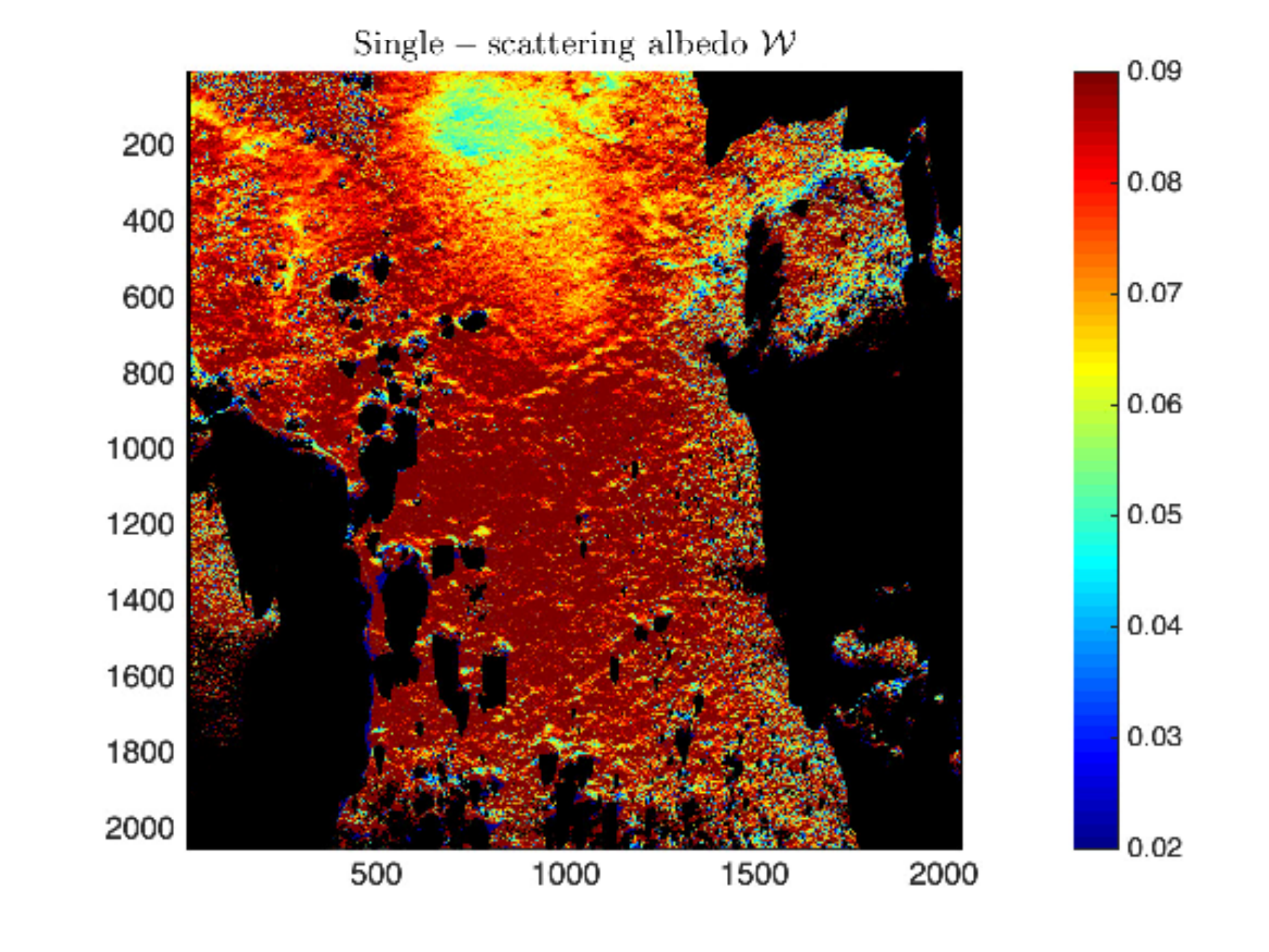}}\\
\scalebox{0.4}{\includegraphics{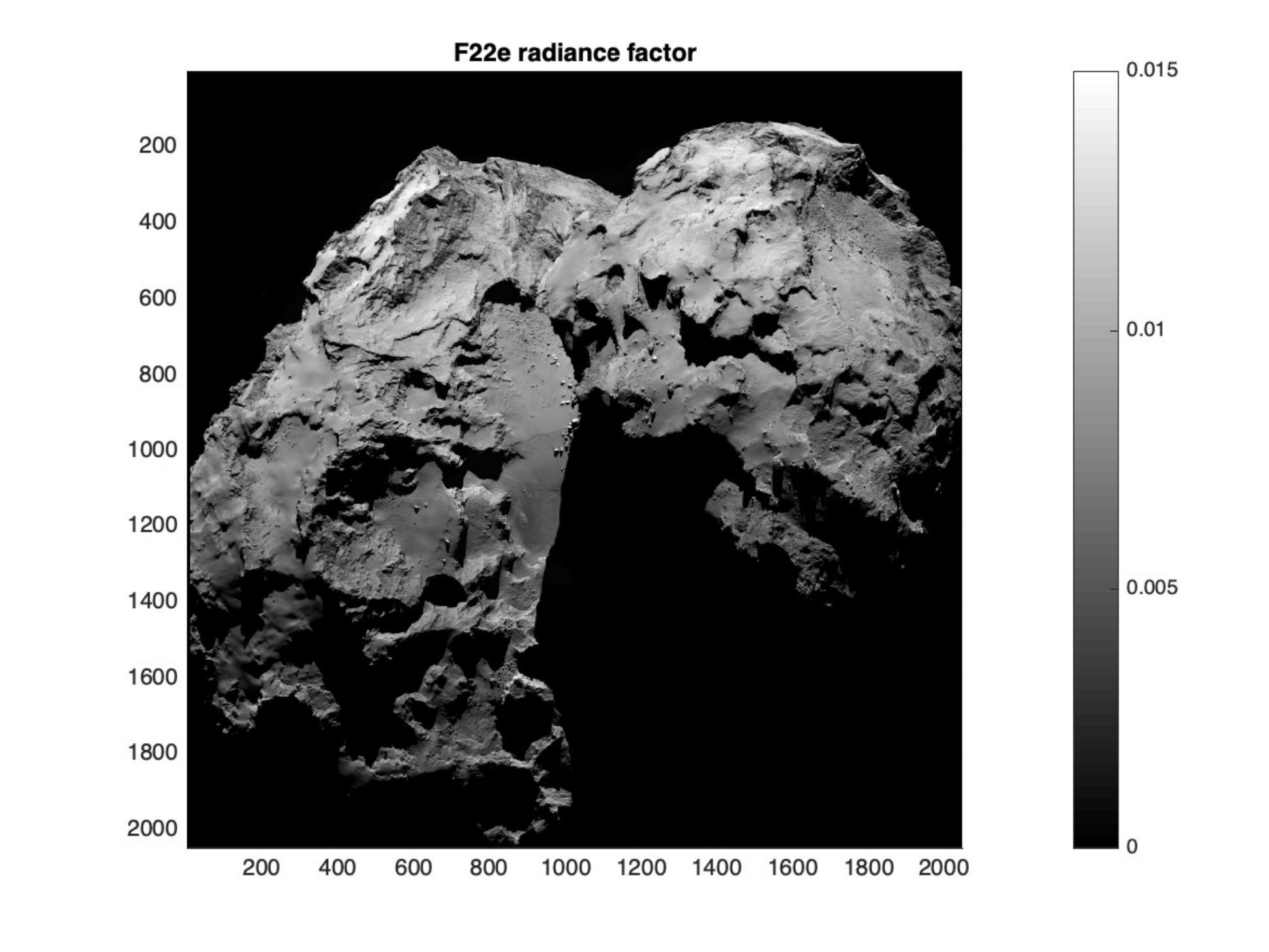}} & \scalebox{0.4}{\includegraphics{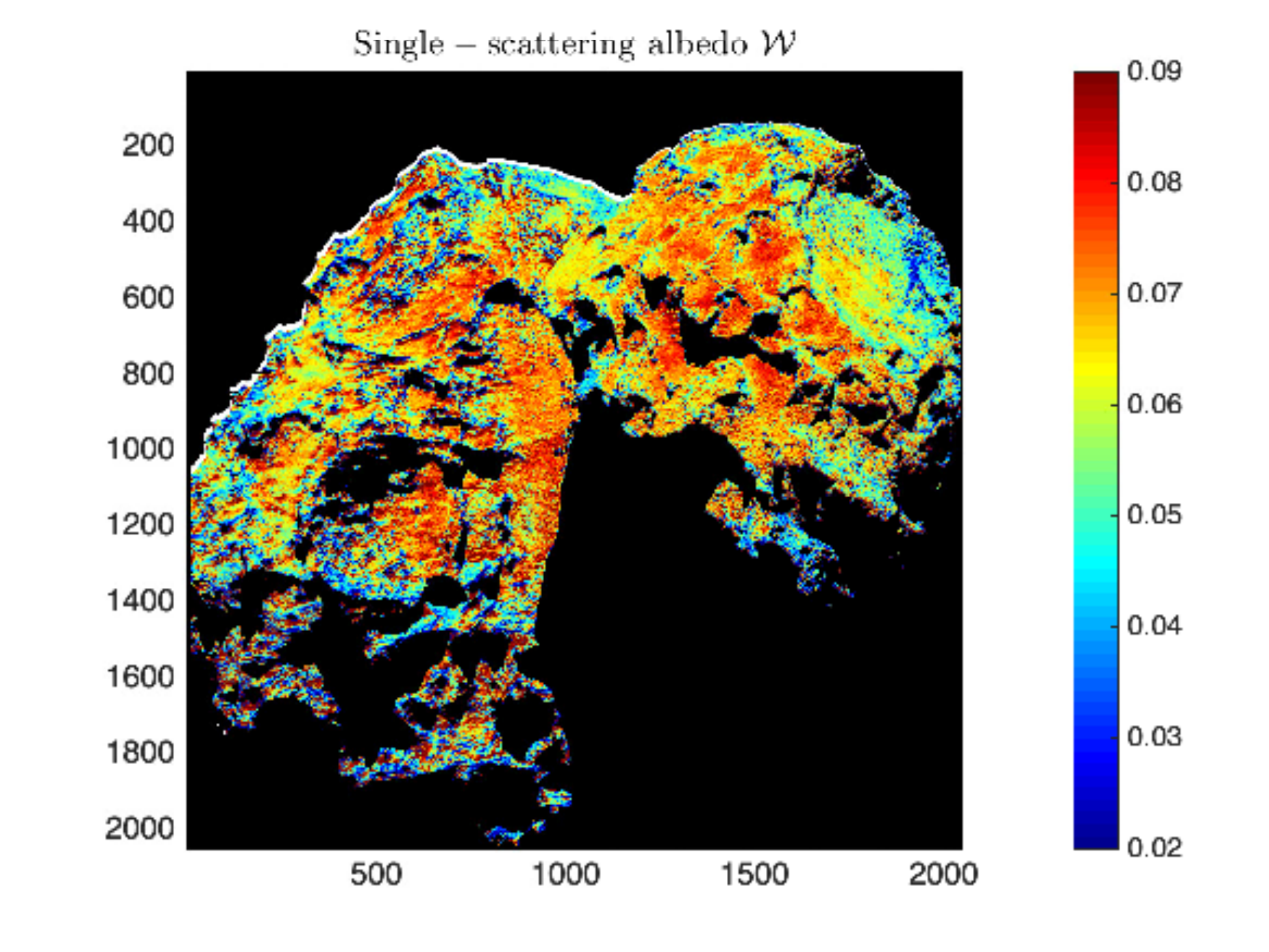}}\\
\end{tabular}
     \caption{\emph{Upper panels:} F22l at $41.2^{\circ}\leq\alpha\leq 43.4^{\circ}$ shows Hapi on 2014 August 30. \emph{Middle panels:} F22k at $90.7^{\circ}\leq\alpha\leq 92.9^{\circ}$, shows Hapi 
on 2014 December 10  (for reference, Aswan on the large lobe is located out--of--frame to the upper right: this view is rotated $\sim 180^{\circ}$ with respect to the others). 
\emph{Lower panels:} F22e at $61.7^{\circ}\leq\alpha\leq 63.7^{\circ}$ shows a panoramic view of the northern hemisphere, including Hapi, on 2015 February 28.}
     \label{fig_image_albedo_F22_01}
\end{figure*}

The part of Hapi with unusually low $\mathcal{W}$--values for smooth terrains (lower Fig.~\ref{fig_image_albedo_F82_04}) happens to be located near 
the North Pole of the comet \citep[see Fig.~1, right panel, in][for context]{pajolaetal19}. This part of Hapi experienced substantial morphological 
changes in the form of expanding shallow pits. A more southern location (here called Pit\#1) started to form around 2014 December 10 and grew during the 
following 2--3 weeks \citep[for before/after images, see Fig.~3A in][]{elmaarryetal17}. A more northern location (here called Pit\#2) started to form 
around 2014 December 30, and stopped evolving at some point between 2015 February 28 and March 17 \citep[the gradual evolution of this feature is shown in 
Figs.~1--3 in][]{davidssonetal22b}. We are here interested in understanding if and how the reflectance properties of this region may have evolved before, during, and 
after the morphological changes. 

\begin{figure*}
\centering
\begin{tabular}{cc}
\scalebox{0.4}{\includegraphics{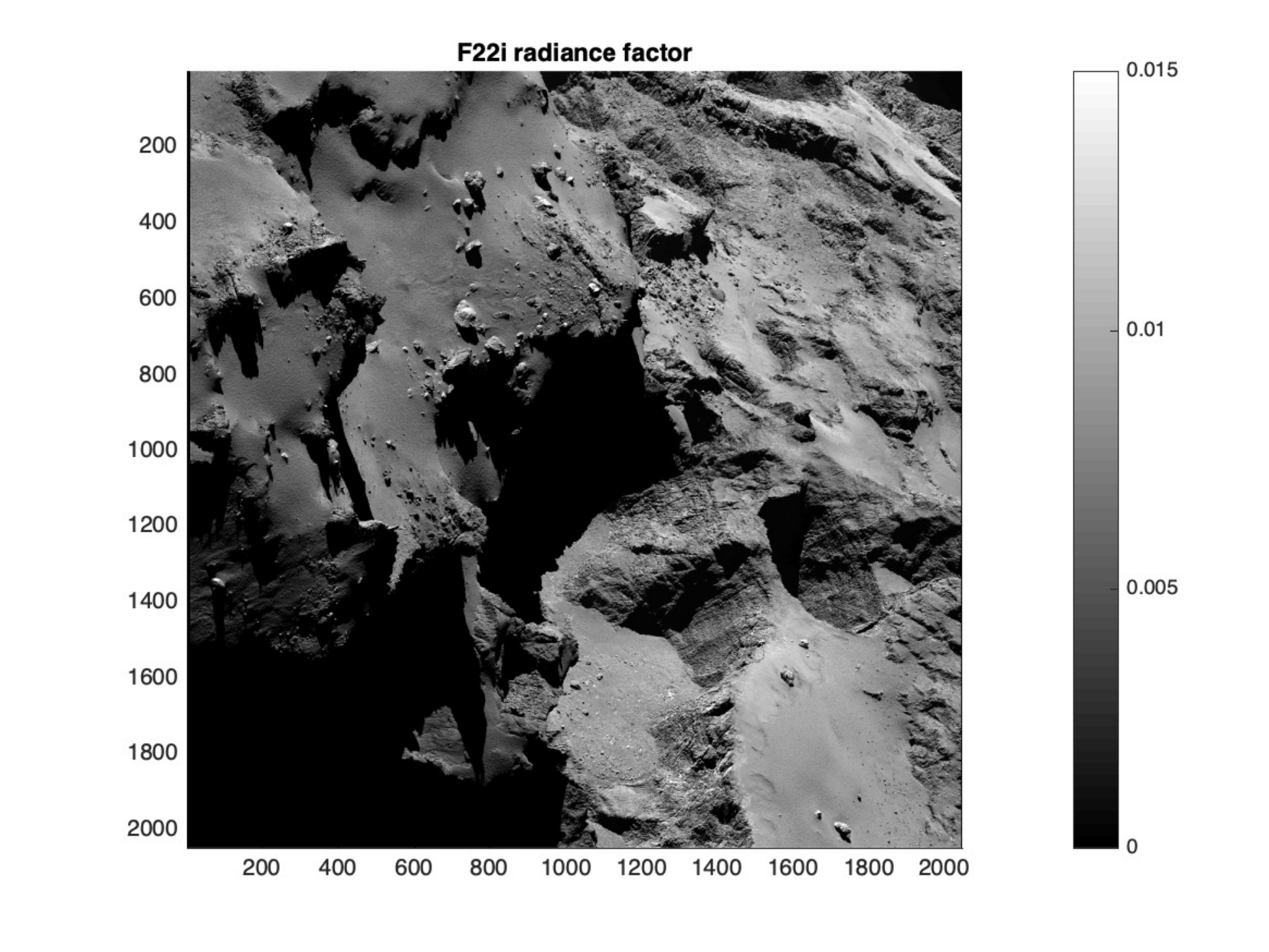}} & \scalebox{0.4}{\includegraphics{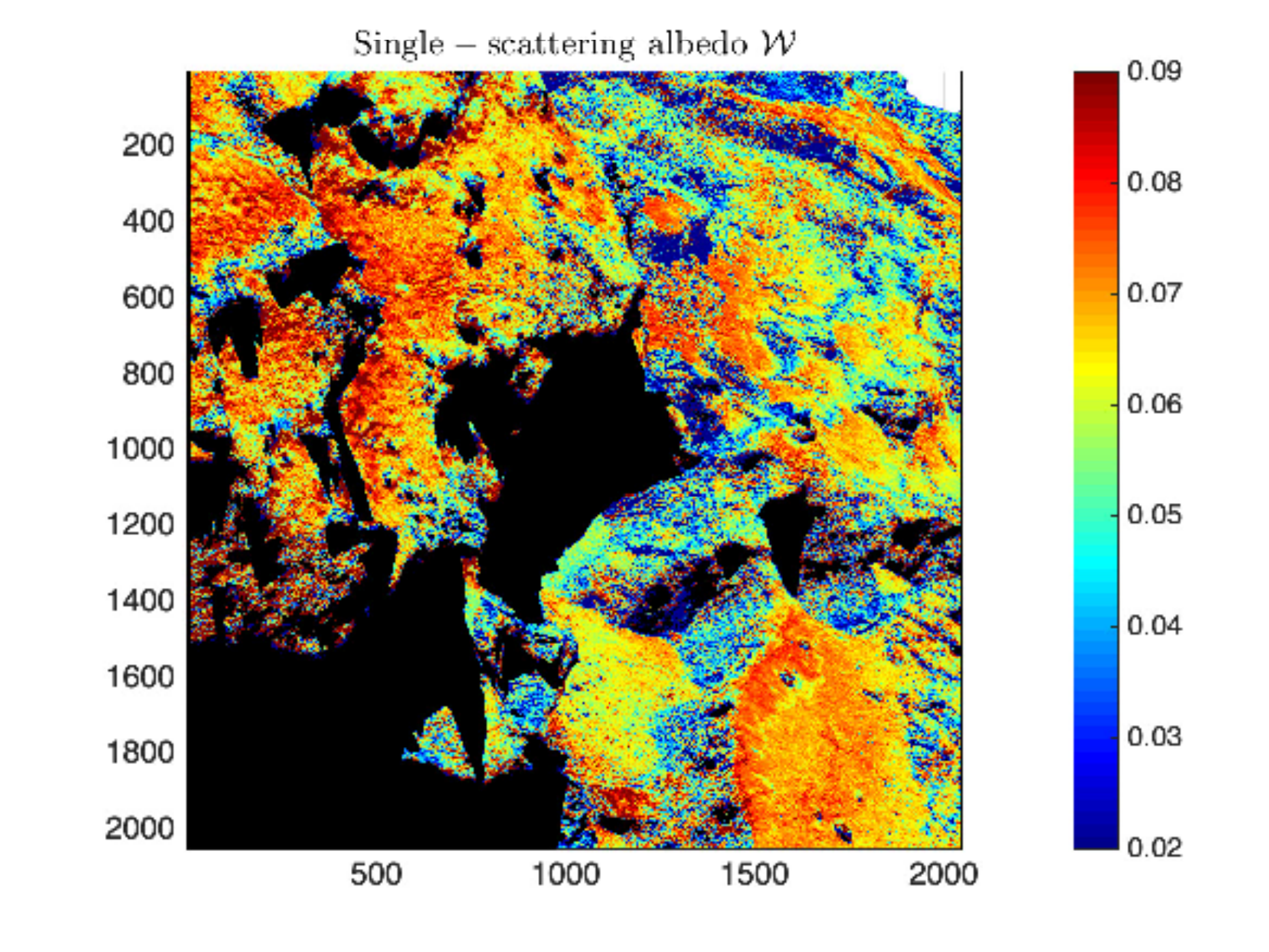}}\\
\scalebox{0.4}{\includegraphics{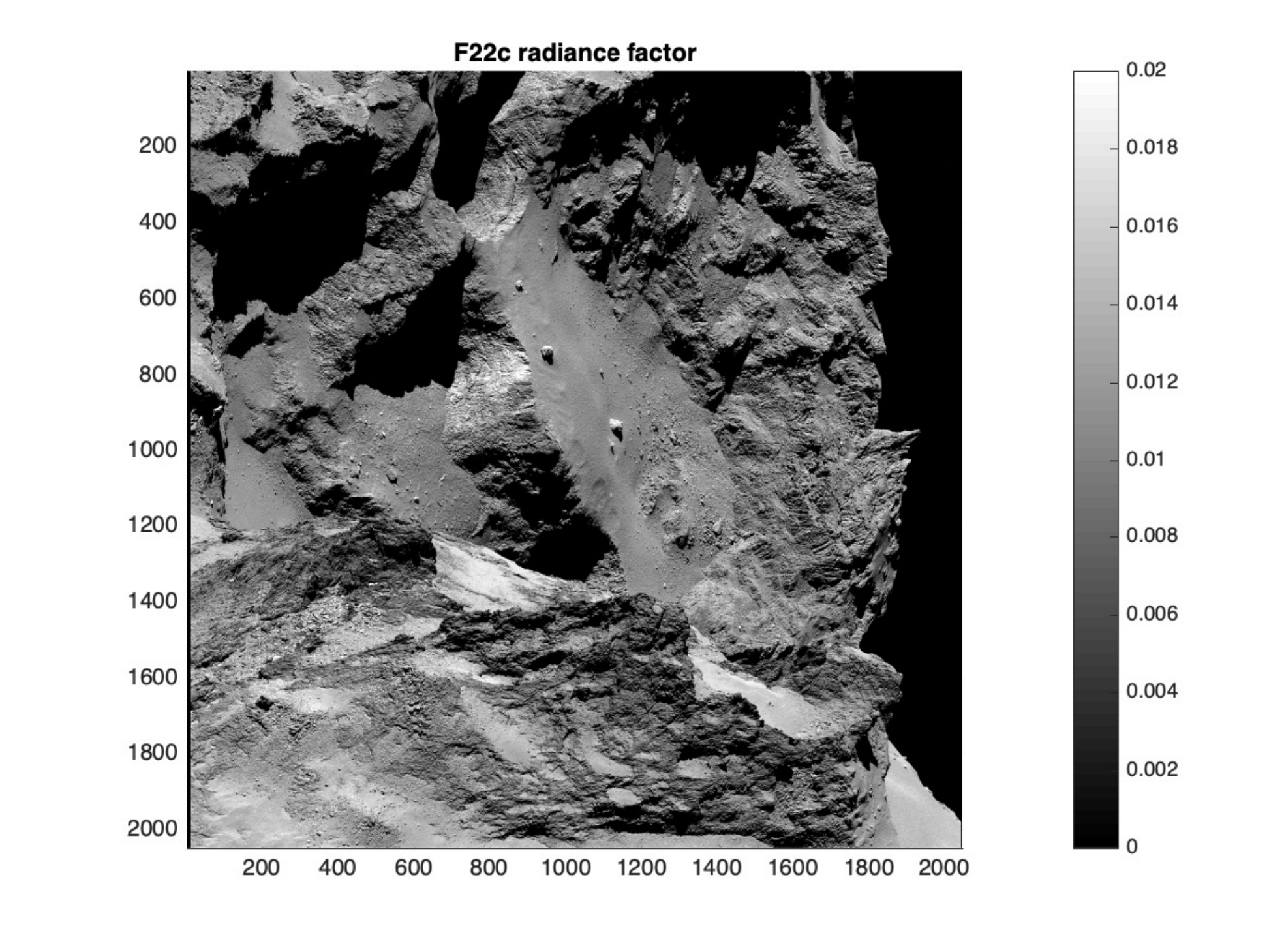}} & \scalebox{0.4}{\includegraphics{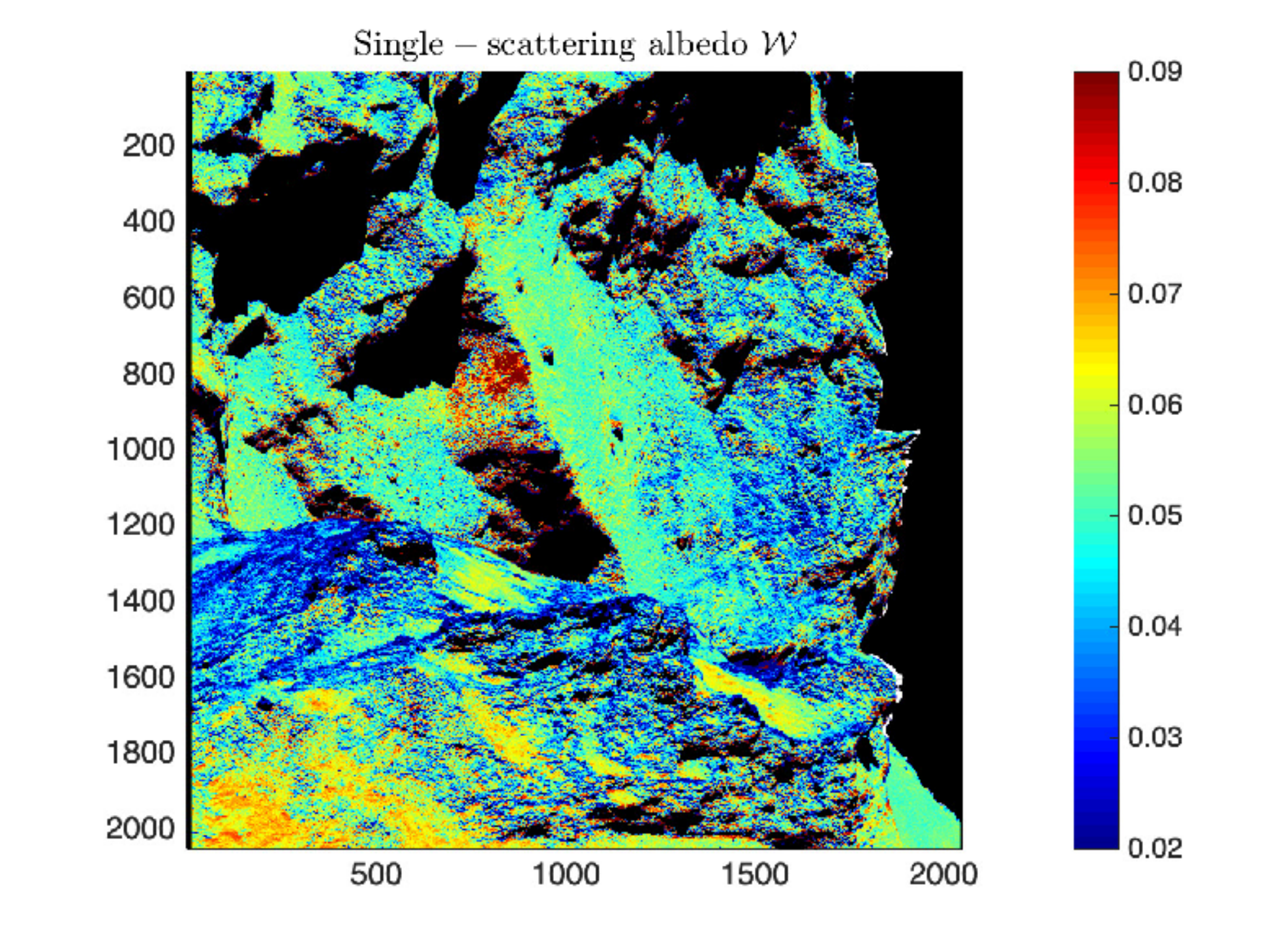}}\\
\end{tabular}
     \caption{\emph{Upper panels:} F22i at $67.2^{\circ}\leq\alpha\leq 69.5^{\circ}$, showing the Aswan cliff prior to collapse, next to the large pit of Seth in the background, with 
a part of Ma'at on the small lobe in the foreground. \emph{Lower panels:} F22c at $46.5^{\circ}\leq\alpha\leq 48.8^{\circ}$ with the Aswan cliff after collapse in the background, and 
the vertical part of Serqet on the small lobe in the foreground.}
     \label{fig_image_albedo_F22_02}
\end{figure*}

Figure~\ref{fig_image_albedo_F22_01} (top panels, F22l) shows parts of Hapi on 2014 August 30, long before any detectable changes. The area of interest is 
roughly confined between pixel coordinates 900--1200 horizontally and 1000--1400 vertically. At this point in time, the region has rather high $\mathcal{W}$ 
values that are typical for smooth terrains, and there is no major albedo variegation. Note, however, another region centred at $\{1600,\,800\}$ that evidently is darker. 
F22k (middle panels) was acquired on 2014 December 10, just days \emph{before} the first major morphological changes started sweeping through this region. It is evident 
that most of Hapi is as bright (red) as the typical smooth terrain, but that our region of interest has become unusually dark (green/yellow). This photometric anomaly has a 
rather sharp quasi--circular spatial confinement (where Pit\#1 later would form), though a less dark (yellow) 'tongue' stretches downwards in the image (where Pit\#2 later would form). 
It therefore seems like the material \emph{first} darkened, \emph{then} initiated pit formation: an important sequence of events we will return to in section~\ref{sec_discussion}. 

The lower panels of Fig.~\ref{fig_image_albedo_F22_01} show a panoramic view of the northern hemisphere (F22e), including the Hapi region, acquired on 2015 February 28. 
At that time, most morphological changes likely had stopped. The region under discussion (near pixel coordinates $\{900,\,1200\}$, just right of the Aswan plateau) is still unusually dark 
(though variegation is not so clear at this relatively low resolution). We also note that F82h (Fig.~\ref{fig_image_albedo_F82_04}, lower panels) was acquired 2015 March 28, 
confirming that the photometric anomaly remained at that time.

We do not believe the anomaly is due to shortcomings of the shape model. Some shape model imperfections are undoubtedly present, because the same shape model (providing $\{i,\,e,\,\alpha\}$ 
for each pixel) was used to analyse all images, and the morphology in this particular region evidently changed with time. However, those changes are small: the depth of the 
pits are $\sim 0.5\,\mathrm{m}$, similar to the $\sim 1\,\mathrm{m}$ resolution of the shape model. The region remains smooth and quasi--flat throughout the \emph{Rosetta} mission and changes to 
local $\{i,\,e,\,\alpha\}$--values at the resolution of the shape model would have been small, if attempts had been made to produce an evolving shape model. We do not know which images 
went into producing the shape model, but it was likely a mixture acquired prior and after the changes (thus representing an average surface). The photometric anomalies are too large to 
be due to small mis--alignments of the shape model facets. Furthermore,  in the radiance factor images themselves, the Pit\#1--2 region is visibly darker compared to its surroundings in December 2014
(middle--left panel of Fig.~\ref{fig_image_albedo_F22_01}), which is not the case in August 2014 (upper--left panel of Fig.~\ref{fig_image_albedo_F22_01}).

Next, we consider the Aswan cliff collapse site.  Figure~\ref{fig_image_albedo_F22_02} shows the Aswan cliff and plateau in the Seth region on the large lobe, both prior (upper panels) and after (lower panels) to the cliff 
collapse that took place in July 2015 \citep{pajolaetal17}. In both cases, there are small--lobe terrain in the foreground (Ma'at in the upper panels and the vertical cliff in 
Serqet in the lower panels). We first note that the consolidated terrain (exposed on the vertical cliffs) all are relatively dark, and with an insignificant degree of internal albedo variegation. The $\mathcal{W}$--value 
of $0.03\pm 0.01$ is similar on the Aswan cliff, on the walls of the large Seth pit \citep[discussed by][]{vincentetal15}, and on the Serqet cliff. As previously pointed out, the contrast between the 
dark consolidated terrain and the brighter smooth terrain is evident (particularly on the patches of airfall on small ledges in Serqet, as well as on the Aswan plateau and in Ma'at). We also point out, that the 
smooth material along the edge to the Aswan cliff is particularly bright, within a wavy region where the airfall layer is particularly thin, presumably due to efficient ejection. 

\begin{figure*}
\centering
\begin{tabular}{cc}
\scalebox{0.4}{\includegraphics{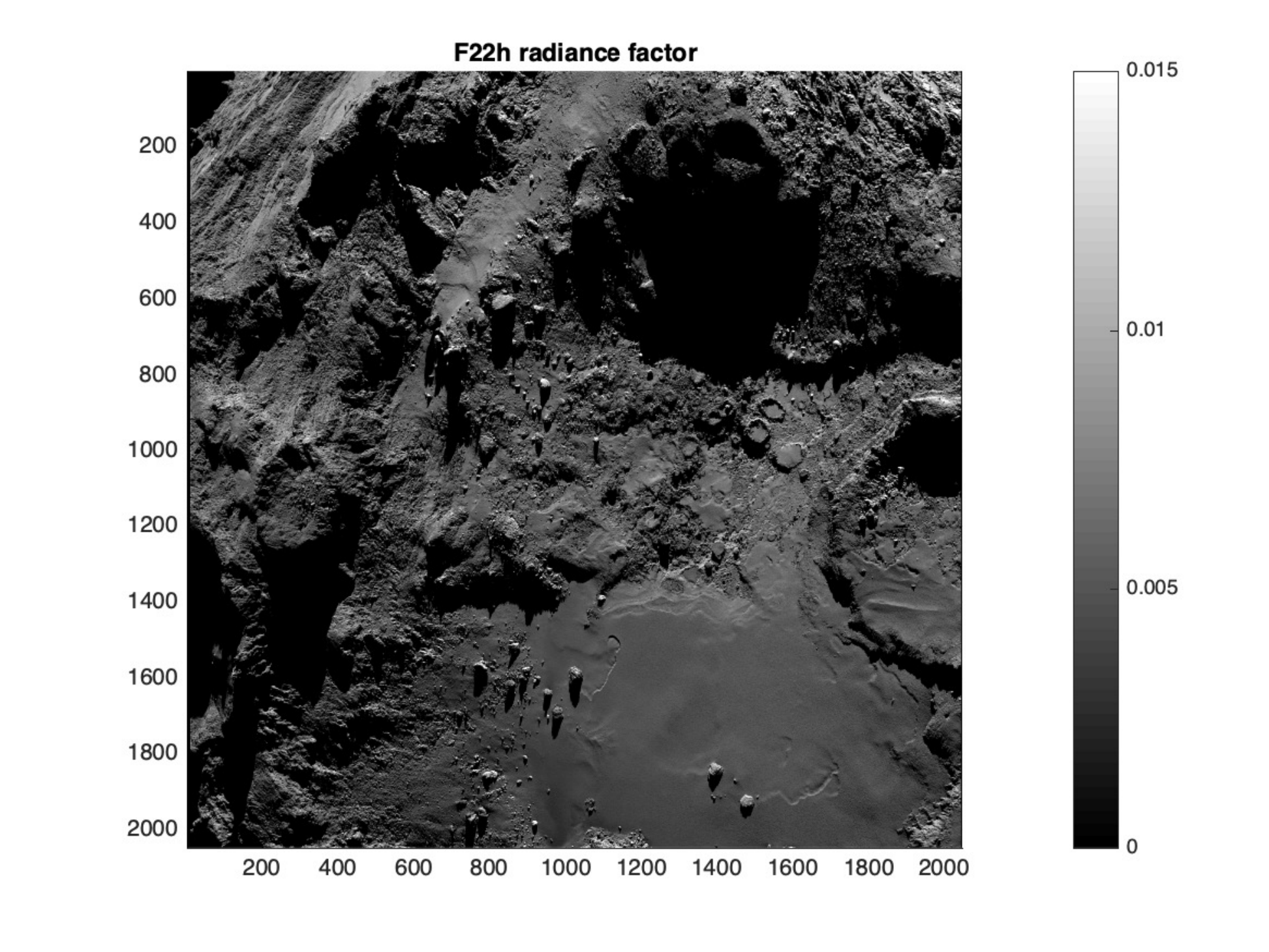}} & \scalebox{0.4}{\includegraphics{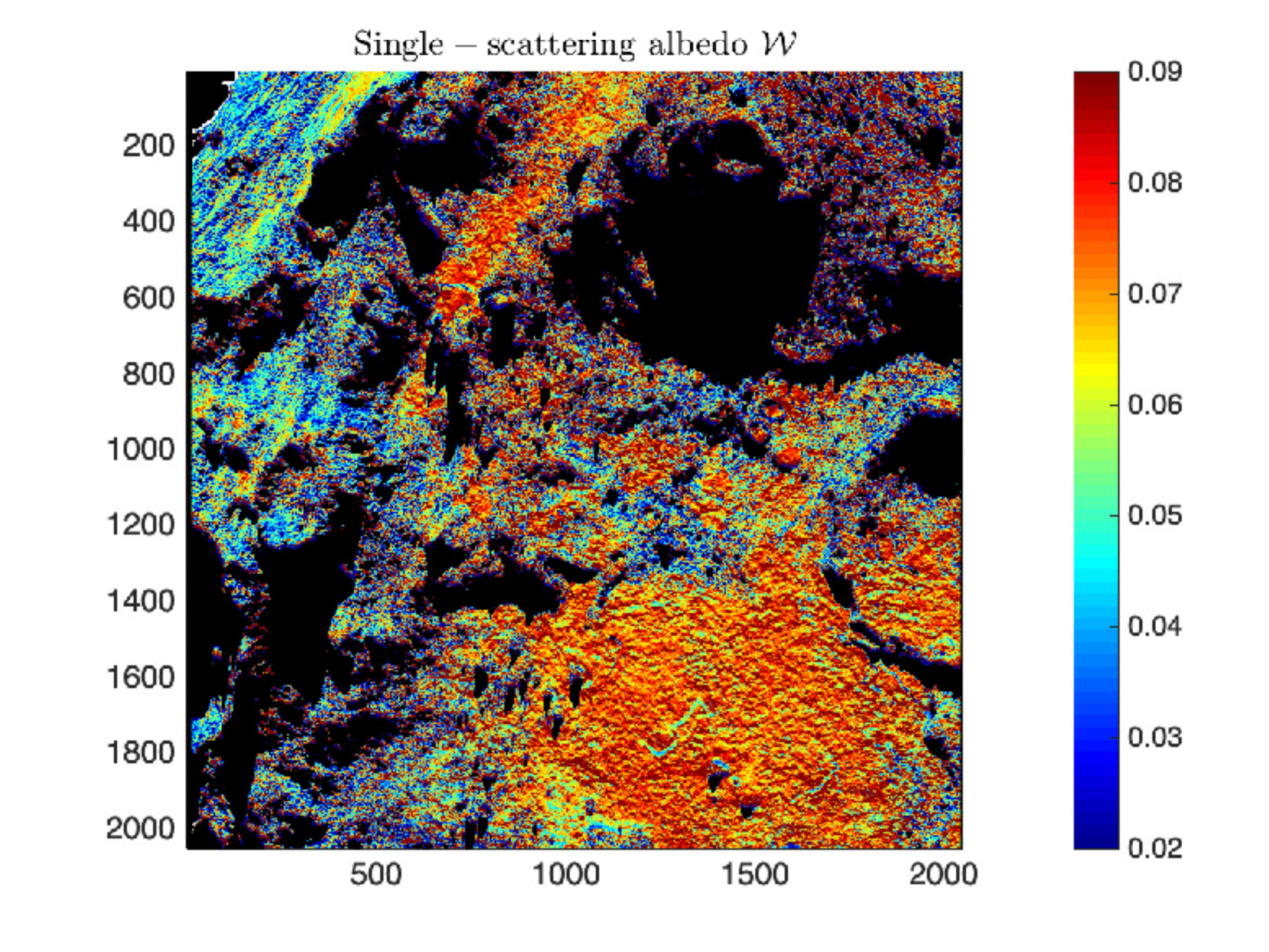}}\\
\scalebox{0.4}{\includegraphics{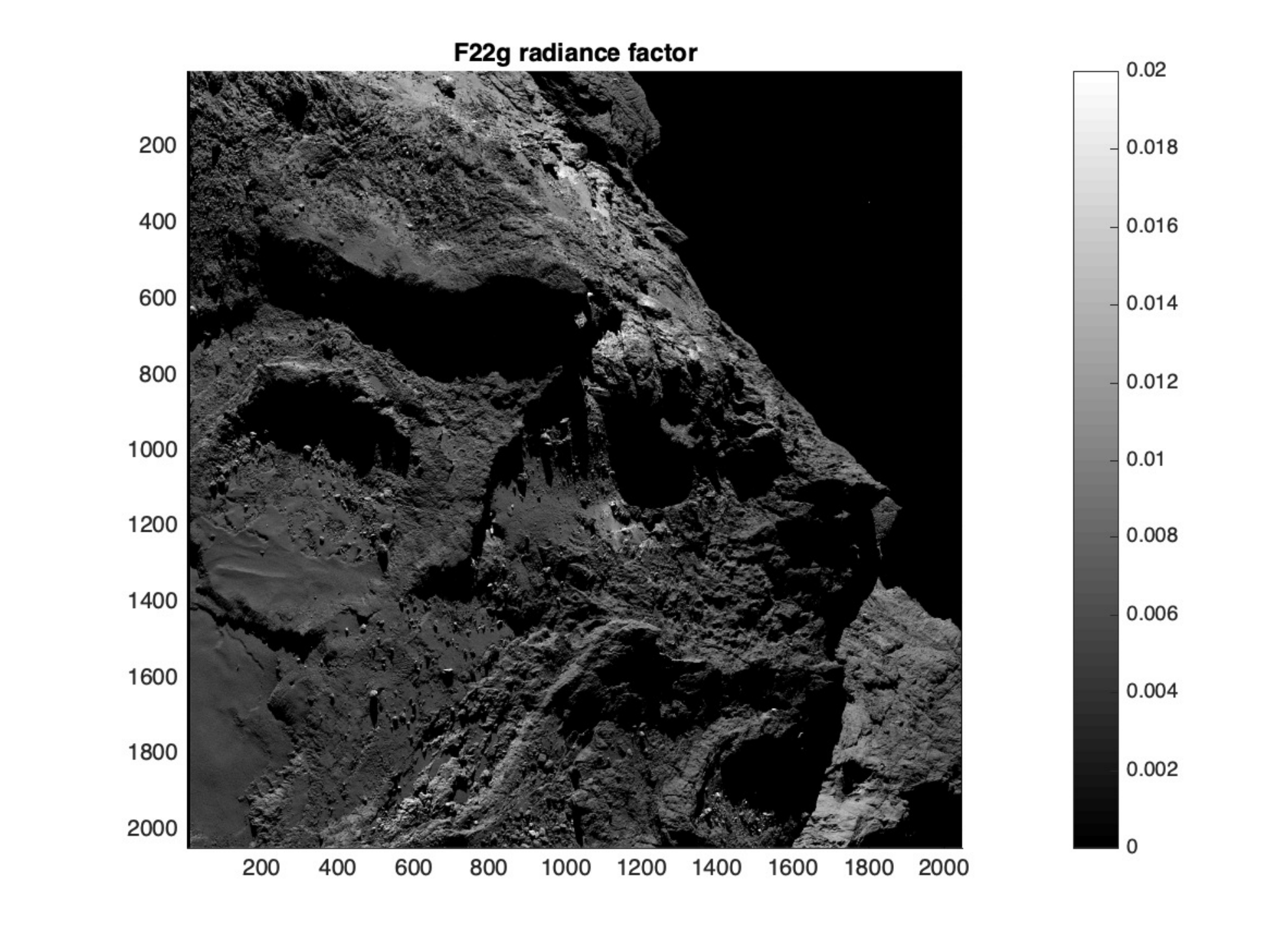}} & \scalebox{0.4}{\includegraphics{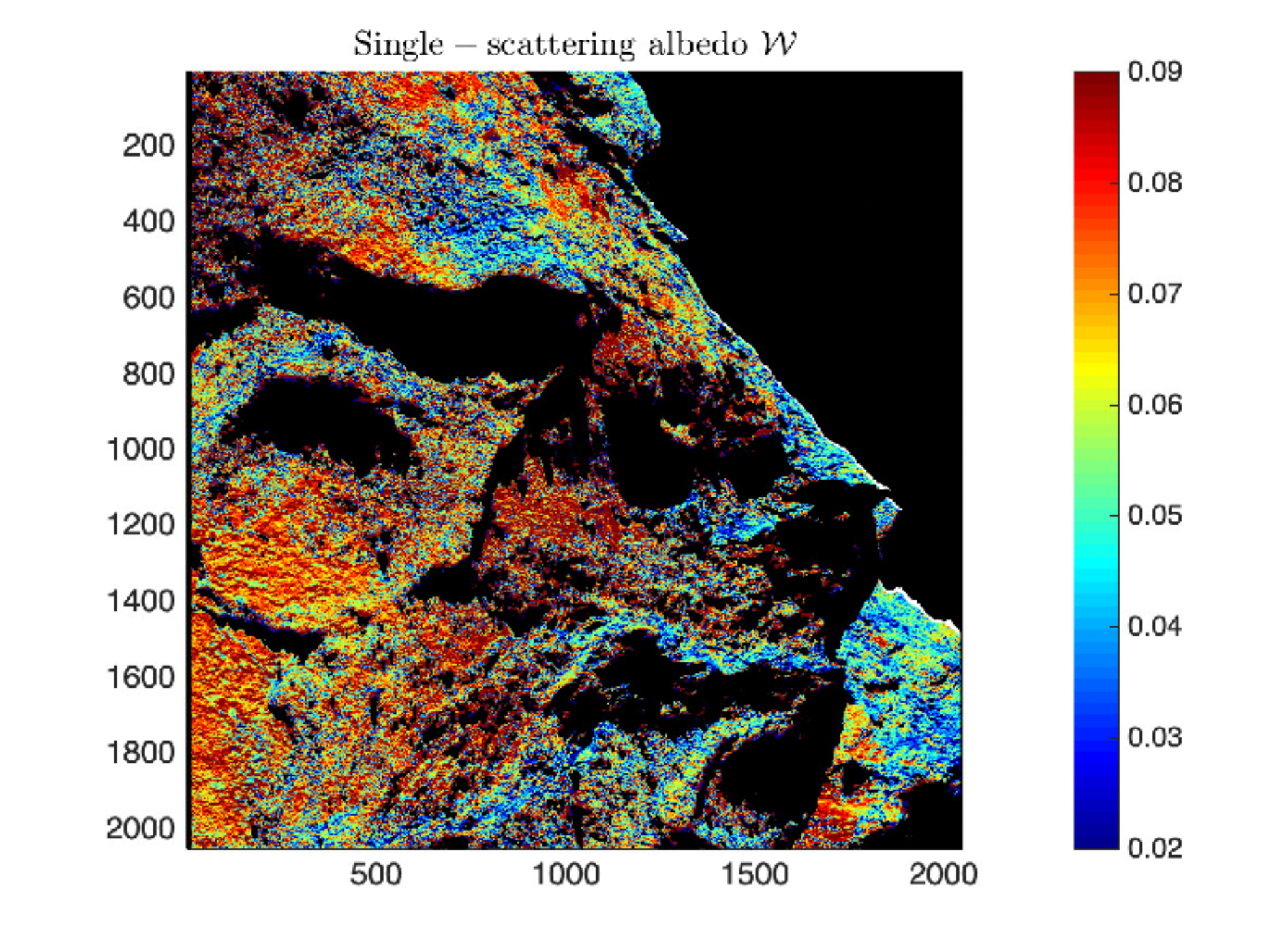}}\\
\end{tabular}
     \caption{\emph{Upper panels:} F22h at $64.6^{\circ}\leq\alpha\leq 66.8^{\circ}$ showing Imhotep, with a piece of Ash in the upper left corner. \emph{Lower panels}: F22g at 
$67.8^{\circ}\leq\alpha\leq 66.8^{\circ}$ showing Imhotep in the lower right corner (the smooth terrain and the accumulation basin), transiting upward to Khonsu, and to the right to Bes. At 
the lower--right corner, note a piece of the small lobe in the background.}
     \label{fig_image_albedo_F22_03}
\end{figure*}

The scar on the cliff wall \citep[removal of a chunk with dimensions $80\times 65\times 12\,\mathrm{m}$;][]{pajolaetal17} near pixel coordinates $\{900,\,800\}$ in Fig.~\ref{fig_image_albedo_F22_02} 
(lower panels) is particularly bright (red) compared to the surrounding walls. We caution that part of this effect could be due to discrepancies between the shape model and the actual topography in the 
two sets of images (that are much more substantial than the previously discussed Hapi case). Yet, there is little doubt that the scar is brighter \citep[the normal albedo that was at least $0.4$ right after 
collapse, was down to $\sim 0.17$ by December 2015, and further darkened to $\stackrel{<}{_{\sim}} 0.12$ by August 2016, two months after F22c was obtained, though there were patches 
as bright at $\sim 0.18$ even at that time;][]{pajolaetal17}. This should be compared to the average normal albedo (and typical range) of $0.07\pm 0.02$ of the nucleus \citep{fornasieretal15}.

We now move on to consider brittle terrain. Distinct from the `rocky' appearance of strongly consolidated terrain, brittle terrain is coarsely crumbled, boulder--rich, and often constitutes 
talus--like mass wasting in the vicinity of more consolidated terrain \citep{thomasetal15a,elmaarryetal15}. Figure~\ref{fig_image_albedo_F22_03} (top panels) shows brittle terrain 
in Imhotep \citep[sub--unit c;][]{thomasetal18}, wedged between consolidated terrain in the upper--left corner (Ash), and smooth terrain (Imhotep sub--unit a) and an accumulation 
basin (Imhotep sub--unit b) in the lower--right corner. The $\mathcal{W}$--map shows the consolidated terrain in Ash and on the rim of the accumulation basin as blue (particularly dark). 
Smooth material in Imhotep sub--unit a, and within the accumulation basin, is substantially brighter (red). Instead, the brittle terrain has a distinct purple hue for the applied colour scale, 
except for a long and narrow ledge that has accumulated reddish airfall material. The purple coloration arises because of numerous small blue and red spots, due to an intimate mixture of 
darker and brighter patches. 

Figure~\ref{fig_image_albedo_F22_03} (lower panels) show an adjacent region (parts of the smooth area and the accumulation basin in Imhotep sub--units a and b are still visible). 
The interesting portion of the image is at the rightmost part of the large lobe (in the foreground of the small lobe visible in the lower right corner of the image). Here, the brittle 
terrain Bes sub--unit a is wedged between two consolidated and distinctively blue (low--$\mathcal{W}$) regions: Khonsu sub--unit c on top (seen nearly edge--on) and Bes sub--unit b below, 
that contains several prominent ridges. Also note that Khonsu sub--unit c transits towards the left into Khonsu sub--unit a, that forms another dark consolidated outcrop just above the 
accumulation basin. The brittle terrain is clearly brighter on average than the consolidated terrain.

\begin{figure*}
\centering
\begin{tabular}{cc}
\scalebox{0.4}{\includegraphics{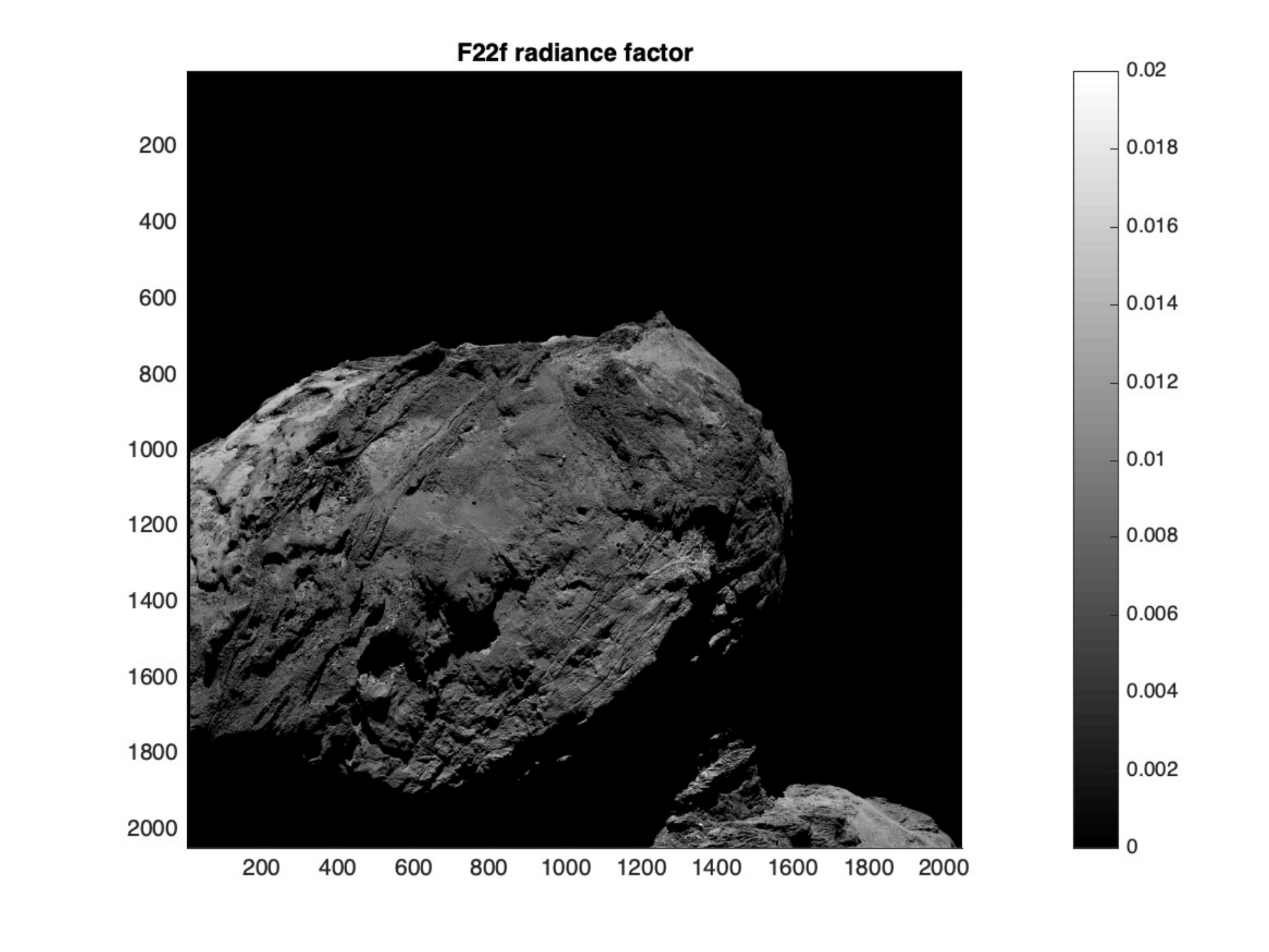}} & \scalebox{0.4}{\includegraphics{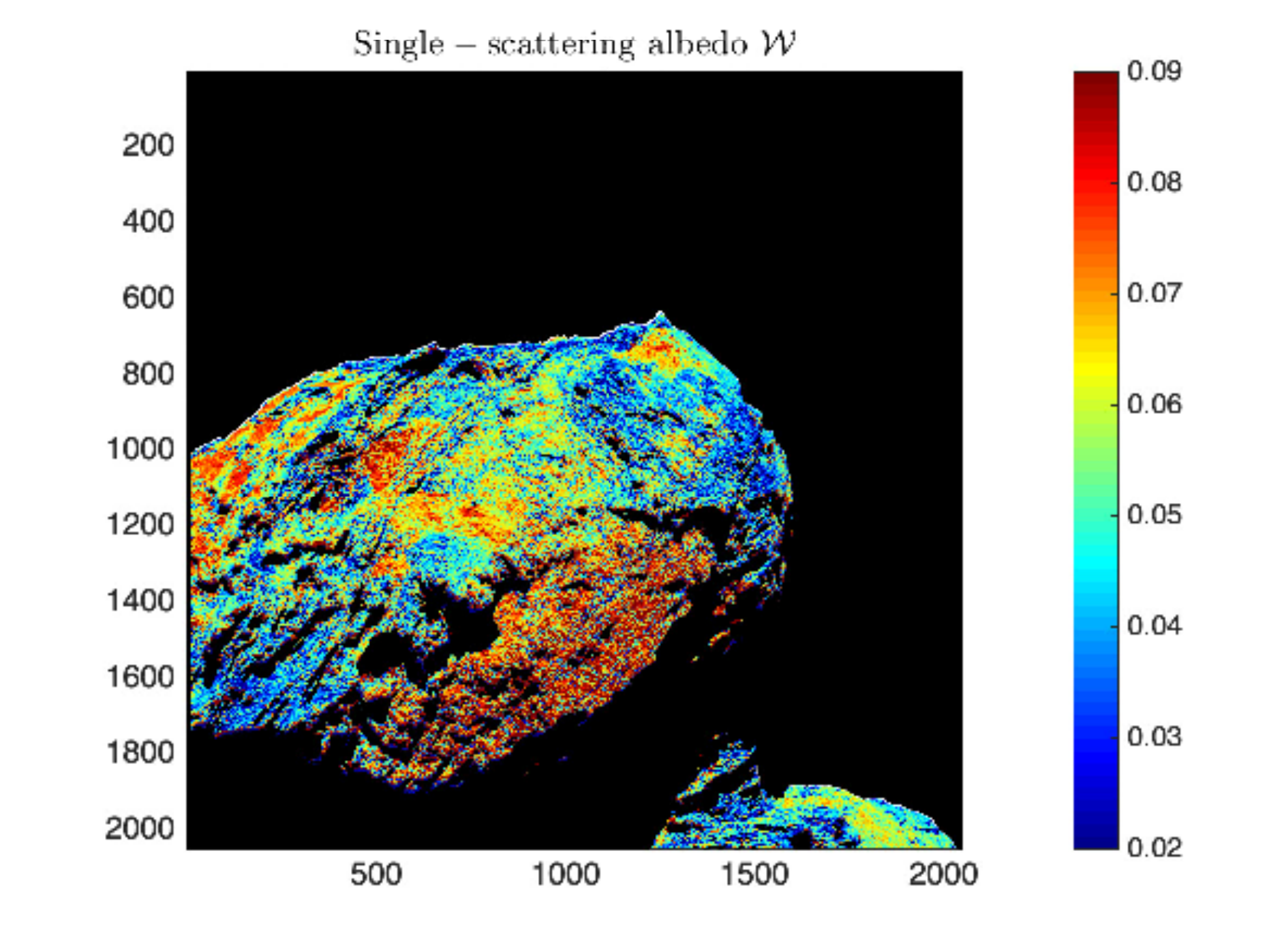}}\\
\scalebox{0.4}{\includegraphics{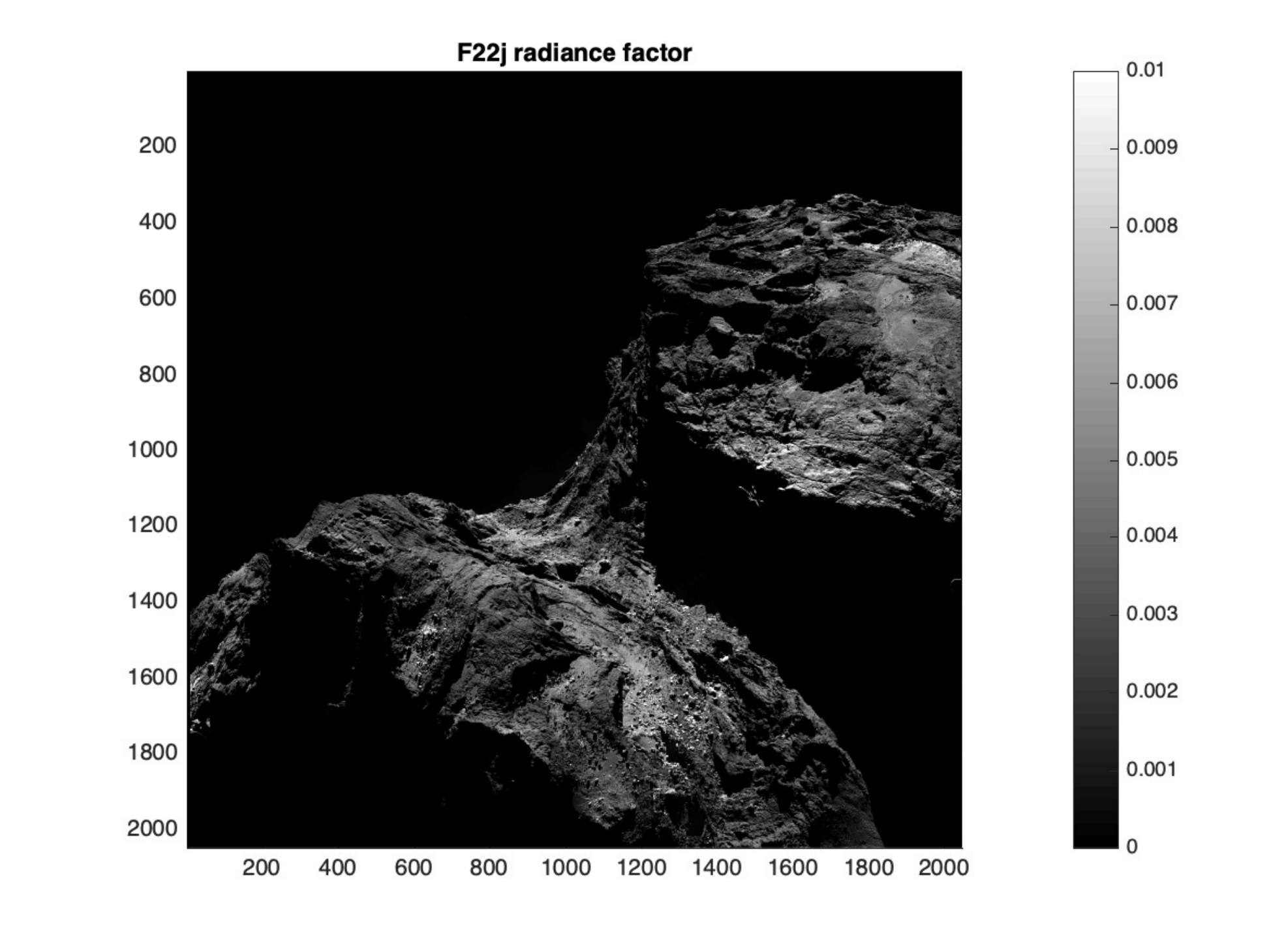}} & \scalebox{0.4}{\includegraphics{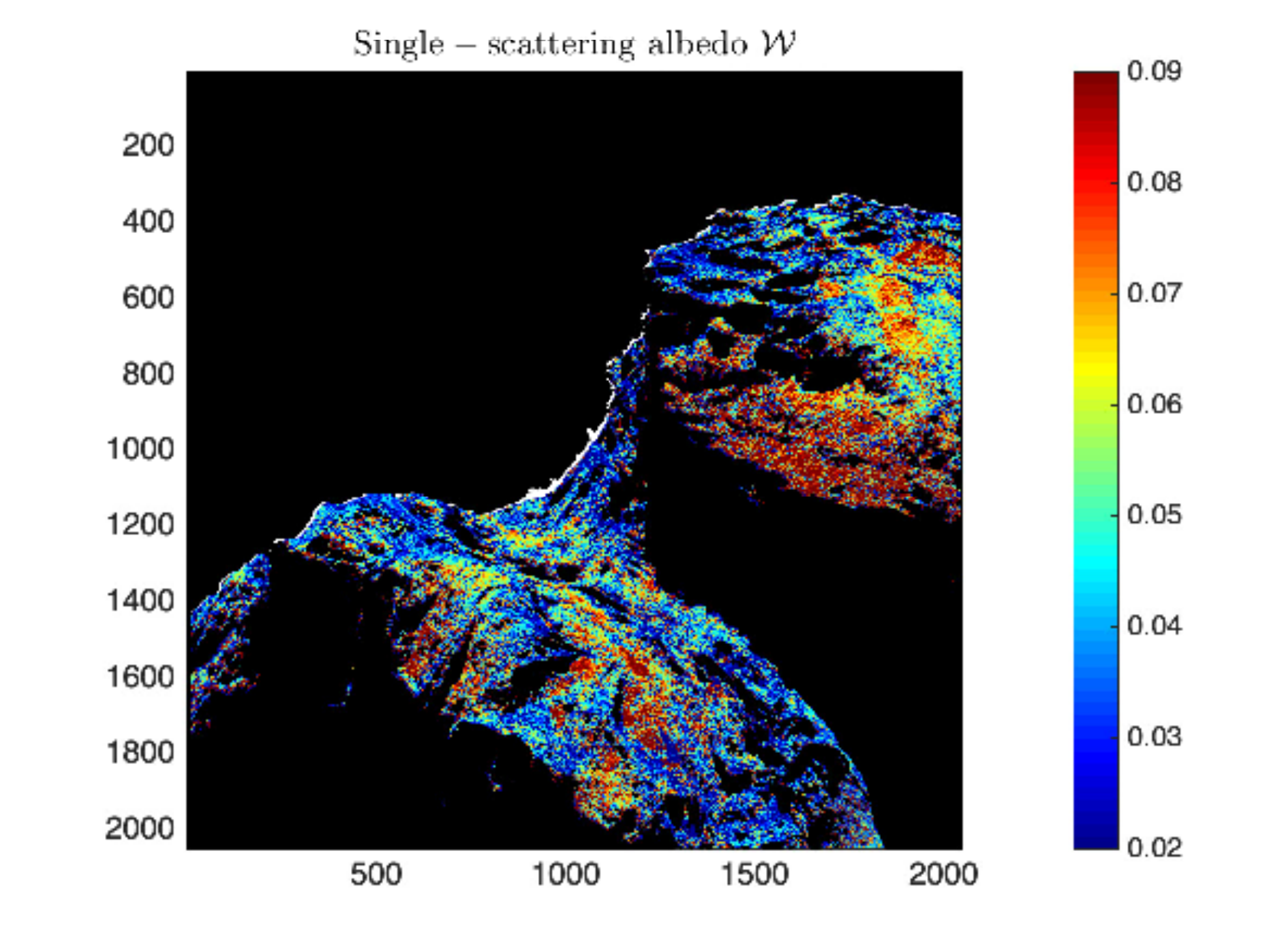}}\\
\end{tabular}
     \caption{\emph{Upper panels:} F22f at $63.6^{\circ}\leq\alpha\leq 65.1^{\circ}$ shows the southern hemisphere of the small lobe: the bright Maftet at the top left,  and the rather 
bright Wosret b at the bottom. Sandwiched between the two (going from left to right): the dark Wosret c, the partially dust--covered (thus brighter) Wosret a, and the dark Bastet a. 
\emph{Lower panels:} F22j at $88.9^{\circ}\leq\alpha\leq 90.8^{\circ}$ showing southern hemisphere sides of both the small lobe (on top) and the large lobe (at the bottom).}
     \label{fig_image_albedo_F22_04}
\end{figure*}

Finally, we consider the southern hemisphere, that almost exclusively consists of consolidated terrain \citep[][]{elmaarryetal16}. Figure~\ref{fig_image_albedo_F22_04} shows the 
small lobe, dominated by the Wosret region (though a small part of Maftet is visible to the upper left, recognisable by its high--$\mathcal{W}$ smooth material coverage, and the rightmost part of 
the small lobe belongs to Bastet). Wosret/Bastet has two distinct regions: 1) an upper band dominated by very dark (blue) consolidated material, except for patches at the centre that consist of brighter (yellow/red) smooth material; 
2) a lower band that also has the morphology of consolidated material, but with a systematically higher albedo, except for a few long and thin dark streaks, some of which are stretching 
the full length of lower Wosret \citep[called sub--unit b by][]{thomasetal18}. The brighter, lower part of Wosret is also clearly seen in Fig.~\ref{fig_image_albedo_F22_04} (lower panels), that also shows large portions of the 
southern--hemisphere large lobe. This part of the nucleus, dominated by the Anhur and Geb regions, consists mostly of dark consolidated terrain, with very few patches of 
bright smooth material. Figure~\ref{fig_image_albedo_F22_04} therefore demonstrates albedo variegation \emph{within} consolidated terrain. This is highly interesting, because consolidated 
terrains are considered the least processed material that is exposed at the surface, suggesting that the nucleus may be heterogeneous on  a global level.

\section{Discussion} \label{sec_discussion}

We begin by comparing our disk--average Hapke--parameter solutions to those of other OSIRIS NAC orange filter fits in the literature (Table~\ref{tab3}). 
We first note that our single--scattering albedo increases, and our cosine asymmetry factor becomes less negative, when we extend the 
considered phase angle interval from $0^{\circ}\leq \alpha\leq 16^{\circ}$ ($a=0$) to $0^{\circ}\leq \alpha\leq 69.6^{\circ}$ ($a=1$). 
In essence, the fit worsens somewhat at low phase (but is still consistent with $Q_{\rm obs}$ to within the standard deviations of the data bins, 
see Fig.~\ref{fig_4pic_8pic_fits}), in order to better accommodate the high--phase data.

\begin{table}
\begin{center}
\begin{tabular}{||l|l|l|l|l|l||}
\hline
\hline
$w$ & $h$ &  $\xi$ & $\bar{\theta}\,\mathrm{[deg]}$ & $\alpha\,\mathrm{[deg]}$ & Reference\\
\hline
0.045 & 0.026 & $-0.41$ & $15$ & 1.3--54 & Fo15, H2002\\
0.034 & 0.061 & $-0.42$ & 28 &  1.3--54 & Fo15, HHS\\
0.046 & 0.064 & $-0.37$ & 15.6 & 1.0--33 & Fe16, H2002\\
0.038 &  0.064 & $-0.37$ & 15.6 & 1.0--33 & Fe16, HHS\\
0.033 & 0.046 & $-0.561$ & -- & 0--16.1 & This work ($a=0$)\\
0.055 & 0.035 & $-0.456$ & 16.2 & 0--69.6 & This work ($a=1$)\\
\hline 
\hline
\end{tabular}
\caption{This table summarises different nominal disk--average solutions for OSIRIS NAC orange--filter ($649.2\,\mathrm{nm}$) images, 
and indicates the phase angle interval used for the fit. Fo15 is \citet{fornasieretal15} and Fe16 is \citet{felleretal16}. H2002 and HHS refer 
to different versions of the Hapke--model \citep[see][for further details]{felleretal16}.}
\label{tab3}
\end{center}
\end{table}

Our nominal ($a=1$) solution has a single--scattering albedo that is somewhat above the range of other fits ($w_1=0.055$, compared to $0.034\leq w\leq 0.046$), 
and our cosine asymmetry factor is lower ($\xi_1=-0.456$, compared to $-0.42\leq\xi\leq -0.37$). However, the other parameters are within the ranges of the 
independent fits ($h_1=0.035$, compared to $0.026\leq h\leq 0.064$, and $\bar{\theta}=16.2^{\circ}$, compared to $15^{\circ}\leq\bar{\theta}\leq 28^{\circ}$). 
There are a number of potential reasons why our $\{w,\,\xi\}$--values differ compared to previous work. First, the applied photometric models are different. 
Ours is more similar to H2002, while HHS contains an additional correction for near--surface porosity \citep[see][for a description of those models]{felleretal16}. 
Second, we assume $B_0=1$, while \citet{fornasieretal15} and \citet{felleretal16} treat the opposition effect amplitude as a free parameter. They obtain values in the range 
$2.50\leq B_0\leq 2.57$, which is surprising because none of the conditions under which $B_0>1$ appear to be fulfilled by cometary grains (see Section~\ref{sec_method}). 

Third, different image sets have been used for the fits (and we have demonstrated that there is substantial albedo variegation, which could make the fitted parameters sensitive to image selection). 
Fourth, the considered phase angle intervals differ.  It is known that the phase angle coverage influences the parameter retrieval, and the parameter pair $\{\xi,\,\bar{\theta}\}$ 
is particularly difficult to disentangle \citep{helfensteinetal88}. It is perhaps significant that our $[w_0,\,w_1]$ range envelopes the other $w$--estimates, and that their upper $\alpha$--values 
are intermediate of our $a=0$ and $a=1$ cases. Overall, we do not consider the discrepancies alarming, and emphasise that the observed range of albedo variegation, 
$0.02 \stackrel{<}{_{\sim}}\mathcal{W}\stackrel{<}{_{\sim}} 0.09$, is substantially wider than the range of suggested disk--average $w$--values in Table~\ref{tab3}.

We therefore move on to consider the physical implications of our findings. The surface material of 67P displays a rather wide range ($\pm 60$ per cent) of intrinsic reflectivity. 
Although part of this variability could be due to differences in local scattering asymmetry, or variations in the degree of 
surface roughness, we collectively discuss the range of reflectance in terms of albedo variegation, for reasons of convenience. We have found a correlation between 
morphology and the single--scattering albedo proxy $\mathcal{W}$. Going from the brightest to the darkest morphologies, we find 
that smooth terrain has the highest $\mathcal{W}$ in general, but there are also examples of relatively dark smooth terrain, that 
appear associated with pit formation and morphological changes. Brittle terrain is darker still, while the strongly consolidated materials 
are the darkest of them all. \citet{moruzzietal22} also report that smooth terrain has higher albedo than consolidated terrain. 

The question is why this albedo variegation exists. One possibility is certainly compositional variations. By studying OSIRIS spectrophotometry, 
\citet{fornasieretal15} discovered that there are three groups of surface materials, in terms of visual spectral slope: 1) low slopes (11--$14\,\%\,{\rm (100\,nm)^{-1}}$) found 
in Hathor, Hapi, Seth, and Ma'at; 2) medium slopes (14--$18\,\%\,{\rm (100\,nm)^{-1}}$) found in Anuket and Serqet; 3) steep slopes ($\geq 18\,\%\,{\rm (100\,nm)^{-1}}$) found 
in Apis, Nut, Maftet, Ma'at, and Bastet. Studies of unusually bright patches and boulders on 67P \citep{pommeroletal15,deshapriyaetal18} have shown that their visual spectral slopes  
are low \citep[sometimes neutral;][]{fornasieretal15,oklayetal16}, and VIRTIS observations have shown that such features display a $3\,\mathrm{\mu m}$ absorption band due to water 
ice \citep{baruccietal16}. The consensus is therefore that low visual spectral slope signifies a larger surface exposure of water ice. This idea is reinforced by the fact that the 
nucleus experiences an overall reduction of spectral slope as it approached perihelion \citep{fornasieretal16}, presumably because the dust mantle becomes thinner and unresolved patches 
of exposed water ice becomes more common. Yet, there is no convincing correlation between visual spectral slope (thus, water abundance) and morphology types. Most steep slopes do occur in consolidated 
terrain, suggesting that many of those have ice--poor surfaces. However, the large Hathor cliff consists of consolidated material, yet it is as spectrally bluish as the smooth material in Hapi \citep{fornasieretal15}. This suggests 
that near--surface water ice may be found in all types of terrain, regardless their morphology, as long as the local illumination conditions allow for the presence of observable amounts of ice. 

The observations by \emph{Rosetta}/VIRTIS revealed the presence of a 2.9--$3.6\,\mathrm{\mu m}$ absorption feature \citep{capaccionietal15}, attributed to a mixture of 
C--H stretching vibrations in aliphatic and aromatic hydrocarbons, OH--groups in carboxylic acids, N--H groups in ammonium ion ($\mathrm{NH_4^+}$) salts, and hydroxylated 
amorphous silicates \citep{capaccionietal15,quiricoetal16,mennellaetal20,pochetal20,raponietal20}. \citet{filacchioneetal16b} found that the band depth varies slightly across 
the nucleus and defined three classes: \#1) shallow band depth ($<10$ per cent) common in Aker, Anubis, Atum, Bastet, Hatmehit, Maftet, and Nut; \#2) medium band depth (10--12 per cent) 
common in Anuket, Apis, Ash, Aten, Imhotep, Khepry, and Ma'at; \#3) deeper band depth ($>12$ per cent) common in Babi, Hapi, Hathor, and Seth. Within each group is a mixture of smooth and 
consolidated terrains. The only aspect that apparently is common is nucleus latitude: class \#1 is predominantly located near the equator and stretching towards latitude $\sim 30^{\circ}\,\mathrm{S}$, 
class \#3 stretches from the equator to high northern latitudes, while class \#2 terrains often are near--equatorial, although some stretch up to $60^{\circ}$ either north or south. 
\citet{filacchioneetal16b} point out that the deeper absorption feature in class \#3 terrains additionally is shifted towards shorter wavelengths, and they attribute both features to 
an enhanced presence of water ice. Indeed, classes \#1 and \#3 roughly correspond to the steep and low slopes of \citet{fornasieretal15}, respectively. Thus, if there is a \emph{compositional} 
component that differs substantially between smooth and consolidated terrain, and strongly affects their reflectances, it does so without affecting the visual spectral slope or the 
NIR absorption feature. In short, we do not find compelling evidence for a compositional origin of the albedo variegation. 

Therefore, we suspect that differences in the \emph{structural} properties of the grain aggregates may be responsible for the albedo variegation. A clue may be provided by \citet{davidssonetal22b}, 
who analysed \emph{Rosetta}/MIRO observations of Pit\#2 acquired in October and November 2014, corresponding to the `tongue shaped' photometric anomaly in the middle panels of 
Fig.~\ref{fig_image_albedo_F22_01}. In October they found that MIRO observations were consistent with a solid--state greenhouse effect in the dust mantle, 
water ice sublimation at a depth of $\sim 0.02\,\mathrm{m}$, $\mathrm{CO_2}$ sublimation at a depth of $0.48\,\mathrm{m}$, a thermal inertia of about 
$20$--$40\,\mathrm{J\,m^{-2}\,K^{-1}\,s^{-1/2}=MKS}$, gas diffusivity corresponding to channels in the millimetre--centimetre range, and millimetre (MM) and sub--millimetre (SMM) 
extinction coefficients of $E_{\rm MM}=25\,\mathrm{m^{-1}}$ and $E_{\rm SMM}=60\,\mathrm{m^{-1}}$. By November, the water ice had withdrawn to a depth of $0.21\,\mathrm{m}$, 
the $\mathrm{CO_2}$ was at $0.44\,\mathrm{m}$ and the overlying dust mantle had experienced strong increases in thermal inertia (to $\sim 110\,\mathrm{MKS}$ at 
depth, though it was merely  $\sim 65\,\mathrm{MKS}$, in the top $8\,\mathrm{mm}$) and extinction coefficients (to $E_{\rm MM}=80\,\mathrm{m^{-1}}$ and 
$E_{\rm SMM}=600\,\mathrm{m^{-1}}$), a drastic reduction of diffusivity (corresponding to micrometre channels), the disappearance of the solid--state greenhouse effect, and the 
emergence of sub--millimetre scattering. The changes in thermal inertia, diffusivity, and optical properties are all consistent with a substantial compaction of the dust mantle, caused 
by a collapse (subsidence) that significantly reduced its porosity. 

We do not find a significant photometric anomaly in August 2014, but it is present in December 2014. We here hypothesise that the observed local darkening in some smooth material 
(associated with pit formation and morphological changes) is caused by the previously mentioned compaction, inferred from MIRO--data analysis. If so, it should have taken place at some point between October and 
November 2014, as suggested by the MIRO observations (unfortunately, OSIRIS did not image the Pit\#1--\#2 regions 
in September--November 2014). For granular media with low albedo and a relatively high porosity, compaction typically leads to an \emph{increase} of 
reflectance \citep{hapke08}, which is opposite to that we propose. However, granular media with low albedo and a porosity of $\psi \stackrel{<}{_{\sim}} 0.5$ instead \emph{darkens} during 
compaction, due to coherent effects between small grains in close contact that start to act as larger optically effective particles \citep{hapke08}. A reduction of reflectivity is often observed 
when grain sizes increase \cite[e.~g.,][]{capaccionietal90,sunetal16,cloutisetal18}. This effect may be particularly strong in comets compared to other small bodies covered by regolith, because 
the vast majority of cometary monomer grains \citep[sizes ranging 0.1--$10\,\mathrm{\mu m}$;][but with smaller grains outnumbering larger ones]{bentleyetal16} are smaller than 
or similar to visual wavelengths, while regolith typically is coarser \citep[e.~g., the median grain size in various lunar regolith 
samples ranges 30--$110\,\mathrm{\mu m}$;][]{mckayetal74}. Therefore, the degree of clumping in cometary granular materials may have significant optical effects. We therefore propose, that the 
observed photometric anomaly in Hapi is caused by a local compaction of surface grains to a sufficiently low porosity that coherent effects have started to dominate the optical photometric 
behaviour of the material.

If this hypothesis is true, the consolidated terrains might consist of even denser grain agglomerates (i.~e., larger optically effective particles), that are darker than average because of particularly strong coherent effects. 
The \emph{Philae} lander, which came to rest at Abydos \citep[in Wosret, close to the boarders with Hatmehit and Bastet, see Fig.~1 in][]{fornasieretal21} performed several measurements on consolidated material 
that suggests it is rather compact. The diurnal temperature curve measured by MUPUS is consistent with a thermal inertia of $85\pm 35\,\mathrm{MKS}$ \citep{spohnetal15}, 
which is higher than estimated for smooth terrains \citep[10--$30\,\mathrm{MKS}$;][]{schloerbetal15}, even in the compacted form mentioned above (if considering the near--surface value of 
$\sim 65\,\mathrm{MKS}$). The most important mechanism to increase the thermal inertia of materials is to reduce their porosity \citep[e.~g.,][]{shoshanyetal02}, and \citet{spohnetal15} suggest that the porosity is 
$0.3\leq\psi\leq 0.65$ at Abydos. Furthermore, SESAME--PP measurements presented by \citet{lethuillieretal16} show that the 409--$758\,\mathrm{Hz}$ dielectric constant 
is $\varepsilon'=2.45\pm 0.2$ within the top metre, which indicates that the porosity (for a carbonaceous chondrite analogue material) should be $\psi<0.5$. Micrometre--sized grains 
of intrinsically dark material, compressed nearly to close--packing density, could reflect light very poorly.

We find that brittle material is less dark than consolidated material. If our hypothesis is correct, the increased reflectivity is a consequence of a lower level of compaction. Because the 
level of compaction is smaller than for consolidated material, it also means that the tensile strength is lower, which could explain why it crumbles and falls apart comparably 
easily. 

There is a genetic relationship between smooth terrains and the other morphologies discussed here. Consolidated terrain is prone to fracturing, as revealed by the 
OSIRIS cameras \citep{elmaarryetal15b}. The consolidated terrain on the southern hemisphere, that is fully illuminated at perihelion and strongly active, fractures into 
macroscopic chunks in the millimetre--decimetre size range, which are ejected into the coma, where OSIRIS documented such large structures in mid--flight 
\citep[e.~g.,][]{rotundietal15,davidssonetal15b}. Some of these chunks land on the less active northern hemisphere as airfall \citep{thomasetal15b,kelleretal15,kelleretal17,davidssonetal21}, that 
has been resolved by the ROLIS camera on \emph{Philae} at Agilkia \citep{mottolaetal15}, as well as by OSIRIS during the final landing of \emph{Rosetta} at Sais \citep{pajolaetal17b}, both 
locations belonging to the Ma'at region on the small lobe. That airfall process is the very \emph{origin} of the smooth terrains. But if smooth--terrain chunks constitute broken--off pieces of 
consolidated terrain material, how come they are brighter than the dark regions they originate from?

We know that the airfall chunks contain abundant water ice when they land. This is evidenced by the strong activity of the northern hemisphere on the inbound orbit, and 
the production of gas and dust from the smooth terrains in Hapi is particularly strong \citep{sierksetal15,hassigetal15,gulkisetal15}. Furthermore, \citet{davidssonetal21} presented thermophysical simulations 
showing substantial water ice retention in coma chunks as small as $1\,\mathrm{cm}$, for flight times as long as $12\,\mathrm{h}$. It is evident that water ice does not mix with the refractories 
up to the very surfaces of the chunks -- if that had been the case, both consolidated and smooth terrains should be much brighter, and display measurable water ice absorption features. In addition, once the 
superficial ice is gone, the smooth material chunks would be darker than the consolidated terrains whence they came, which is the opposite of what we observe. Nevertheless, we 
suggest that the removal of ice from their interiors, causes a structural change that leads to brightening. Specifically, the removal of water ice reduces the cohesion of the chunks, whose 
structural integrity then only depends on the van~der~Waals forces between refractory grains in the dusty mantles that surround the icy cores. The vapour pressure of escaping gases 
might expand the chunk by puffing it up. Doing so increases the porosity of the chunk, which would lead to a reduction of the coherent effect and an increase of the reflectivity (as long as the 
porosity remains at $\psi \stackrel{<}{_{\sim}}0.5$).

\begin{table*}
\begin{center}
\begin{tabular}{||l|l|l|l|l|l|l|l|l||}
\hline
\hline
Object & $w$ & $h$ &  $\xi$ & $\bar{\theta}\,\mathrm{[deg]}$ & $h_{\rm c}$ & $B_0$ & $B_{0,{\rm c}}$ & Reference\\
\hline
19P/Borrelly & $0.020\pm 0.004$ & $0.0084\pm 0.005$ & $-0.45\pm 0.05$ & $20\pm 5$ & -- & $1.0\pm 0.5$ & -- & \citet{burattietal04}\\
67P & $0.055$ & $0.035$ & $-0.456$ & 16 & -- & 1.0 & -- & This study\\
Phoebe & 0.07 & 0.04 & $-0.20$ & 33 & -- & $>1.0$ & -- & \citet{simonellietal99}\\
 & & & & & & & & \citet{burattietal08}\\
Moon & 0.25 & 0.05 & $-0.25$ & 20 & -- & 1.0 & -- & \citet{buratti85}\\
 & & & & & & & & \citet{hillieretal99}\\
Vesta & 0.49 & 0.076 & $-0.23$ & 8 & -- & 1.66 & -- & \citet{lietal13}\\
Charon & 0.72 & 0.150 & $-0.09$ & 23 & 0.037 & 0.001 & 0.536 & \citet{burattietal19}\\
Rhea & 0.861 & 0.08 & $-0.29$ & $13\pm 5$ & -- & 1.37 & -- & \citet{verbiscerveverka89}\\
  & 0.989 & 0.0004 & +0.2 & 33 & -- & 1.8 & -- & \cite{ciarnielloetal11}\\
Pluto & 0.95 & 0.108 & $-0.21$ & 19 & $\leq 0.00146$ & 0.33 & 0.99 & \citet{hillieretal21}\\
Europa & 0.964 & 0.0016 & $-0.15\pm 0.04$ & 10 & -- & 0.5 & -- & \citet{buratti85}\\
 & & & & & & & & \citet{domingueetal91}\\ 
Iapetus, low  & & &   & 6 & & & & \citet{leeetal10}\\ 
albedo hemisphere  & & & & & & & &\\ 
\hline 
\hline
\end{tabular}
\caption{A comparison of Hapke parameters for a selection of comets, asteroids, and satellites. Hapke's later models have separate parameters for the shadow--hiding ($h,\,B_0$) and coherent backscatter
portions of the opposition surge ($h_{\rm c},\,B_{0,{\rm c}}$). We note that \citet{masoumzadehetal19} found evidence of a  shadow--hiding but not a coherent backscatter opposition effect on 67P. Our omittance 
of the latter is therefore motivated.}
\label{tab4}
\end{center}
\end{table*}

In Wosret we also observe albedo variegation within consolidated terrain, specifically, the southern half is measurably brighter than the northern half of Wosret (except where the latter 
locally is covered by patches of smooth material) and neighbouring Bastet. These results feed into an ongoing discussion about layering within the comet, and their spectrophotometric properties, 
that we first need to summarise briefly. One of the most surprising discoveries at 67P was the prominent cliffs and plateaus that form systems of terraces of various sizes, 
that are found in numerous locations on both lobes \citep{sierksetal15}. By mapping the location of plateaus, \citet{massironietal15} demonstrated that their placement is 
consistent with two sets of concentric shells, one centred at the core of the large lobe, and the other centred at the core of the small lobe. Each ellipsoidal shell would presumably 
have been located interior to the original shape of each lobe. But because of excavation, through massive ejection events or gradual sublimation, many layers that originally 
were buried at depths of several hundreds of metres, are now exposed at the surface. The cuesta--like morphology on the current nucleus was attributed to differential 
sublimation, caused by varying abundances of volatiles within each layer by \citet{massironietal15}. The description of the internal stratification, based on their terrace--like surface expressions, 
was further discussed and extended by \citet{penasaetal17}.

Two studies have focused on the spectrophotometric properties of the nucleus in the context of the aforementioned stratification: \citet{ferrarietal18} considered the northern part of the big lobe, 
and \citet{tognonetal19} studied the southern parts of both lobes. Both studies divided exposed layers into an `inner class' and an `outer class' based on their elevations above the 
respective lobe centre. The large lobe has the inner class at approximately 1.6--$2.2\,\mathrm{km}$ elevation and the outer class at 2.2--$2.5\,\mathrm{km}$ elevation, while the small lobe 
has the corresponding classes at 1.2--$1.6\,\mathrm{km}$ and 1.6--$1.8\,\mathrm{km}$ elevations, respectively. On the northern large lobe, the inner class primarily corresponds to brittle 
terrain in Imhotep, while the outer class corresponds to consolidated terrain in Atum, Ash, and Apis. On the southern small lobe, the inner class primarily corresponds to Nieth, Anuket, and 
the Wosret sub--unit b (i.~e., its southern part), while the outer class includes Bastet, Maftet, and the northern half of Wosret. On the southern large lobe, Anhur contains inner--class terrain, 
while Geb and parts of Bes constitute outer class terrain. No correlation between spectral slope and inner/outer classes are found on the northern large lobe \citep{ferrarietal18}. For the southern 
hemisphere, the slopes are nearly the same for both classes, except at $\lambda\geq 0.9\,\mathrm{\mu m}$, where the inner class has somewhat smaller spectral slopes \citep{tognonetal19}.

There is, however, one property that differs systematically: the reflectance of inner class terrains is higher than that of the outer class terrains by 60--90 per cent \citep{ferrarietal18,tognonetal19}. 
Our study confirms this important result. A detailed comparison between elevation maps (with respect to the lobe centres) presented in these papers \citep{penasaetal17,ferrarietal18,tognonetal19} and the terrains in 
Figs.~\ref{fig_image_albedo_F22_03} and \ref{fig_image_albedo_F22_04} confirms that the purple regions (comparably high $\mathcal{W}$) correspond to relatively low elevations, 
and the blue regions (lowest $\mathcal{W}$) correspond to relatively high elevations. Specifically: 1) the brittle terrains in Imhotep sub--unit c belong to some of the deepest 
parts of the nucleus, stratigraphically located $\sim 500\,\mathrm{m}$ below the regions of Ash in the same image (Fig.~\ref{fig_image_albedo_F22_03}, upper panels); 2) the border area between 
Bes sub--units a and b (relatively high $\mathcal{W}$) is stratigraphically located $\sim 300\,\mathrm{m}$ below the low--$\mathcal{W}$ joint between Khonsu sub--unit c and Bes sub--unit a 
 (Fig.~\ref{fig_image_albedo_F22_04}, upper panels); 3) the purple (higher--$\mathcal{W}$) Wosret sub--unit b is stratigraphically located $\sim 700\,\mathrm{m}$ below the blue (lower--$\mathcal{W}$) 
Wosret sub--unit c and Bastet sub--unit a (Fig.~\ref{fig_image_albedo_F22_04}, lower panels).

If our albedo--porosity hypothesis is valid, the observed increase of brightness with increasing depth under the original surface, would mean that the porosity increases 
systematically with depth.  That is to say, a porous and brighter core is surrounded by a more compacted and darker shell (although the transition might be more gradual 
than suggested by a core--mantle structure).  In this context, the observations of \emph{Rosetta}/CONSERT of the interior of the small lobe \citep{kofmanetal15} are interesting. 
A detailed analysis of the radio waves transmitted through portions of the small lobe, allowed for the determination of the dielectric constant, that was found to decrease with 
depth \citep{ciarlettietal15,ciarlettietal18}. Specifically, they found that the dielectric constant decreased from $\varepsilon'=1.7$ near the surface, to $\varepsilon'=1.3$ at 
a depth of $\sim 150\,\mathrm{m}$. That suggests a gradual increase of porosity and/or ice abundance with depth \citep{ciarlettietal15,ciarlettietal18}. We note that \citet{ferrarietal18} 
discussed several possible explanations for the difference in reflectance between inner and outer classes, including a potential correlation 
between brightness, porosity, and depth, without favouring any particular explanation.

In accordance with our interpretation, the differential erosion that is responsible for the creation of terraces, is not due to variation in ice abundance 
(i.~e., rapidly eroding layers are ice--rich) but due to porosity and strength (i.~e., rapidly eroding layers are more porous and weaker then slowly eroding layers). Alternatively, 
the particle--sizes in inner--class regions may be smaller than in outer--class regions, without necessarily having different bulk porosity. 

Table~\ref{tab4} compares our disk--average solution to those of a selection of comets, asteroids, and satellites (arranged in order of increasing single--scattering albedo). 
The width of the opposition peak (measured by $h$) of 67P is not as wide as for bright objects, but similar to those of darker ones. Our value for the cosine asymmetry of 67P 
is substantially lower than for most brighter objects, but is very similar to that of Comet 19P/Borrelly \citep{burattietal04}. Macroscopic roughness, which incorporates all size--scales 
within the geometric optics limit, from clumps of particles to mountains and craters, offers clues to geophysical processes on planetary surfaces. The mean slope angle of 
$\bar{\theta}=16^{\circ}$ for 67P is somewhat low for a small body. Previous analyses of planetary surfaces with a dearth of rough facets showed that Europa, the low albedo 
hemisphere of Iapetus, and some asteroids exhibit small slope angles, although asteroidal roughness is typically in the 18--$29^{\circ}$ range \citep{burattietal04}. Europa's surface 
has undergone recent resurfacing with a resulting low mean roughness \citep{buratti85,domingueetal91}. The low--albedo hemisphere of Iapetus exhibits an extremely smooth 
surface, which \citet{leeetal10} attribute to infilling of rough facets by Saturn's Phoebe ring. We attribute the rather low roughness of the northern hemisphere of 67P to the airfall process 
\citep{thomasetal15b,kelleretal15,kelleretal17,davidssonetal21}, in which dust particles from active vents are reaccreted back onto the comet's surface.

\section{Conclusions} \label{sec_conclusions}

We have analysed a number of images of Comet 67P acquired by the OSIRIS NAC camera in the orange filter. 
A standard photometric model has been fitted to the data in order to obtain disk--average parameters. Discrepancies between 
the radiance factors of the images with respect to the best fit average model have then been used as a measure of the 
albedo variegation of the nucleus, expressed by the single--scattering albedo proxy $\mathcal{W}$. Our main conclusions are summarised as follows.

\begin{enumerate}
\item Albedo variegation on Comet 67P is substantial and falls within the range $0.02 \stackrel{<}{_{\sim}}\mathcal{W}\stackrel{<}{_{\sim}} 0.09$.
\item Most smooth terrains have higher than average $\mathcal{W}$. 
\item Certain areas within smooth terrain appear to darken prior to the onset of morphological changes (manifested as expanding 
shallow pits and moving escarpments).
\item Most consolidated terrains have lower than average $\mathcal{W}$. 
\item Brittle and consolidated material that constitutes stratigraphically low terrain (in the context of the onion--like layering of each nucleus lobe) 
is brighter (has higher $\mathcal{W}$--values) than similar but stratigraphically higher terrain. 
\item We do not find compelling evidence for systematic compositional differences between consolidated, brittle, and smooth material. We therefore 
propose that the observed albedo variegation is caused by variations in porosity, and therefore, the coherent effect (due to compaction, small bright particles 
start acting as larger and darker optically effective particles). 
\end{enumerate}

\section*{Acknowledgements}

Parts of this research were carried out at the Jet Propulsion Laboratory, California Institute of Technology, under a 
contract with the National Aeronautics and Space Administration. The authors acknowledge funding from 
NASA grant 106843--231402.02.11.01.02 awarded by the \emph{Rosetta} Data Analysis Program. We are indebted to 
Dr. Xian Shi, Max Planck Institute for Solar System Research, G\"{o}ttingen, Germany, for providing the software used to 
read *.img files. We thank Dr. Pedro J. Guti\'{e}rrez from Instituto de Astrof\'{i}sica de Andaluc\'{i}a, Granada, Spain, 
for valuable discussions, and Dr. Stefano Mottola, DLR, Berlin, Germany, for carefully reviewing our paper. 
OSIRIS was built by a consortium led by the Max--Planck--Institut f\"{u}r Sonnensystemforschung, G\"{o}ttingen, Germany, in collaboration 
with CISAS, University of Padova, Italy, the Laboratoire d'Astrophysique de Marseille, France, the Instituto de Astrof\'{i}sica de Andaluc\'{i}a, 
CSIC, Granada, Spain, the Scientific Support Office of the European Space Agency, Noordwijk, The Netherlands, the Instituto Nacional 
de T\'{e}cnica Aeroespacial, Madrid, Spain, the Universidad Polit\'{e}chnica de Madrid, Spain, the Department of Physics and Astronomy 
of Uppsala University, Sweden, and the Institut f\"{u}r Datentechnik und Kommunikationsnetze der Technischen Universit\"{a}t Braunschweig, 
Germany. The support of the national funding agencies of Germany (DLR), France (CNES), Italy (ASI), Spain (MEC), Sweden (SNSB), and the 
ESA Technical Directorate is gratefully acknowledged. We thank the \emph{Rosetta} Science Ground Segment at ESAC, the \emph{Rosetta} Mission Operations 
Centre at ESOC and the \emph{Rosetta} Project at ESTEC for their outstanding work enabling the science return of the \emph{Rosetta} Mission.\\

\noindent
\emph{COPYRIGHT}.  \textcopyright\,2022. California Institute of Technology. Government sponsorship acknowledged.

\section*{Data Availability}

The data underlying this article will be shared on reasonable request to the corresponding author.

\bibliography{MN-22-1886-MJ_R1.bbl}

\appendix

\section{Roughness} \label{appendix01}

We here provide the expressions $\mu_0'(\bar{\theta},\,i,\,e,\,\alpha)$, $\mu'(\bar{\theta},\,i,\,e,\,\alpha)$, and $S(\bar{\theta},\,i,\,e,\,\alpha)$, used in equation~(\ref{eq:01}). 
They were originally formulated by \citet{hapke84}. The auxiliary parameters $\Psi$ and $f$ are given by
\begin{equation} \label{eq:A01}
\cos\Psi=\frac{\cos\alpha-\cos i\cos e}{\sin i\sin e}
\end{equation}
and
\begin{equation} \label{eq:A02}
f(\Psi)=\exp\left(-2\tan\frac{\Psi}{2}\right).
\end{equation}
Two cases need to be considered, depending on whether the inclination $i$ is smaller or larger than the emergence angle $e$. 
In case $i<e$ we have
\begin{equation} \label{eq:A03}
S(\bar{\theta},\,i,\,e,\,\alpha)=\frac{\mu'\mu_0}{\mu'^0\mu_0'^0}\frac{1}{\sqrt{1+\upi\tan^2\bar{\theta}}}\left\{1-f+\frac{f}{\sqrt{1+\upi\tan^2\bar{\theta}}}\frac{\mu_0}{\mu_0'^0}\right\}^{-1},
\end{equation}
where
\begin{equation} \label{eq:A04}
\begin{array}{c}
\displaystyle\mu_0'=\frac{1}{\sqrt{1+\upi\tan^2\bar{\theta}}}\Bigg\{\cos i+\sin i\tan\bar{\theta}\times\\
\\
\displaystyle\frac{\cos\Psi\exp\left(-\frac{\cot^2\bar{\theta}\cot^2e}{\upi}\right)+\sin^2\frac{\Psi}{2}\exp\left(-\frac{\cot^2\bar{\theta}\cot^2i}{\upi}\right)}{2-\exp\left(-\frac{2\cot\bar{\theta}\cot e}{\upi}\right)-\frac{\Psi}{\upi}\exp\left(-\frac{2\cot\bar{\theta}\cot i}{\upi}\right)}\Bigg\},
\end{array}
\end{equation}
and
\begin{equation} \label{eq:A05}
\begin{array}{c}
\displaystyle \mu'=\frac{1}{\sqrt{1+\upi\tan^2\bar{\theta}}}\Bigg\{\cos e+\sin e\tan\bar{\theta}\times\\
\\
\displaystyle \frac{\exp\left(-\frac{\cot^2\bar{\theta}\cot^2e}{\upi}\right)-\sin^2\frac{\Psi}{2}\exp\left(-\frac{\cot^2\bar{\theta}\cot^2i}{\upi}\right)}{2-\exp\left(-\frac{2\cot\bar{\theta}\cot e}{\upi}\right)-\frac{\Psi}{\upi}\exp\left(-\frac{2\cot\bar{\theta}\cot i}{\upi}\right)}\Bigg\},
\end{array}
\end{equation}
and
\begin{equation} \label{eq:A06}
\mu_0'^0=\frac{1}{\sqrt{1+\upi\tan^2\bar{\theta}}}\left\{\cos i+\sin i\tan\bar{\theta}\frac{\exp\left(-\frac{\cot^2\bar{\theta}\cot^2i}{\upi}\right)}{2-\exp\left(-\frac{2\cot\bar{\theta}\cot i}{\upi}\right)}\right\},
\end{equation}
and
\begin{equation} \label{eq:A07}
\mu'^0=\frac{1}{\sqrt{1+\upi\tan^2\bar{\theta}}}\left\{\cos e+\sin e\tan\bar{\theta}\frac{\exp\left(-\frac{\cot^2\bar{\theta}\cot^2e}{\upi}\right)}{2-\exp\left(-\frac{2\cot\bar{\theta}\cot e}{\upi}\right)}\right\}.
\end{equation}
In case $i\geq e$, the expressions for $\mu_0'^0$ and $\mu'^0$ remain the same. However, $S$, $\mu_0'$, and $\mu'$ change as follows,
\begin{equation} \label{eq:A08}
S(\bar{\theta},\,i,\,e,\,\alpha)=\frac{\mu'\mu_0}{\mu'^0\mu_0'^0}\frac{1}{\sqrt{1+\upi\tan^2\bar{\theta}}}\left\{1-f+\frac{f}{\sqrt{1+\upi\tan^2\bar{\theta}}}\frac{\mu}{\mu'^0}\right\}^{-1},
\end{equation}
and
\begin{equation} \label{eq:A09}
\begin{array}{c}
\displaystyle \mu_0'=\frac{1}{\sqrt{1+\upi\tan^2\bar{\theta}}}\Bigg\{\cos i+\sin i\tan\bar{\theta}\times\\
\\
\displaystyle \frac{\exp\left(-\frac{\cot^2\bar{\theta}\cot^2i}{\upi}\right)-\sin^2\frac{\Psi}{2}\exp\left(-\frac{\cot^2\bar{\theta}\cot^2e}{\upi}\right)}{2-\exp\left(-\frac{2\cot\bar{\theta}\cot i}{\upi}\right)-\frac{\Psi}{\upi}\exp\left(-\frac{2\cot\bar{\theta}\cot e}{\upi}\right)}\Bigg\},
\end{array}
\end{equation}
and
\begin{equation} \label{eq:A10}
\begin{array}{c}
\displaystyle \mu'=\frac{1}{\sqrt{1+\upi\tan^2\bar{\theta}}}\Bigg\{\cos e+\sin e\tan\bar{\theta}\times\\
\\
\displaystyle \frac{\cos\Psi\exp\left(-\frac{\cot^2\bar{\theta}\cot^2i}{\upi}\right)+\sin^2\frac{\Psi}{2}\exp\left(-\frac{\cot^2\bar{\theta}\cot^2e}{\upi}\right)}{2-\exp\left(-\frac{2\cot\bar{\theta}\cot i}{\upi}\right)-\frac{\Psi}{\upi}\exp\left(-\frac{2\cot\bar{\theta}\cot e}{\upi}\right)}\Bigg\}.
\end{array}
\end{equation}

\bsp	
\label{lastpage}
\end{document}